\RequirePackage{lineno}

\documentclass[onecolumn,floatfix,superscriptaddress,a4paper,
               showpacs,showkeys,nofootinbib,preprint]{revtex4}

\usepackage{epsfig}
\usepackage{latexsym}
\usepackage{xspace}
\usepackage{hyperref}
\usepackage[latin2]{inputenc}
\usepackage{indentfirst}
\usepackage{enumerate}

\usepackage{amsmath}
\usepackage{amssymb}
\usepackage[english]{babel}
\usepackage{url}

\newcommand{\eq}[1]{\begin{align} #1 \end{align}}

\begin{document}
\today
\vspace*{1cm}

\title{Recent Developments in the Study of Deconfinement in Nucleus-Nucleus Collisions }
\author{M. Gazdzicki}
 \affiliation{Goethe--University, Frankfurt, Germany}
 \affiliation{Jan Kochanowski University, Kielce, Poland}
\author{M.I. Gorenstein}
 \affiliation{Bogolyubov Institute for Theoretical Physics, Kiev, Ukraine}
 \affiliation{Frankfurt Institute for Advanced Studies, Frankfurt, Germany}
\author{P. Seyboth}
\affiliation{Max-Planck-Institut fuer Physik, Munich, Germany}
 \affiliation{Jan Kochanowski University, Kielce, Poland}

\begin{abstract}
Deconfinement refers to the creation of a state of quasi-free quarks
and gluons in strongly interacting matter.
Model predictions and experimental evidence for
the onset of deconfinement in nucleus-nucleus collisions were discussed in our first
review on this subject.
These results motivated further experimental and theoretical studies.
This review addresses two subjects.
First, a summary of the past, present and future experimental
programmes related to discovery and study of properties of
the onset of deconfinement are
presented.
Second, recent progress is reviewed on analysis methods
and preliminary experimental results for new strongly intensive fluctuation measures
are discussed,
which are relevant for current and future studies of
the onset of deconfinement and searches for the critical point
of strongly interacting matter.
\end{abstract}

\pacs{12.40.-y, 12.40.Ee}

\keywords{Onset of deconfinement, nucleus-nucleus collisions, event-by-event flcutuations}

\maketitle

\newpage

\tableofcontents
\newpage

\vspace{3cm}

\section{Introduction}

\begin{figure}[!htb]
\includegraphics[width=0.80\textwidth]{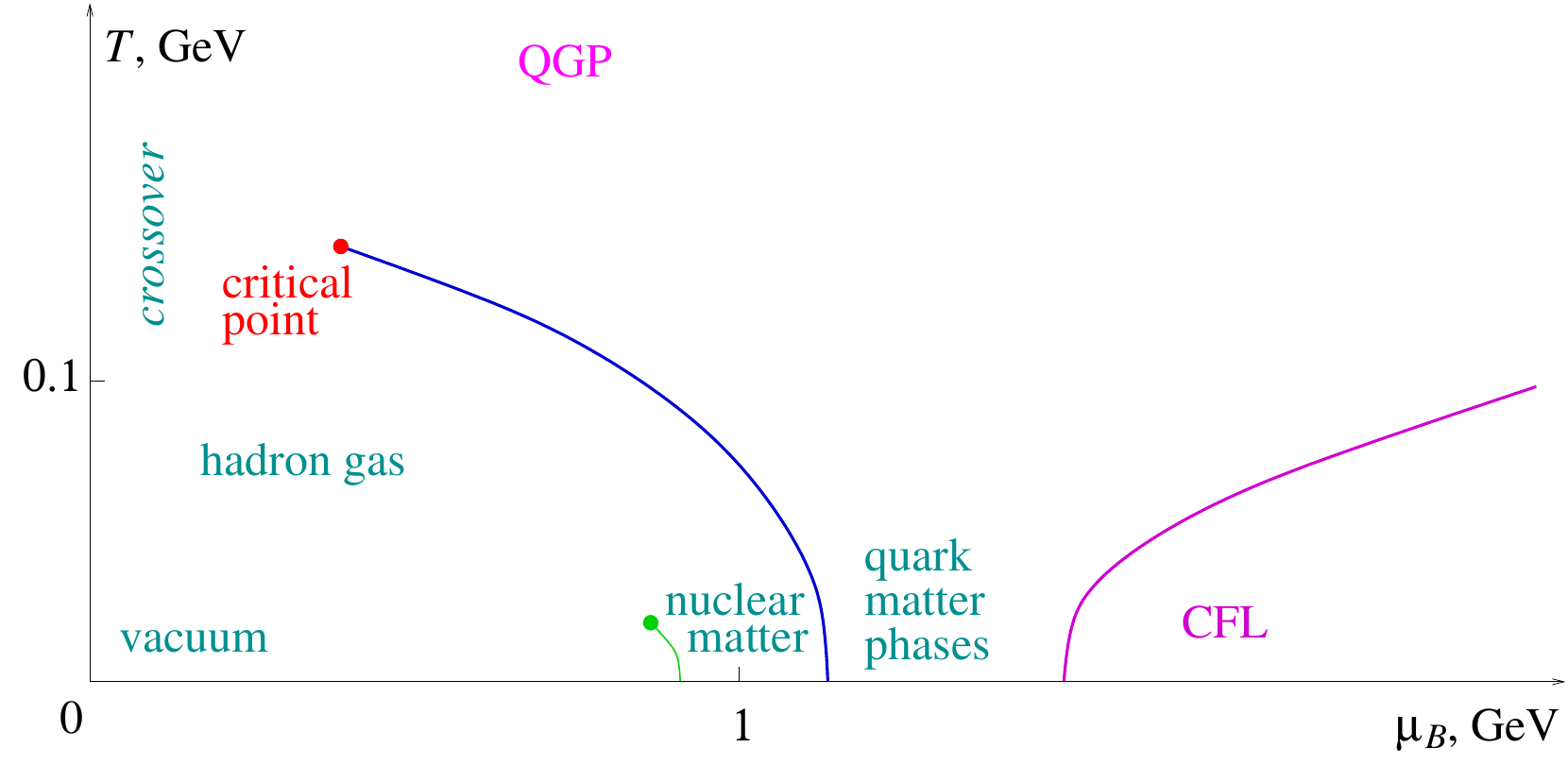}
\caption{\label{phase_diag}
Semi-quantitative sketch of the phase diagram of strongly interacting
matter~\cite{Stephanov_2006}.
}
\end{figure}

Ordinary hadron matter is believed to be composed of hadrons in which constituent
quarks and gluons are confined. At high temperature and/or pressure hadrons will
be so densely packed that they are expected to dissolve into a new phase of
quasi-free quarks and gluons, the quark-gluon-plasma (QGP). Theoretical considerations
lead to a phase diagram of strongly interacting matter as shown schematically in
Fig.~\ref{phase_diag}~\cite{Stephanov_2006}. The existence region of hadrons at
low temperature $T$ and high baryochemical potential $\mu_B$ is believed to be separated
from the QGP at high $T$ by a first order phase boundary, which ends with decreasing
$\mu_B$ in a critical point and then turns into a crossover transition. Experimentally
the phase diagram has been explored by studying the final states produced in
nucleus-nucleus collisions. The location of the freeze-out points is obtained from
fitting the measured ratios of particle yields to a statistical model with $T$ and $\mu_B$
as parameters~\cite{Becattini_2004}. It was found that with increasing collision energy
$T$ increases and $\mu_B$ decreases along a smooth freeze-out curve~\cite{Cleymans_2006}.

Experimental searches for QGP signals
started at the Super Proton
Synchrotron (SPS) of the European Organization for Nuclear
Research (CERN) and the Alternating Gradient Synchrotron (AGS) of
Brookhaven National Laboratory (BNL) in the mid 1980s. Today they
are pursued also at much higher collision energies at the
Large Hadron Collider (LHC) at CERN and at the Relativistic
Heavy Ion Collider (RHIC) at BNL.

Most probably the QGP is formed at the early stage
of heavy ion collisions already at the top SPS energy~\cite{qgp_sps}.
Unambiguous evidence of the QGP state is, however,
nearly impossible
to obtain since quantitative predictions of proposed QGP
signals~\cite{qgp_signals,qgp_signals_1,qgp_signals_2}
from the theory of strong interactions are difficult.
For this reason the NA49 Collaboration at the CERN SPS has
searched over the past years for signs of the onset of QGP
creation in the energy dependence of hadron production properties.
This search was motivated by a statistical model~\cite{GaGo}
showing that the onset of deconfinement should lead to rapid
changes of the energy dependence of numerous experimentally
detectable properties of the collisions, all appearing in a common
energy domain. The predicted features were
observed~\cite{:2007fe,:2007fe1} and dedicated experiments now continue
detailed studies in the energy region of the onset of
deconfinement.
Model predictions and experimental evidence for
the onset of deconfinement were discussed in our first
review on this subject~\cite{review}.

The present review addresses two subjects which have seen significant
advances over the last years.

First, present and future experimental
programmes related to the discovery and study of the onset of
deconfinement are  summarized.
We start with a historical sketch of the NA49 beam energy
scan with central Pb+Pb collisions which led to the
discovery of the onset of deconfinement.
Next we review the experimental programmes which are presently under
way: NA61/SHINE at the CERN SPS, BES at the BNL RHIC and
Heavy Ions at the CERN LHC. We close the chapter by summarizing the
well advanced work on two new large facilities,
MPD at JINR NICA and CBM at FAIR SIS-100.

Second, we review recent progress on analysis methods of
event-by-event fluctuations in nucleus-nucleus collisions,
which are relevant for the current and future studies of
the onset of deconfinement and the search for the critical point.
The experimental and theoretical progress in studies
of event-by-event fluctuations seems to have the highest
priority in the field of heavy ion collisions.
After more than 40 years of experimental efforts we have now
rich and relatively well understood
experimental data on single particle spectra in Pb+Pb and
p+p interactions from several GeV to several TeV.
But, mostly due to the incomplete acceptance of detectors and
not well developed data analysis tools,
results on event-by-event fluctuations are not yet mature.
High-quality and systematic measurements of event-by-event
fluctuations are needed mostly for two reasons:\\
-- to test statistical and dynamical models - the simplest tests
at the level of fluctuations are still missing and \\
--  to study properties of the phase transition, in particular
properties of the transition at the onset of deconfinement
and search for the critical point of strongly interacting matter.

\section{Review of experimental programmes}

In this section  the  present and future experimental
programmes related to the study of properties of
the onset of deconfinement are reviewed.
First, however, the history of the programme which led to the
discovery of the onset of deconfinement, namely
the beam energy scan programme
with central Pb+Pb collisions at the CERN SPS
is briefly recalled.

\subsection{First Pb-beam Programme at the CERN SPS -
Observation of the Phase Transition
in Nucleus--Nucleus Collisions }

During the period 1999-2002 experimental data on Pb+Pb
collisions were recorded by several experiments (NA49, NA45, NA57, NA50 and NA60)
first at the top SPS energy of 158$A$~GeV. The primary aim of this program was to
find evidence for the QGP in these reactions. Results were consistent with the
signals expected for the production of this new state of matter~\cite{qgp_signals}
but not unique. This led to the extension of the program
to lower beam energies (20$A$, 30$A$, 40$A$, 80$A$~GeV) in order to search
for a clear onset of deconfinement which was predicted to be at the low SPS
energies. Only NA49 participated in the full energy scan. The history and
the basic results of the SPS energy scan programme are reviewed in the following.

\vspace{0.2cm}
\noindent
\subsubsection{A Brief History of Ideas}
In the mid  90s numerous results on collisions of
light nuclei at the BNL AGS (beams of Si at 14.6$A$~GeV)
and the CERN SPS  (beams of O and S at 200$A$~GeV)
were obtained.
The experiments with heavy nuclei (AGS: Au+Au at 11.6$A$~GeV,
SPS: Pb+Pb at 158$A$~GeV) were just starting.
This was the time when the first look at the energy dependence
of hadron production in nucleus--nucleus (A+A) collisions
at high energies became possible.
Compilations, on pion production~\cite{GaRo1} and
on strangeness production~\cite{GaRo2} resulted in a clear
conclusion: the energy dependence of hadron multiplicities
measured in A+A collisions and p+p interactions are
very different.
Furthermore the data on A+A collisions suggested
that there is a significant
change in the energy dependence of  pion and strangeness
yields which is located between the top AGS and SPS energies.
Based on the statistical approach to strong interactions~\cite{Fe,La}
it was conjectured~\cite{Ga,Ga1,Ga2} that the change is related to the
onset of deconfinement during the early stage of the A+A
collisions.
Soon after, following this hypothesis, a quantitative model was developed,
the Statistical Model of the Early Stage (SMES)~\cite{GaGo}.
It assumes creation of the early-stage matter according to the
principle of  maximum entropy.
Depending on the beam energy $E$
the matter is in the confined ($E <$ 30$A$~GeV),
mixed (30$A$ $< E <$ 60$A$~GeV) or
deconfined ($E >$ 60$A$~GeV) phases.
The phase transition is assumed to be of the first  order.

\vspace{0.2cm}
\noindent
\subsubsection{A Brief History of Experiments}
Based on these ideas the NA49 Collaboration
proposed in 1997 to study hadron production in Pb+Pb
collisions at 30$A$~GeV~\cite{na49_add1}.
At this energy the SMES  predicted a sharp
maximum of the strangeness to pion
ratio as a characteristic signal of the onset of
deconfinement.
Following this request a 40$A$~GeV Pb--beam
was delivered to NA49 in 1998 as a test.
The 5 weeks long run at 40$A$~GeV took place in
1999\footnote{This program was started with 40$A$~GeV instead of
the originally requested 30$A$~GeV due to technical
SPS reasons.}.
Data were registered by NA49, NA45, NA50 and
NA57.
The success of this first run at low SPS energy and the
exciting preliminary results shown by NA49 justified
a continuation of the program.
In 2000 a beam at 80$A$~GeV was delivered for 5 days to
NA49 and NA45 and
the energy scan program was completed in 2002. Results of the analysis
suggested that there is a significant
change in the energy dependence of  pion and strangeness
yields which is located between the top AGS and SPS energies.
Based on the SMES~\cite{GaGo}
it was concluded~\cite{review} that the change is related to the
onset of deconfinement at the early stage of A+A
collisions.
\begin{figure}[!htb]
\includegraphics[width=0.45\textwidth]{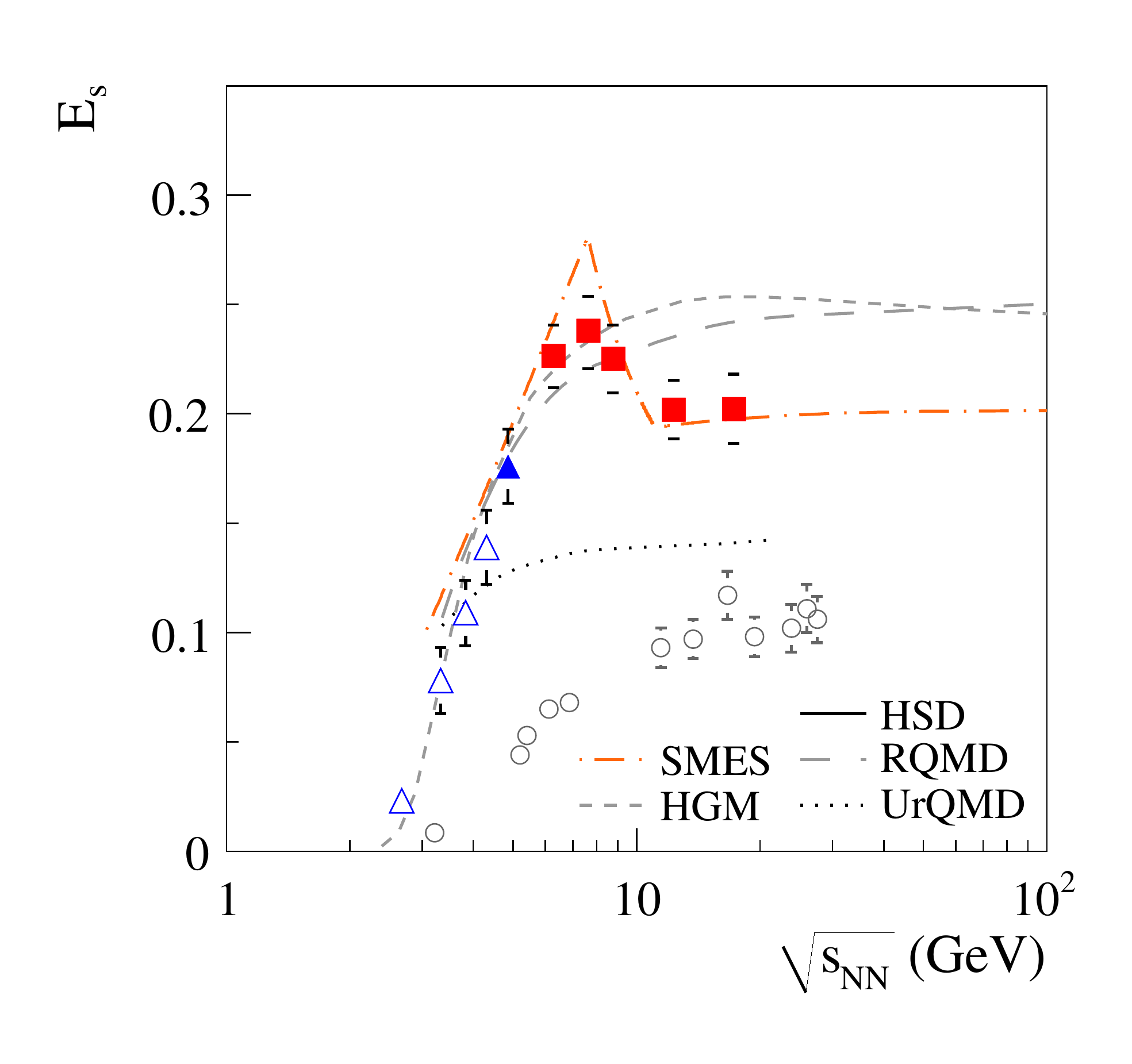}
\includegraphics[width=0.45\textwidth]{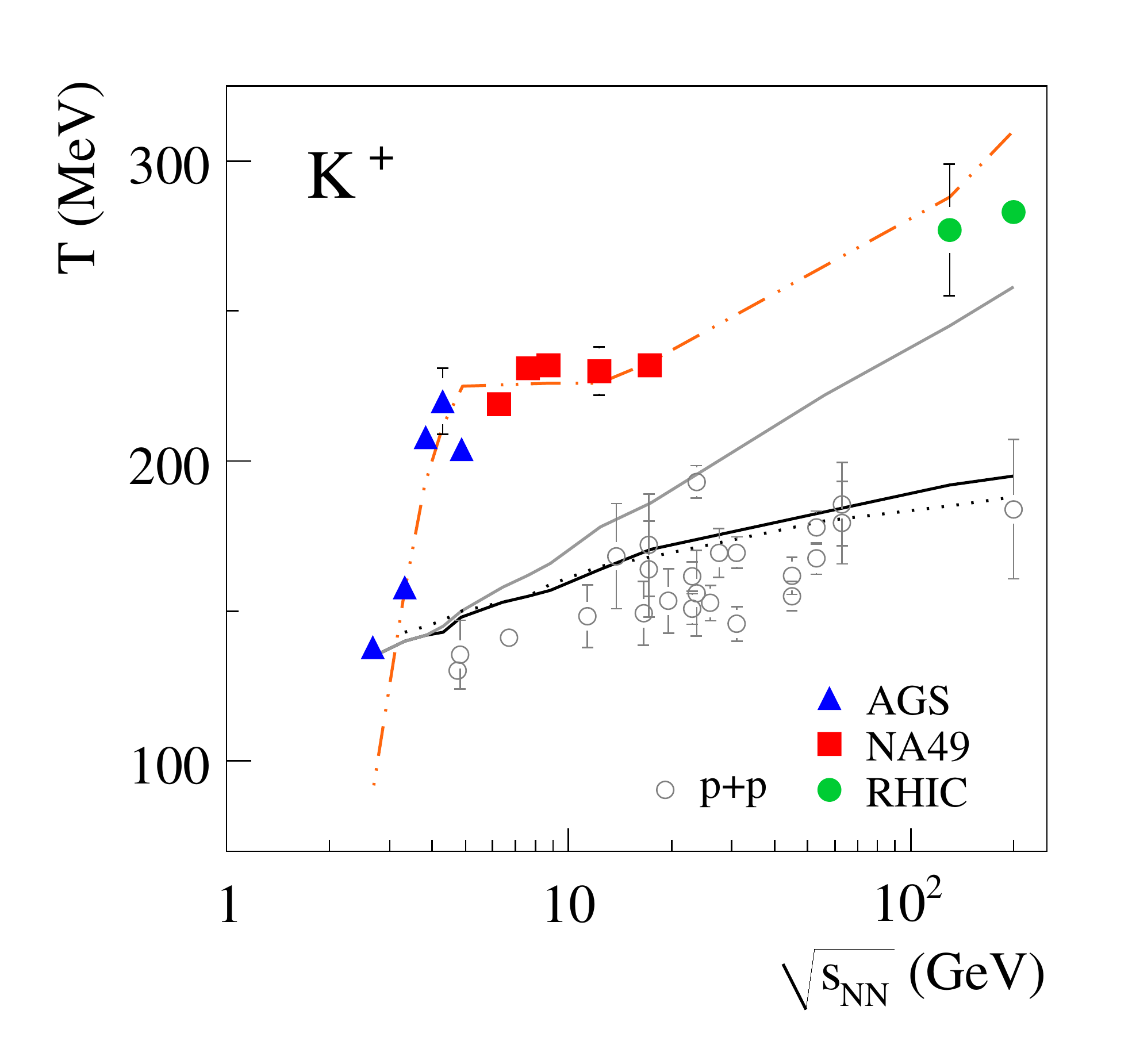}
\caption{\label{onset}
Dependence of hadron production properties in central
Pb+Pb (Au+Au) collisions (closed symbols) and p+p interactions
(open symbols) on the nucleon-nucleon cms energy $\sqrt{\text{s}_{\text{NN}}}$.
Left: relative strangeness production as measured by the ratio
$\text{E}_S = (\langle \Lambda \rangle) + \langle \text{K} + \bar{\text{K}} \rangle) / \langle \pi \rangle$.
Right: inverse slope parameter T of the transverse mass spectra
of K$^{+}$ mesons. The results for Pb+Pb (Au+Au) reactions are
compared with various models. For details see text.
}
\end{figure}

Results from this programme were published
\cite{NA49_Rfluct1,NA49_Phipt,NA49_Rfluct,na49results3,na49results4,NA49_omega,na49results6,na49results7,na49results8,na49results9}
and mostly concern the energy dependence of hadron production in central
Pb+Pb collisions at 20$A$, 30$A$, 40$A$, 80$A$ and 158$A$~GeV.
Key results and their comparison with models are summarized in
Fig.~\ref{onset} for the energy dependence of relative strangeness production
measured by the E$_S$ ratio ($left$) and the inverse slope parameter T of
the transverse mass spectra of K$^+$ mesons ($right$).
The rapid changes in the SPS energy range for central Pb+Pb collisions (solid squares)
suggest the onset of new physics in heavy ion collisions at low SPS energies.
The energy dependence of the same observables measured in p+p interactions
(open symbols in Fig.~\ref{onset}) is
very different from that measured in central Pb+Pb (Au+Au)
collisions and does not show any anomalies.

Relative strangeness production rises to a peak and then drops to a plateau
at a value predicted for deconfined matter~\cite{GaGo}.
The energy dependence of the T parameter shows
the step behaviour typically expected for a $1^{st}$ order phase transition~\cite{vHove}
and the resulting softness of the equation of state in the mixed-phase regime~\cite{Shuryak}.

Models which do not assume the deconfinement phase transition
(HGM \cite{HGM}, RQMD \cite{RQMD}, UrQMD \cite{UrQMD} and HSD \cite{HSD,HSD1})
fail to describe the data.
The introduction of a $1^{st}$ order phase transition at
low SPS energies (SMES \cite{GaGo}
and hydrodynamic evolution of the fireball~\cite{Gorenstein:2003cu,brasil,brasil1})
allow to describe the measured structures in the energy dependence of T and E$_S$
(see dash-dotted curves in Fig.~\ref{onset}).


There are attempts to explain the results on nucleus-nucleus
collisions by (model dependent) extrapolations of the
results from proton-nucleus interactions~\cite{fischer,cole}.
As detailed enough p+A data exist only at the top AGS and SPS
energies
the extrapolations are possible only at these two energies and thus there are
no predictions of the energy dependence of the quantities
relevant for the onset of deconfinement (see Fig.~\ref{onset}).
The underlying models are based on the assumption that particle yields in
the projectile hemisphere are due to production from
excited projectile nucleon(s).
This assumption is, however, in contradiction to the recent
results at RHIC~\cite{bialas} and SPS~\cite{mixing} energies
which clearly demonstrate a strong mixing of the projectile and
target nucleon contributions in the projectile hemisphere.
Furthermore, qualitative statements on similarity or differences between
p+A and A+A reactions may be strongly misleading because of
a trivial kinematic reason. The center of mass system in
p+A interactions moves toward the target nucleus A with increasing
A, whereas its position is A-independent for A+A collisions.
For illustration several typical examples are considered.
The baryon longitudinal momentum distribution in A+A collisions
gets narrower with increasing A and, of course, it remains
symmetric in the center of mass system.
In p+A collisions it shrinks in the proton hemisphere and
broadens in the target hemisphere (for example see slide 5
in Ref.~\cite{fischer}).
Thus a naive comparison would lead to the conclusion
that A+A collisions are qualitatively similar to p+A
interactions if the proton hemisphere is considered, or
that they are qualitatively different if the target hemisphere
is examined.
One encounters a similar difficulty when discussing
the A-dependence of particle ratios in a limited
acceptance.
For instance the kaon to pion ratio is independent of A
in p+A interactions if the total yields are considered~\cite{bialkowska},
it is however strongly A-dependent
in a limited acceptance (e.g. see slides 10 and 11
in~Ref.~\cite{fischer}).
Recent string-hadronic models~\cite{RQMD,UrQMD,HSD,HSD1}
take all trivial kinematic effects into account and
parameterize reasonably well p+p and p+A results.
Nevertheless they fail to reproduce the A+A data
(see e.g. Fig.~\ref{onset}).

\subsection{Present Second Generation Heavy--Ion Experiments}

\subsubsection{System Size and Energy Scan Programme at the CERN SPS -
Study of Properties of the Phase Transition
in Nucleus--Nucleus Collisions }

Already in 2003 it was realized~\cite{EoI} that
further progress in understanding effects
related to the onset of deconfinement can only be achieved
by a new comprehensive study of hadron production
in proton-proton, proton-nucleus and nucleus-nucleus collisions.

The two most important open questions are:
\begin{itemize}
\item
   what is the nature of the transition from the anomalous energy dependence
   measured in central Pb+Pb collisions at SPS energies to the smooth dependence
   measured in p+p interactions ?
\item
   is it possible to observe the predicted signals of the onset of
   deconfinement in fluctuations~\cite{mg_fluct,mg_fluct1}
   and anisotropic flow~\cite{Kolb:2000sd,Kolb:2000sd1} ?
\end{itemize}

These questions motivated the present ion programme
at the CERN SPS devoted to an extensive system size and energy scan.
Moreover, the programme received further urgency from the possibility
to discover the critical point of strongly interacting matter.
The programme  was proposed to CERN in November 2006~\cite{proposal}.
Based on this proposal, pilot data taking
took place in September 2007.
The first physics data with hadron beams
were recorded in 2009 and with ion beams (secondary $^7$Be beams) in 2011.

NA61/SHINE~\cite{NA61facility} is the only experiment
which records data within this programme. It is
a multi-purpose facility to study hadron production in hadron-proton,
hadron-nucleus and nucleus-nucleus collisions at
the CERN
SPS.
NA61/SHINE has greatly profited from the long development of the
CERN proton and ion sources, the accelerator chain, as well as
the H2 beam-line of the CERN North Area. The latter
has recently been modified to also serve as a fragment separator
as needed to produce the Be beams for NA61/SHINE.
Numerous components of the NA61/SHINE set-up were inherited from
its predecessors, in particular, the last one, the NA49 experiment.
Important upgrades increased the data taking rate by a factor 10
and significantly improved the capability for defining the
centrality of the collisions.

\begin{figure}[t]
\centering
\includegraphics[width=0.5\textwidth]{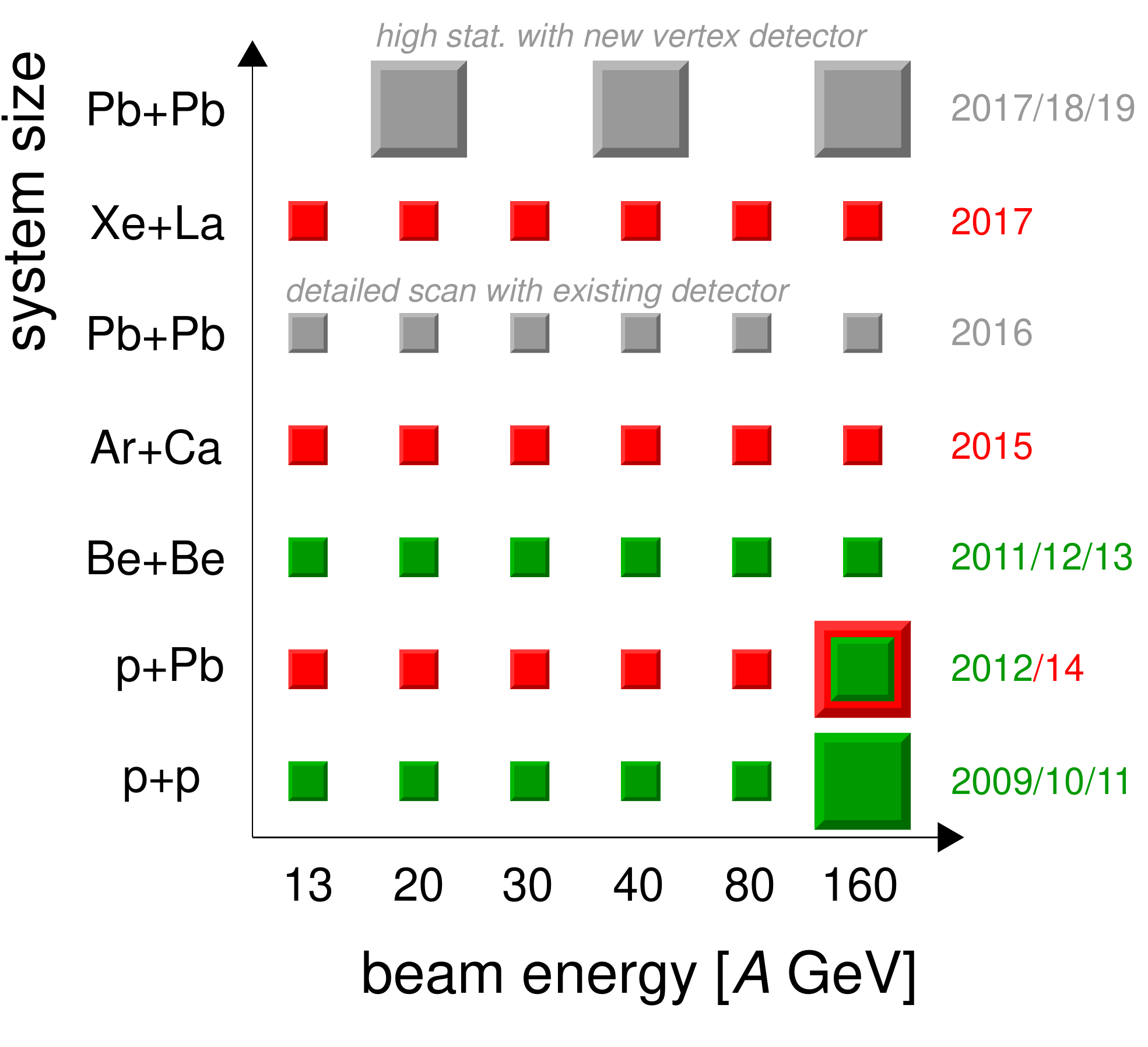}
\caption{
The NA61/SHINE data taking schedule for the ion program
and its proposed extension for the period 2016--2019 (in gray).
The proposed energy scan with Pb+Pb collisions in 2016
is needed to significantly decrease statistical
and systematic uncertainties of results on Pb+Pb
collisions by superseding the NA49 results with new NA61/SHINE
measurements.
The high statistics runs for Pb+Pb collisions with the new
vertex detector are suggested in the period 2017-2019.
They will  allow for precise measurements of multi-strange
hyperons and for first measurements of the open charm production
at the top SPS energies.
}
\label{fig:na61_plans}
\end{figure}

The status and plans of data taking of NA61/SHINE are
summarized in Fig.~\ref{fig:na61_plans}.
First results were already obtained for inclusive spectra and fluctuations
of identified hadrons produced in p+p and $^7$Be+$^9$Be
interactions~\cite{SR2013}.

Two possible extensions of the NA61/SHINE ion programme
are proposed.
First, a detailed energy scan with Pb+Pb collisions
at beam energies of 13$A$, 19$A$, 30$A$, 40$A$, 75$A$, 150$A$~GeV
is planned for 2016.
Within a single data-taking period (42 days) data at
all energies can be recorded with a typical event statistics
of about 10 times the one of the NA49 data.
This, together with the important detector upgrades,
will allow to significantly decrease statistical
and systematic uncertainties of results on Pb+Pb
collisions by superseding the NA49 results with new NA61/SHINE measurements.
Second, high-statistics measurements of Pb+Pb
collisions at 19$A$, 40$A$ and 150$A$\,GeV/c beam energy
are foreseen in the period 2017--2019 after
upgrading the NA61/SHINE apparatus by adding a vertex
detector.
They should allow to determine the
energy dependence of rare processes, in particular, production of
D$^0$ mesons at the top SPS energies and multi-strange hyperons in the full
SPS energy range.

Moreover measurements of di-muon production in the SPS
energy range is considered by two proto--collaborations,
NA60+ and CHIC~\cite{chic}.

\subsubsection{Beam Energy Scan Programme at the BNL RHIC}

In parallel to the NA61/SHINE programme the Beam
Energy Scan (BES) Programme at the BNL RHIC was proposed~\cite{critRHIC}.
Both programmes
were motivated by the NA49 results on the onset
of deconfinement and the possibility to observe the critical
point (CP) of strongly interacting matter.

The present goal of the BES programme related to the
onset of deconfinement is summarized in ref.~\cite{cpodGO}.
The aim is to find whether (and where in collision energy)
the key QGP signatures observed at
the top RHIC energy will turn off.
This may indicate below which energy the system stays in the hadron gas phase.
The disappearance of a single signature would not be enough
to claim the onset of deconfinement,
because there are other phenomena not related to deconfinement
which may cause a similar effect.
However, the modification or disappearance of several signatures
simultaneously would definitely
provide a compelling case. The particular observables
identified as the essential drivers of this
part of the run are:
constituent quark number scaling (NCQ), hadron suppression in central collisions
characterized by the ratio R$_{cp}$ between particle yields in peripheral and central collisions,
untriggered pair correlations in the space of pair separation in azimuth and
pseudo-rapidity, local parity violation in strong interactions and strong
fluctuations as possible indications of the CP.

\begin{figure}[!htb]
\includegraphics[width=0.95\textwidth]{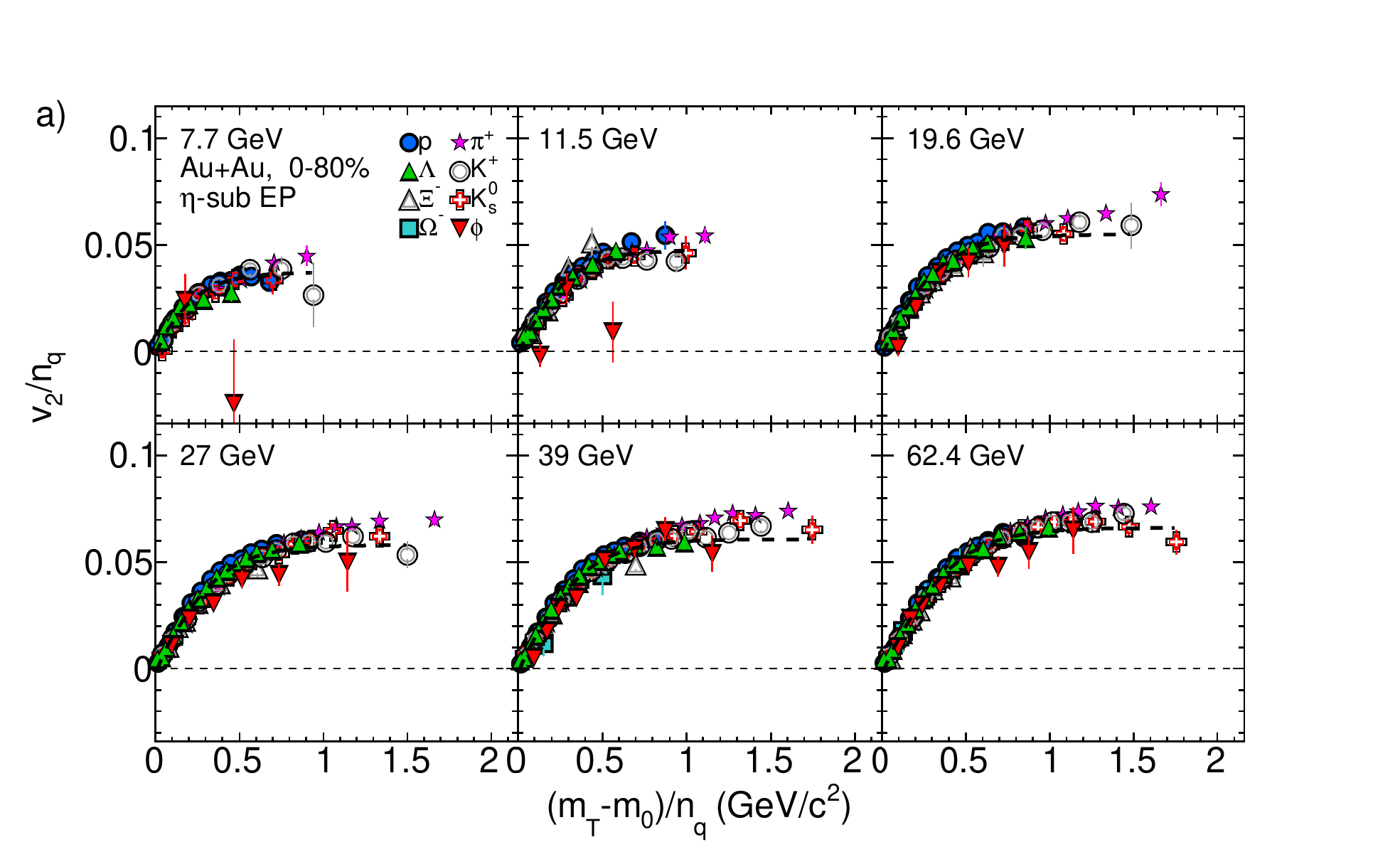}
\caption{\label{fig:cqns}
The Number-of-Constituent Quark (NCQ) scaled elliptic flow, $v_{2}/n_{q}$ versus $(m_{T}-m_{0})/n_{q}$,
for 0--80\% central Au+Au collisions for selected particles and the various collision energies
of the RHIC beam energy scan~\cite{BES_v2}. Here $n_{q}$ denotes the number of
constituent quarks in particle of type q and m$_0$ and m$_{\text{T}}$ the particle rest mass
and transverse mass, respectively. The dashed lines show the results of a
parameterisation fitted simultaneously to all particles except the pions.
}
\end{figure}

In 2010 and 2011 RHIC completed phase I of the BES programme
recording data on Au+Au collisions at nucleon-nucleon c.m.s. collision energies
$\sqrt{\text{s}_{\text{NN}}}$ = 7.7, 11.5, 19, 27 and 39 GeV.
This is complemented by the data collected earlier at higher energies
(62, 130 and 200 GeV).
The phase II of the BES~\cite{BES-II} is foreseen for 2017-2018 after the RHIC and
STAR detector upgrades.
Data taking on Au+Au collisions with the STAR detector is planned
in the collider mode and, parasitically, in the fixed target mode
down to a collision energy of 3~GeV.

Results from the BES phase I~\cite{BES_v2} show that elliptic flow,
quantified by v$_2$, scales according to the
NCQ
down to the lowest energy (see Fig.~\ref{fig:cqns}).
Flow develops at the very early stage of the reaction
from the anisotropic pressure in the almond-shaped interaction region
of non-central collisions. If the fireball is in a partonic state at this
stage, flow is thought to be imparted to the constituents from which the
observed hadrons are coalescing. This picture would lead to the
observed NCQ scaling.

\begin{figure}[!htb]
\includegraphics[width=0.6\textwidth]{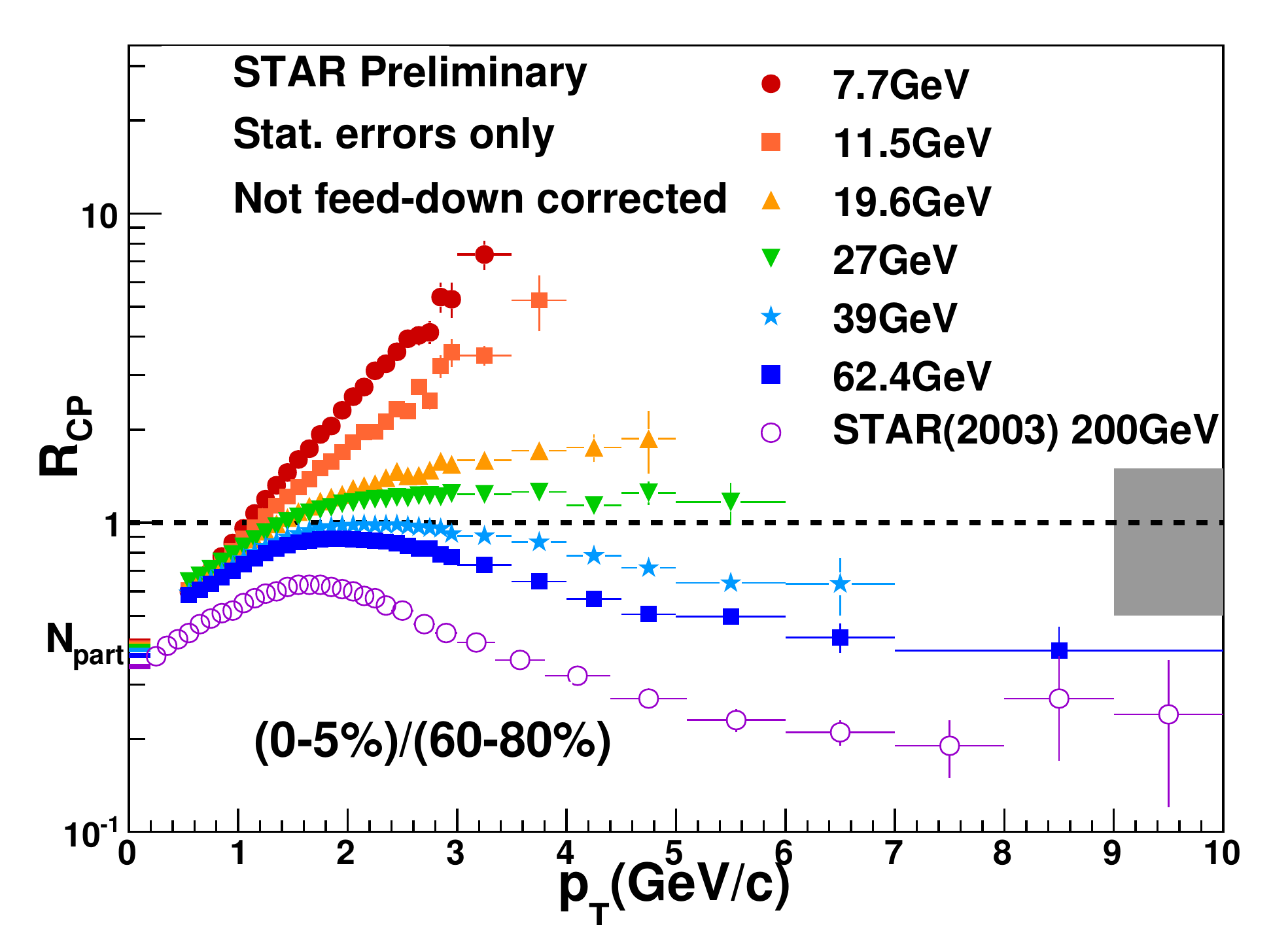}
\caption{\label{fig:rcp}
The STAR preliminary results of the energy dependence of charged
particles R$_{CP}$~\cite{BES_Rcp}. The error
bars on points reflect statistical errors.
The systematic error (gray box) is independent of energy and p$_T$,
and combines error on N$_{coll}$ scaling
and on the yields due to the background.
}
\end{figure}

The ratio R$_{cp}$ is widely used to characterize the energy loss of partons
in the fireball matter. When the fireball is in the QGP phase
the high gluon density is expected to lead to strong attenuation of
the parton energy (jet quenching) and a resulting suppression of
particle spectra at large transverse momenta p$_{\text{T}}$.
Preliminary results on R$_{cp}$~\cite{BES_Rcp} (see Fig.~\ref{fig:rcp})
show a monotonic evolution with collision energy from enhancement at
low energy (the Cronin effect) to suppression at high energy. Thus, the BES did not
observe any threshold-like changes, in addition
to those reported by NA49.

\begin{figure}[!htb]
\includegraphics[width=0.6\textwidth]{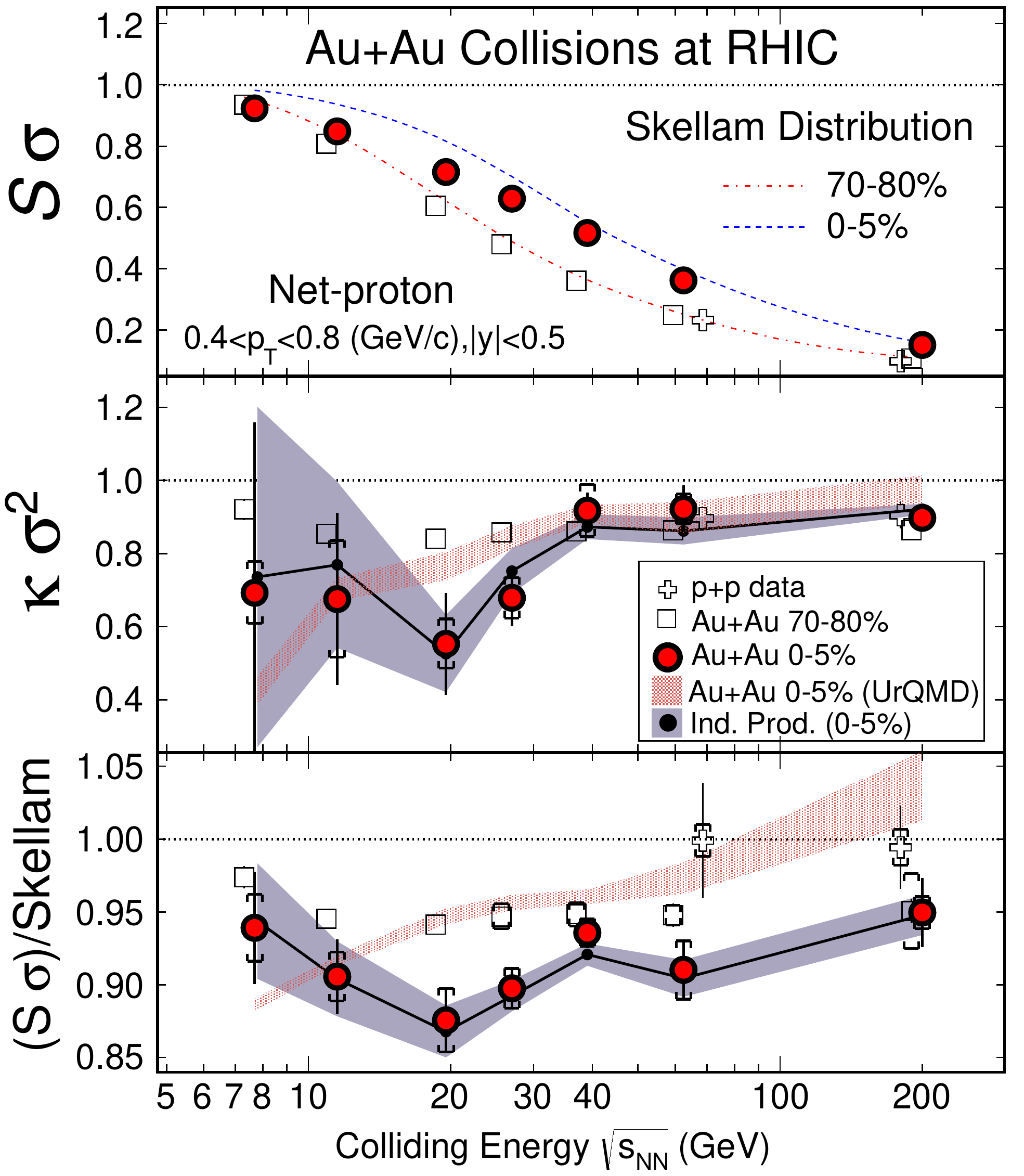}
\caption{\label{fig:netp}
Collision energy and centrality dependence of the products of the variance $\sigma$,
the skewness $S$ and the kurtosis $\kappa$ of the net-proton multiplicity
distribution in Au+Au collisions~\cite{BES_netp}. The experimental results are compared to
the reference Skellam distribution as well as predictions of the UrQMD model. For a
hadron-resonance gas $\kappa\sigma$ and $S\sigma$/Skellam are close to unity.
}
\end{figure}

It was pointed out in Ref.~\cite{Stephanov_netp} that fluctuations of the net-proton
multiplicities are a suitable experimental observable of the effects of a CP.
The higher moments of the multiplicity distribution are expected to be especially sensitive~\cite{Step,Step1,Step2}.
The STAR collaboration recently published results from the
BES programme~\cite{BES_netp}
for the variance $\sigma$, the skewness $S$ and the kurtosis $\kappa$ of the
net-proton multiplicity in Au+Au collisions (see Fig.~\ref{fig:netp}).
Measurements of $\kappa\sigma$ and $S\sigma$ are compared to the hadron-resonance gas
model, the UrQMD model, and the reference Skellam distribution
(the distribution of a random variable defined as the difference of two
independent random variables following Poisson distributions).
The values of $S\sigma$ are close to the Skellam distribution and
there appears to be a decreasing trend with decreasing energy for $\kappa\sigma$
which is qualitatively reproduced by the UrQMD model. No firm conclusion
with regard to the CP can be drawn at present, but more precise data are
expected from the future BES~II program.

It is important to note that the preliminary results from BES
confirm the NA49 measurements relevant for the evidence
for the onset of deconfinement. This will be illustrated in
the following section.

\subsubsection{LHC: Collider and Fixed Target Experiments on Hot QGP}

In 2011 first results on central Pb+Pb collisions at the LHC were released.
These data were recorded at $\sqrt{s_{NN}}$~=~2.76~TeV, a nucleon-nucleon cms
energy which is
approximately 360 times higher than that of the onset of deconfinement.
Nevertheless, these results are important to verify the interpretation
of the NA49 results.

\begin{figure}[!htb]
\begin{center}
\begin{minipage}[b]{0.95\linewidth}
\includegraphics[width=0.45\linewidth]{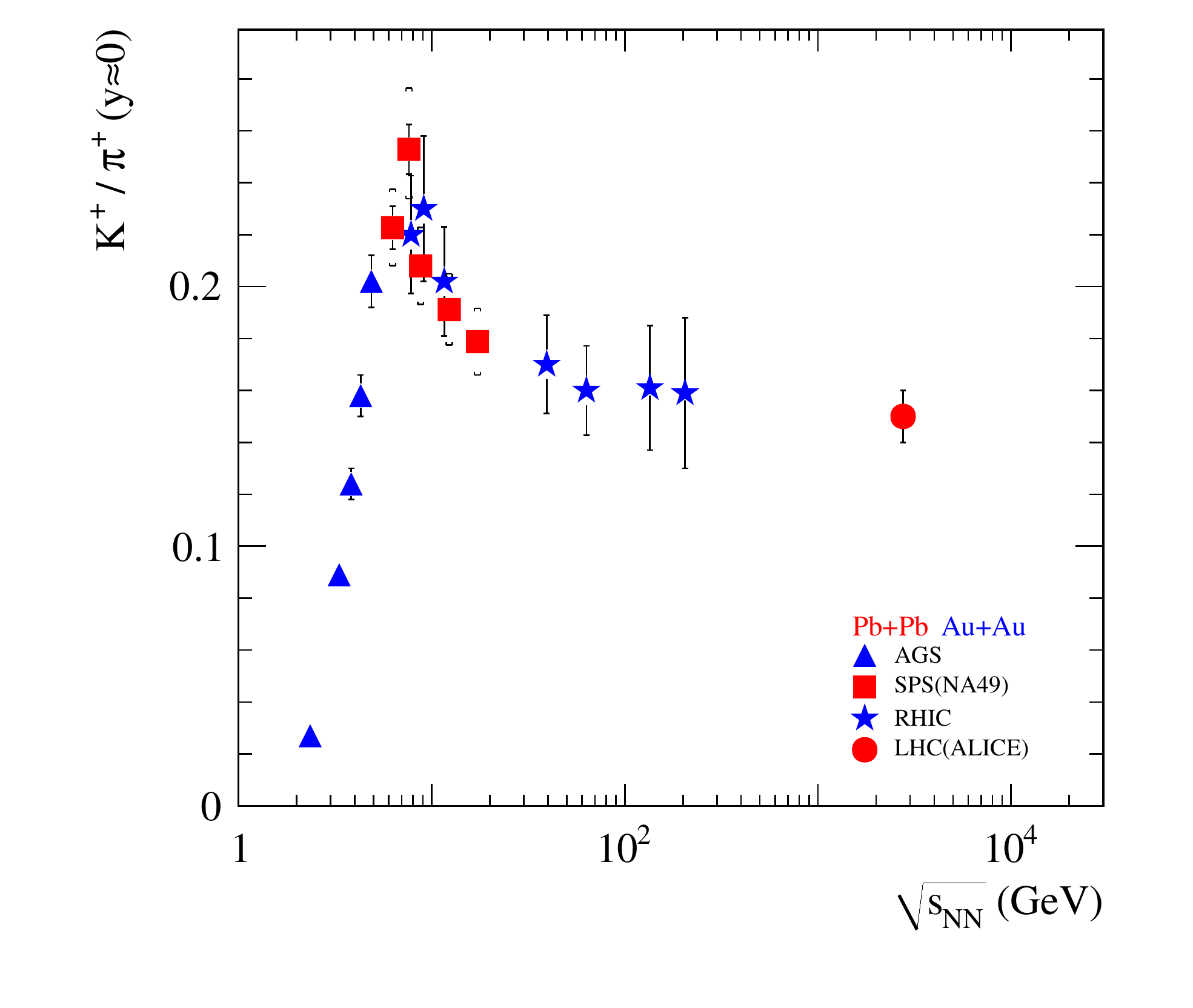}
\includegraphics[width=0.45\linewidth]{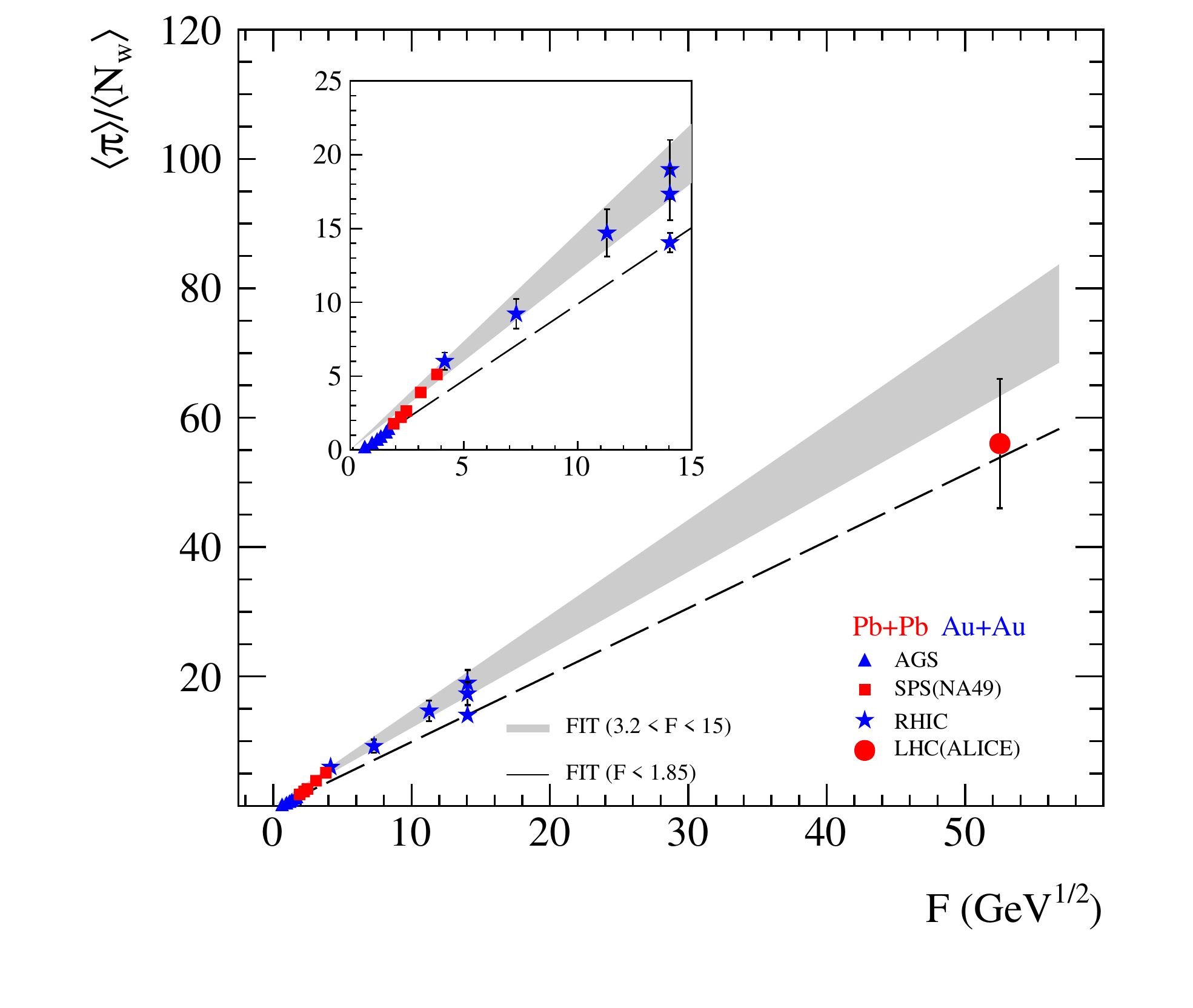}
\includegraphics[width=0.45\linewidth]{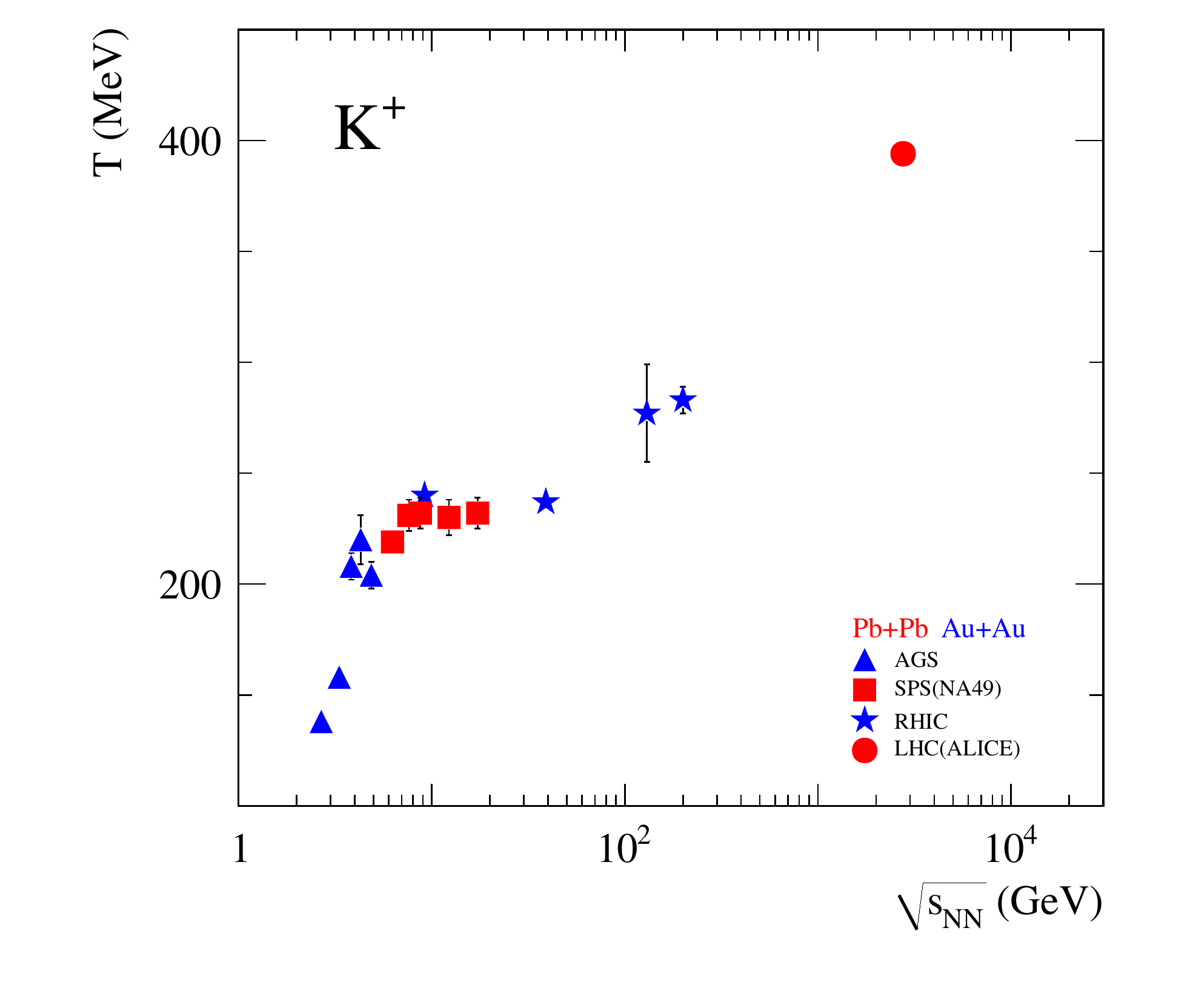}
{\hspace*{0.2 cm}
\includegraphics[width=0.45\linewidth]{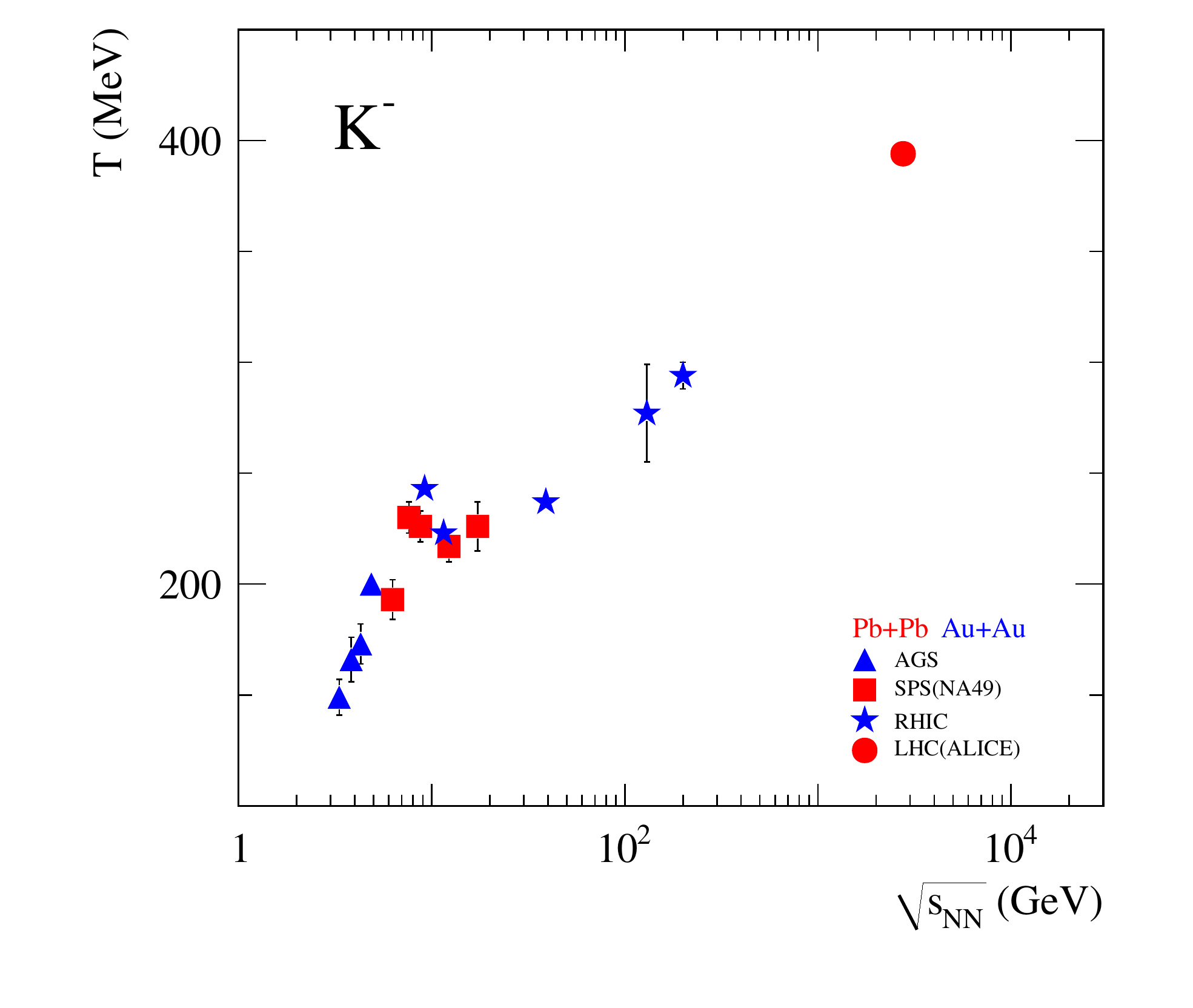}}
\end{minipage}
\caption{\label{fig:evidence_2011}
Heating curves of strongly interacting matter, status mid--2010.
Hadron production properties (see Ref.~\cite{review} for details)
are plotted
as a function of collision energy ($F \approx \sqrt{s_{NN}}$) for central Pb+Pb (Au+Au)
collisions:
$top-left$ -- the $\langle {\rm K}^+ \rangle /\langle \pi^+\rangle$ ratio,
$top-right$ -- the mean pion multiplicity per participant nucleon,
bottom -- the inverse slope parameter  of the
transverse mass spectra of K$^+$ ($left$) and K$^-$ ($right$) mesons.
The new LHC and RHIC data are included in the
{\it horn} ($top-left$), {\it kink} ($top-right$) and {\it step} ($bottom$)
plots.
The K$^+/\pi^+$ ratio is measured by ALICE~\cite{schukraft} and
STAR~\cite{kumar} at mid-rapidity only and thus the {\it horn}
plot is shown here for the mid-rapidity data.
The observed changes of the energy dependence for central Pb+Pb (Au+Au)
collisions are related to:
decrease of the mass of strangeness carries and the ratio of
strange to non-strange degrees of freedom ({\it horn}: top-left plot),
increase of entropy production ({\it kink}: top-right plot),
weakening of transverse ({\it step}: bottom plots)
expansion at the onset of deconfinement.
}
\end{center}
\end{figure}

The updated plots on the energy dependence of hadron production properties are
shown in Fig.~\ref{fig:evidence_2011}~\cite{anar}.
They include LHC data and results from the RHIC BES
programme.
The RHIC results~\cite{kumar,kumar1} confirm the NA49 measurements at
the onset energies.
The LHC data~\cite{schukraft,schukraft1} demonstrate that
the energy dependence of hadron production properties
shows rapid changes only at the low SPS energies.
 A smooth evolution is observed between
the top SPS (17.2~GeV) and the current LHC (2.76 TeV) energies.
This agrees with the
interpretation of the NA49 structures as due to the onset of deconfinement.
Above the onset
energy only a smooth change of the quark-gluon plasma properties
with increasing collision
energy is expected.
Thus currently there seems to be no strong physics motivation for
measurements in the gap between the top RHIC and LHC energies.

A possibility to use proton and ion beams extracted from
the LHC for future fixed target experiments is discussed
since several years~\cite{AFTER}.
This would allow to extend the collision energy range
of fixed target experiments up to about 70~GeV per nucleon--nucleon pair.
This extension may be important in the search for the critical
point of strongly interacting matter, if the current NA61/SHINE
programme did not lead to its discovery at the SPS energies.

\subsection{Future Experimental Programmes}

MPD~\cite{mpd} at NICA together
with CBM~\cite{cbm} at SIS-100/300 belong to the
third generation of experiments  planned to study A+A
collisions at energies of the CERN SPS.

\subsubsection{NICA: Collider Studies of the Properties of the Mixed Phase}

Exploration of the transition
between the confined and deconfined phases of strongly interacting matter
is the top priority of the NICA  programme~\cite{nica}.

The first round of NICA experiments intends to concentrate~\cite{nica_wp}
on a variety of diagnostic observables that have already been employed
in experimental programmes at SPS and  RHIC (the beam energy scans);
these are based on general considerations about phase changes
and the associated different degrees of freedom.
The MPD detector at NICA will be optimized for the study of
fluctuations and correlations of bulk event properties
and a primary goal will be to measure the excitation functions and
the dependence of fluctuations and correlations on centrality and system size.

The observables include event-by-event fluctuations of multiplicity
and transverse momentum of charged particles and identified
particles (p, K, $\pi$) as well as the corresponding joint distributions.
Correlation studies address long-range angular correlations
like the transverse Fourier components $v_1$ and $v_2$ of identified particles
(p, K, $\pi$, $\Lambda$), their antiparticles, and light clusters,
as well as three-body correlations
which are the basis for studying the chiral magnetic effect.
Short-range two-particle correlations will be employed to
measure the size and internal dynamics of the freeze-out stage.
Particular care will be taken to provide as large as possible coverage
in rapidity and transverse momentum.
Measurements are planned as a function of collision energy
for the following systems:
\begin{itemize}
 \item p+p collisions,
 \item d+d collisions with a possibility to off-line select
  reactions with (p,p), (p,n), and (n,n) spectators,
 \item d+Pb collisions, and
 \item collisions of identical heavy nuclei, such as Pb+Pb.
\end{itemize}
For later also the study of collisions of identical nuclei of intermediate mass
is envisaged. Clearly,
when measuring correlations, the detector will simultaneously collect
centrality-selected high-precision data
on double differential spectra of identified hadrons.
Thus, freeze-out conditions can be precisely established
for collisions in the transition domain.

In a second stage of the experiment measurements of open-charm hadrons,
di-leptons, and di-photons are also considered at NICA.
The first data from the NICA programme are expected to come in 2017.

\subsubsection{FAIR: Fixed Target Studies of High Baryon Density Matter}

The Compressed Baryon Matter (CBM) experiment~\cite{cbm} is under construction at
the GSI FAIR accelerator complex~\cite{fair} which is expected to start delivering
heavy ion beams from the SIS100 ring with energies up to 11$A$~GeV in 2018.
The planned addition of SIS300 would extend the energy range to 35$A$~GeV
after 2025. CBM will study strongly interacting matter from the region of
the onset of deconfinement, where the highest initial stage baryon density
is achieved~\cite{max_rhoB}, down to beam energies of 2$A$~GeV. Thus CBM will cover the
energy domain of past AGS experiments and the low energy range of the SPS
using a state-of-the-art detector.
The experiment will explore a region of the phase diagram where the CP
and new forms of baryon-rich matter (e.g. quarkyonic matter) might be
found.

CBM is designed to measure with unprecedented sensitivity spectra of hadrons,
including multi-strange baryons and charmed mesons, as well as lepton pairs
produced in p+p, p+A and A+A collisions. The research program
focuses on the following observables~\cite{cbm}:
\begin{itemize}
 \item Open and hidden charm production: measurements of yields and transport properties
       (anisotropic and radial flow) which are sensitive to the properties of
       the fireball medium (confined or partonic).
 \item Di-lepton pairs: study of in-medium properties of vector mesons
       via low-mass pairs (chiral restoration),
       search for virtual photon emission from a partonic phase in
       the intermediate mass region,
       study of the production properties of charmonia in the high-mass region
       (absorption in partonic matter).
 \item Yields and phase space distributions of hadrons (including multi-strange baryons):
       these are sensitive to the properties of the produced matter e.g. whether
       confined or deconfined, or possibly in an exotic state like quarkyonic matter.
 \item Hypernuclei: production expected via coalescence of $\Lambda$ hyperons
       and nucleons providing information on hyperon-nucleon and hyperon-hyperon
       interactions which play an important role in neutron star models.
 \item Fluctuations and correlations: location  and properties of the
       onset of deconfinement  and search for
       the critical point.
\end{itemize}
The CBM programme  can be expected to provide essential new
knowledge on the properties of baryon-rich matter starting from 2018.


\section{Progress on Analysis Methods of Event-by-Event Fluctuations}
The study of event-by-event (e-by-e) fluctuations in high-energy
nucleus-nucleus collisions opens new possibilities to
investigate properties of strongly interacting matter.
 (see, e.g.
Ref.~\cite{Koch:2008ia} and references therein).
Specific
fluctuations can signal the onset of deconfinement when the
collision energy becomes sufficiently high to create the
QGP
at the initial stage of A+A
collision~\cite{mg_fluct,mg_fluct1}.  By measuring the fluctuations, one may also observe
effects caused by dynamical instabilities when the expanding
system goes through the 1$^{st}$ order transition line between the
QGP
and the hadron-resonance gas~\cite{fluc2,fluc2a}. Furthermore,
the critical point of strongly interacting matter
may be signaled by a characteristic
fluctuation pattern~\cite{fluc3,fluc3a,fluc3b,Koch:2005vg,Koch:2005pk}.
Therefore, e-by-e
fluctuations are an important tool for the study of properties of the
onset of deconfinement and the search for the
CP
of strongly interacting matter.
However, mostly due to the incomplete acceptance of detectors and
difficulties to control e-by-e the number of interacting
nucleons as well as not well
adapted data analysis tools, results on e-by-e fluctuations are not
yet mature. Even the simplest tests of statistical and dynamical models
at the level of fluctuations are still missing.

In this section progress in two areas related to
the study of e-by-e
fluctuations are reported.
First, the role of fluctuations in the number of interacting
nucleons is discussed and strongly intensive quantities
are introduced.  Second, a novel procedure, the identity method,
is described for analyzing fluctuations of identified hadrons under typical experimental
conditions of incomplete particle identification.

\subsection{Fluctuations in the Number of Participants}\label{NPFLCT}

In each A+A collision only a fraction of all 2$A$ nucleons
interact. These are called participant nucleons and are denoted as
$N_P^{proj}$ and $N_P^{targ}$ for the projectile and target
nuclei, respectively. The corresponding nucleons, which do not
interact, are called the projectile and target spectators,
$N_S^{proj} = A - N_P^{proj}$ and $N_S^{targ} = A - N_P^{targ}$.
The fluctuations in high energy A+A collisions are dominated by
e-by-e
variations of the impact parameter. However, even for
a fixed impact parameter the participant numbers,
$N_P^{proj}$ and $N_P^{targ}$, fluctuate from event to event. This is
due to the fluctuations of the initial states of the colliding
nuclei and the probabilistic character of the interaction process.
The fluctuations of $N_P^{proj}$ and $N_P^{targ}$
usually lead to a large and uninteresting
background. In order to minimize its contribution fixed-target
experiments select samples of collisions with a fixed
number of projectile participants. This selection is achieved by
the measurement of $N_S^{proj}$ in each individual A+A collision
by a calorimeter which covers the projectile fragmentation domain.
The advantage of this event selection is that it depends only on
fluctuations of the initial state and is independent of
the internal properties of the fireball created in the collisions.
In collider experiments the number of spectators is
impossible to measure because charged nuclear fragments
follow the directions of the circulating beams. In these experiments
events are typically selected using quantities derived from the produced
hadrons, e.g. charged hadron multiplicity or transverse energy at mid-rapidity.
This event selection has a significant disadvantage.
Namely, it depends on the properties of the created fireball
and may lead to a correlation between the quantity used for the event selection
and the fluctuation quantities under study.

It was argued in Ref.~\cite{BrFl} that any centrality selection in
A+A collisions is equivalent to the geometrical selection via impact parameter $b$.
Indeed, different centrality selection criteria result in approximately the same
average values of physical quantities, e.g. average hadron multiplicities.
However, they lead to rather different fluctuations of these quantities, e.g. different
values of the scaled variances and correlations of hadron multiplicities.
This was explicitly demonstrated in Ref.~\cite{hauer}, where results for three
different  centrality
selections -- via impact parameter $b$, via the number of participants $N_P$, and via
the charged particle multiplicities in the mid-rapidity window -- were compared using the HSD
transport approach to A+A collisions.

Even in fixed target experiments
in which events with $N_P^{proj} = const$  can be selected
the number of target participants fluctuates considerably. Hence, an
asymmetry between projectile and target participants is
introduced, i.e. $N_P^{proj}$ is constant by constraint, whereas
$N_P^{targ}$ fluctuates.

\begin{figure}[t!]
\begin{center}
\includegraphics[width=0.6\textwidth]{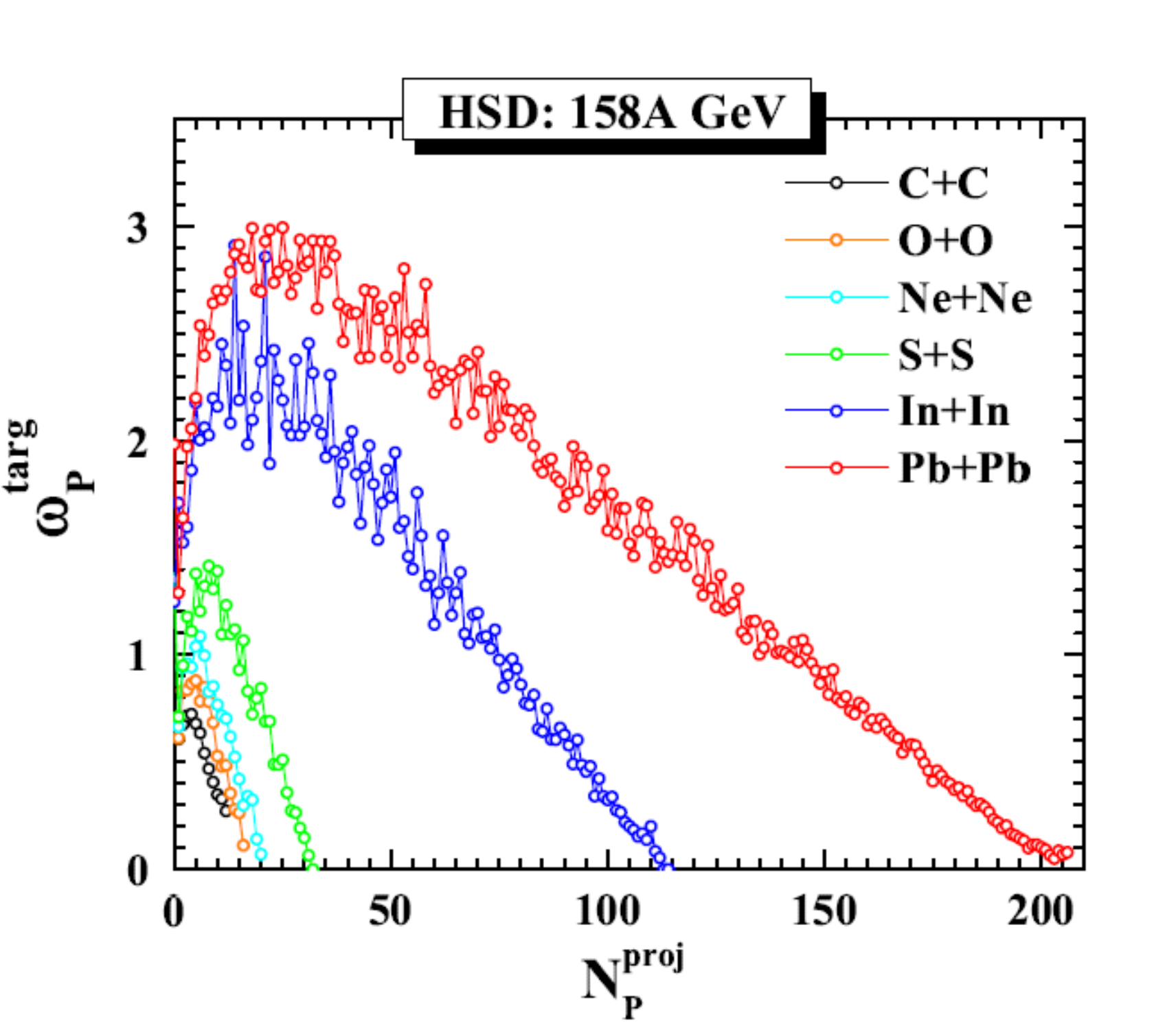}
\end{center}
\caption{The scaled variance $\omega_P^{targ}$  for the
fluctuations of the number of target participants, $N_P^{targ}$
as a function of the number of projectile participant, $N_P^{proj}$~\cite{KLGB}.
HSD simulations of $\omega_P^{targ}$ are shown as a function of
$N_P^{proj}$ for different colliding nuclei (Pb+Pb, In+In, S+S,
Ne+Ne, O+O and C+C) at $E_{lab}$=158~AGeV.} \label{fig:w_targ}
\end{figure}

At fixed values of the numbers of participants $N_P^{proj}$ and
$N_P^{targ}$ we introduce
the probability $W_i(N_i; N_ P^{targ},N_P^{proj})$ for producing
$N_i$ (the index $i$ corresponds to the type of particle species, e.g.
$i=-,+,ch$, i.e. negative, positive, all charged particles) final hadrons.
At fixed $N_P^{proj}$
the averaging procedure is defined as
\eq{\label{dynav} \langle \cdots \rangle~\equiv~
\sum_{N_P^{targ}\ge 1}^{A}\sum_{N_i\ge 0} \cdots ~W_P(N_P^{targ};
N_P^{proj})~ W_i(N_i; N_ P^{targ},N_P^{proj})~,}
where $W(N_P^{targ}; N_P^{proj})$  is the probability for a given
value of $N_ P^{targ}$ in a sample of events with fixed number of
the projectile participants $N_P^{proj}$.
The variance can be written as
\eq{ Var(N_i)~ \equiv~\langle N_i^2\rangle~-~\langle
N_i\rangle^2~=
~\omega_i^*~\langle N_i\rangle~+~\omega_P~n_i~\langle N_i\rangle~,
\label{VarNi} }
where
the scaled variance $\omega_i^*$ corresponds to the fluctuations
of $N_i$ at fixed values of $N_P^{proj}$ and $N_P^{targ}$, $n_i
\equiv \langle N_i\rangle /\langle N_P \rangle$~, and
$N_P=N_P^{targ}+N_P^{proj}$ is the total number of participants.
Eq.~(\ref{VarNi}) is based on two assumptions:
first, that $\omega_i^*$ does not depend on $N_P$, and
second, that the average multiplicities $N_i$ are
proportional to the number of participating nucleons.
For the scaled variances, $\omega_i$, one finds
\eq{ \label{omegai} \omega_i~\equiv~\frac{Var(N_i)}{\langle
N_i\rangle}~=~\omega_i^*~+~\omega_P~n_i~. }
The average values are $\langle N_P^{targ}\rangle\cong
N_P^{proj}$, thus, $\langle N_P^{targ}\rangle\cong \langle N_P
\rangle /2$.
Therefore, the scaled variance $\omega_P$ for the total number of
participants in Eq.~(\ref{omegai}) equals to
$\omega_P=\omega_P^{targ}/2$ as only half of the total number
of participants
fluctuate.
The value of $\omega_P^{targ}$ depends on $N_P^{proj}$, as shown
in Fig.~\ref{fig:w_targ} by calculations using the HSD model for
collisions of light, medium and large size nuclei
at beam energy of 158$A$~GeV~\cite{KLGB}.

Equation~(\ref{omegai}) corresponds to the so-called
model of multiple independent sources (MIS).
The model assumes that particles are produced by identical
and independent sources. The numbers of sources are taken to be
proportional to the number of projectile and target participant
nucleons.
The physical nature of the sources  can be different,
e.g.  wounded nucleons~\cite{WNM}, strings
and  resonances,
or the fluid cells at chemical freeze-out in hydrodynamical models.
Equation~(\ref{omegai}) gives the final multiplicity fluctuations as a sum
of two terms: the fluctuations from one source, $\omega_i^*$, and
the contribution due to the fluctuations of the number of sources,
$\omega_P~n_i$.

\begin{figure}[t!]
\begin{center}
\includegraphics[width=0.5\textwidth]{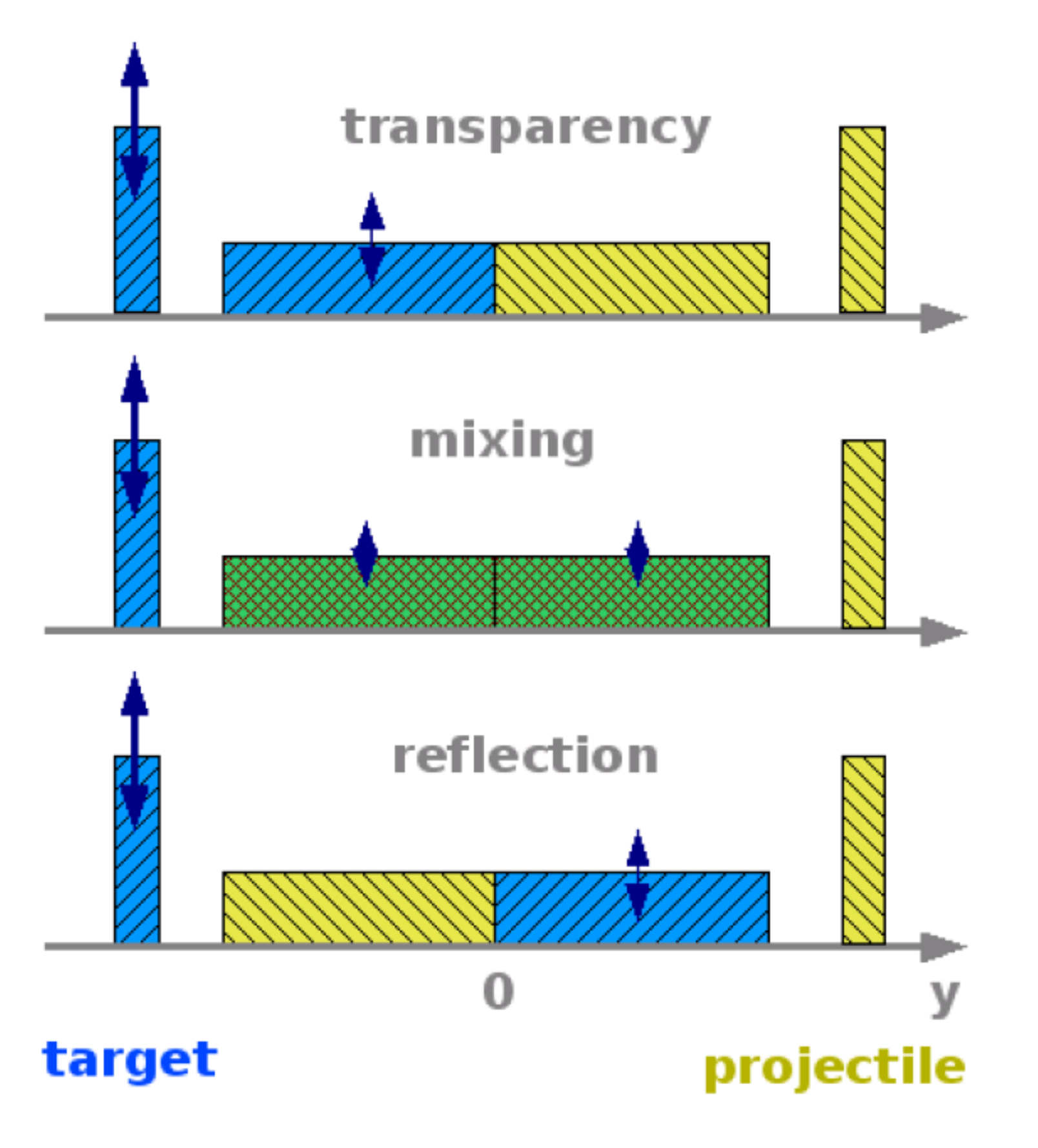}
\end{center}
\vspace{-0cm}
\caption{The schematic sketch of the rapidity distributions of the baryon
number or the particle production sources (horizontal rectangles)
in nucleus-nucleus collisions resulting from the transparency,
mixing and reflection models. The spectator nucleons are indicated
by the vertical rectangles. In the collisions with a fixed number
of projectile spectators only matter related to the target shows
significant fluctuations (vertical arrows). See Ref.~\cite{mixing}
for more details.} \label{fig:sketch}
\end{figure}

Models of hadron production in relativistic A+A
collisions can be divided into three basic groups:
transparency, mixing, and reflection models (see
Ref.~\cite{mixing}). The first group assumes that the final
longitudinal flow of the hadron production sources related to
projectile and target participants follows in the direction of the
projectile and target, respectively. One calls this group
transparency (T-)models. If the hadron production sources
from projectile and target are mixed, these models are
called mixing (M-)models. Finally, one may assume that the
initial flows are reflected in the collision process. The
projectile related matter  then flows in the direction of the
target and the target related matter flows in the direction of the
projectile. This class of models is referred to as reflection
(R-)models. The rapidity distributions resulting from the T-, M-,
and R-models are sketched in Fig.~\ref{fig:sketch}.

An asymmetry between the projectile and target participants
introduced by the experimental selection procedure in a fixed-target
experiment can be used to distinguish between projectile related
and target related final state flows of hadron production sources.
The multiplicity fluctuations measured in the target momentum
hemisphere clearly are larger than those measured in the
projectile hemisphere in T-models. The opposite relation is
predicted for R-models, whereas for M-models the fluctuations in
the projectile and target hemispheres are expected to be the same.
In collisions with a fixed number of $N_P^{proj}$ and
fluctuating number of $N_P^{targ}$
the amount of mixing of projectile and target
related matter in the final state of the collisions can be
distinguished by an analysis of fluctuations.
%
\begin{figure}[t!]
\begin{center}
\centering
\includegraphics[width=0.6\textwidth]{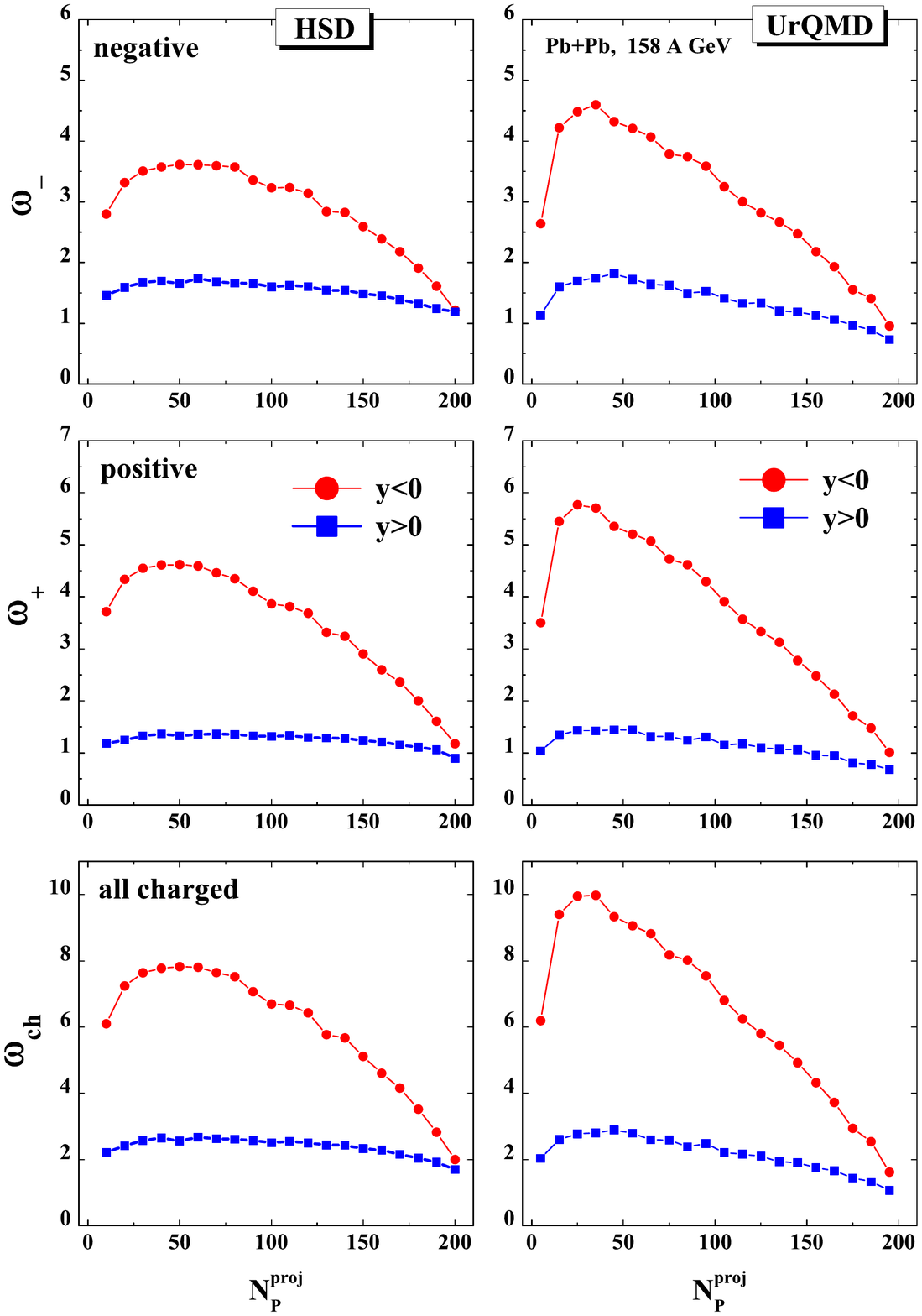}
\end{center}
\caption{The scaled variance $\omega_i$ of the multiplicity
 distribution for particles produced in the projectile (boxes)
 and target (circles) hemisphere in simulations of
 Pb+Pb collisions at beam energy of 158$A$~GeV
 using the HSD ($left$) and UrQMD ($right$) models~\cite{Konch:2006}.}
\label{fig:proj-targ}
\end{figure}

As an illustrative example the fluctuations of the particle multiplicities in the
projectile ($y>0$) and target ($y<0$) hemispheres are now considered
within the transport models  HSD and UrQMD (see Ref.~\cite{Konch:2006}).
As one can see from Fig.~\ref{fig:w_targ}, the number of target participants,
$N_P^{targ}$, fluctuates considerably in samples with
$N_P^{proj} = const$~.
Because of this asymmetry between projectile and
target participants the transport
models give very different results for the particle number
fluctuations (quantified by the scaled variance $\omega$ of
the multiplicity distribution) in the projectile and target hemispheres. As clearly
seen from Fig.~\ref{fig:proj-targ} the particle number
fluctuations in the target hemisphere are much stronger  than
those in the projectile hemisphere. Note that there is also a strong
$N_P^{proj }$-dependence of  $\omega_i$ in the target hemisphere,
which is almost absent in the projectile hemisphere.
Thus, the fluctuations of $N_P^{targ}$ have a small influence on
the final multiplicity fluctuations in the projectile hemisphere,
but they contribute very strongly to those in the target
hemisphere. These results demonstrate that the HSD and UrQMD transport
models belong to the T-type. Note that they do not have enough
mixing effects to explain the data in A+A collisions~\cite{mixing}.

Some combination of the system size $N_P$ and collision energy
$E_{lab}$ might move the chemical freeze-out point close to the QCD
CP.
One then expects an increase of multiplicity
fluctuations with respect to their `background values'.
Fig.~\ref{fig:w_targ} presents from the HSD model the scaled variance
$\omega_P^{targ}$ for C+C, O+O, Ne+Ne, S+S, In+In, and Pb+Pb
collisions at 158~AGeV as a function of $N_P^{proj}$.
Why does one need to study central collisions for ions of small and
intermediate mass number $A$
instead of peripheral Pb+Pb collisions in a search
of the CP ?
Figure~\ref{fig:w_targ} explains this issue.
At fixed $N_P^{proj}$ the average total number of participants,
$N_P\equiv N_P^{proj}+N_P^{targ}$, is equal to $\langle N_P\rangle
\cong 2 N_P^{proj}$, and, thus, it fluctuates as
$\omega_P=0.5 \omega^{targ}_P$. Then, for example, the value of
$N_P^{proj}\cong 30$ corresponds to almost zero participant number
fluctuations, $\omega_P\cong 0$, in S+S collisions while
$\omega_P$ becomes large and is close to 1 and 1.5 for In+In and
Pb+Pb collisions of the same $N_P^{proj}$, respectively.
Even if $N_P^{proj}$ is fixed exactly, the
sample of peripheral collision events in the heavy-ion case
contains large fluctuations of the participant number:  this would
`mask' the CP
signals. As also seen from
Fig.~\ref{fig:w_targ}, the picture becomes actually
more complicated if the atomic mass  number $A$ is  small. In
this case, the number of participants from the target
fluctuates significantly even for the largest fixed value of
$N_P^{proj}=$A.

\subsection{Strongly Intensive Quantities  $\Delta$ and $\Sigma$}\label{SIQ}

A significant increase of transverse momentum and
multiplicity fluctuations is expected in the vicinity of the
CP
\cite{fluc3,fluc3a,fluc3b}. One can probe different regions of the
phase diagram by varying the collision energy and size of
colliding nuclei \cite{size}. The possibility to observe signatures
of the critical point inspired the energy and system size scan
program of the NA61/SHINE Collaboration at the CERN
SPS \cite{Ga:2009} and the BES  program
of the STAR and PHENIX Collaborations at the
BNL RHIC
\cite{RHIC-SCAN}. In these studies one measures and then compares
e-by-e
fluctuations in collisions of different nuclei at
different collision energies. The average sizes of the created
physical systems and their fluctuations are expected to be
rather different (see previous section~\ref{NPFLCT}). This strongly affects the
observed fluctuations, i.e. the measured quantities would
not describe the local physical properties of the system but
rather reflect the system size fluctuations. For instance, A+A
collisions with different centralities may produce a system with
approximately the same local properties (e.g. the same
temperature and baryonic chemical potential) but with the volume
changing significantly from interaction to interaction. Note that
in high energy collisions the average volume  of created matter
and its variations from collision to collision usually cannot be
controlled experimentally.
Therefore, a suitable choice of
statistical tools for the study of e-by-e fluctuations is really
important.

Intensive quantities are defined within the grand canonical
ensemble of statistical mechanics. They depend on temperature and
chemical potential(s), but they are independent of the system
volume.
Strongly intensive quantities introduced in Ref.~\cite{GG:2011} are, in addition,
independent of volume fluctuations.
They are the appropriate
measures for studies of e-by-e
fluctuations
in A+A collisions
%
and can be defined from two,
{\it extensive state quantities} $A$ and $B$. Here, we call $A$
and $B$ {\it extensive} when the first moments of their
distributions for the ensemble of possible states are proportional
to volume. They are referred to as {\it state quantities} as they
characterize the states of the considered system, e.g. final
states
of A+A collisions or
micro-states of the grand canonical ensemble. For example, $A$ and
$B$ may stand for multiplicities of kaons and pions, respectively, in a
particular state.

%
%
%

There are two families of strongly intensive quantities which
depend on the second and first moments of $A$ and $B$ and thus
allow to study state-by-state fluctuations~\cite{GG:2011}:
 \eq{\label{Delta-AB}
 \Delta[A,B]
 ~&=~ \frac{1}{C_{\Delta}} \Big[ \langle B\rangle\,
      \omega[A] ~-~\langle A\rangle\, \omega[B] \Big]~,
 \\
  \Sigma[A,B]
 ~&=~ \frac{1}{C_{\Sigma}}\Big[
      \langle B\rangle\,\omega[A] ~+~\langle A\rangle\, \omega[B] ~-~2\left(
      \langle AB \rangle -\langle A\rangle\langle
      B\rangle\right)\Big]~,
 \label{Sigma-AB}
 }
where
 \eq{\label{omega-AB}
 \omega[A] ~\equiv~ \frac{\langle A^2\rangle ~-~ \langle A\rangle^2}{\langle
 A\rangle}~,~~~~
 \omega[B] ~\equiv~ \frac{\langle B^2\rangle ~-~ \langle B\rangle^2}{\langle
 B\rangle}~,
 }
and averaging $\langle \ldots \rangle$ is performed over the
ensemble of multi-particle states.
The normalization factors
$C_{\Delta}$ and $C_{\Sigma}$ are required to be proportional to
the first moments of any extensive quantities.

In Ref.~\cite{GGP:2013} a specific choice of the $C_{\Delta}$ and
$C_{\Sigma}$ normalization factors was proposed. It  makes the quantities
$\Delta[A,B]$ and $\Sigma[A,B]$ dimensionless and  leads to
$\Delta[A,B] = \Sigma[A,B] = 1$ in the independent particle
model,  as will be shown below.

From the definition of $\Delta[A,B]$ and $\Sigma[A,B]$ it follows that
$\Delta[A,B] = \Sigma[A,B] = 0$ in the case of absence of
fluctuations of $A$ and $B$, i.e., for $ \omega[A] = \omega[B] =
\langle AB \rangle -\langle A\rangle \langle B \rangle = 0 $. Thus
the proposed normalization of $\Delta[A,B]$ and $\Sigma[A,B]$
leads to a common scale on which the values of the fluctuation
measures calculated for different state quantities $A$ and $B$ can
be compared.

There is an important  difference between the $\Sigma[A,B]$ and
$\Delta[A,B]$ quantities. Namely, in order to calculate
$\Delta[A,B]$ one  needs to measure only the first two moments:
$\langle A\rangle$, $\langle B\rangle$ and $\langle A^2\rangle$,
$\langle B^2\rangle$. This can be done by  independent
measurements of the distributions $P_A(A)$ and $P_B(B)$. The
quantity $\Sigma[A,B]$ includes the correlation term, $\langle AB
\rangle -\langle A \rangle \langle B \rangle$, and thus
requires, in addition,  simultaneous measurements of $A$ and $B$
in order to obtain the joint distribution $P_{AB}(A,B)$. Under exchange
of $A$ and $B$ the quantities $\Sigma[A,B]$ and $\Delta[A,B]$ have
the property $\Sigma[A,B]=\Sigma[B,A]$
and $\Delta[A,B]= -~ \Delta[B,A]$. Using the last relation one can
always define $\Delta \geq 0$  by exchanging the $A$ and $B$
quantities.

\subsubsection{$\Delta$ and $\Sigma$ in the Independent Particle Model }\label{IPM}

The Independent Particle Model (IPM) assumes that:
\begin{enumerate}[(i)]
\setlength{\itemsep}{1pt} \item the state
quantities $A$ and $B$ can be
expressed as
\eq{\label{ind-part}
A~=~\alpha_1~+\alpha_2~+\ldots~+
\alpha_{N}~,~~~~~B~=~\beta_1~+\beta_2~+\ldots~+ \beta_{N}~,
}
where  $\alpha_j$ and $\beta_j$ denote single particle
contributions to $A$ and $B$, respectively, and $N$ is the number
of particles;
\item inter-particle correlations are absent, i.e. the probability
of any multi-particle state is the product of probability
distributions $P(\alpha_j,\beta_j)$ of single-particle states, and
these probability distributions are the same for all $j=1,\ldots,
N$ and independent of $N$~:
\eq{ \label{dist-alpha} P_N (\alpha_1,\beta_1,\alpha_2,\beta_2,
\dots, \alpha_N,\beta_N) = {\cal P}(N)\times
P(\alpha_1,\beta_1)\times  P(\alpha_2,\beta_2)\times \cdots \times
P(\alpha_N,\beta_N)~, }
where ${\cal P}(N)$ is an arbitrary multiplicity distribution of
particles.
\end{enumerate}

It can be shown \cite{GGP:2013} that within the
IPM the average values of the first and second moments of $A$ and
$B$ are equal to:
 \eq{\label{A}
 &\langle A \rangle~=~\overline{\alpha} ~\langle N\rangle~,~~~~
 \langle A^2 \rangle~=~\overline{\alpha^2}~\langle N \rangle ~+~
 \overline{\alpha}^2
 ~\left[\langle N^2\rangle ~-~\langle N\rangle\right]~,\\
&\langle B \rangle~=~\overline{\beta}~\langle N\rangle~,~~~~
 \langle B^2 \rangle~=~\overline{\beta^2}~\langle N \rangle ~+~
 \overline{\beta}^2
 ~\left[\langle N^2\rangle ~-~\langle N\rangle\right]~,\label{B} \\
 &\langle AB \rangle~=~
\overline{\alpha\,\beta}~\langle N \rangle ~+~
 \overline{\alpha}\,\cdot\,\overline{\beta}
 ~\left[\langle N^2\rangle ~-~\langle
 N\rangle\right]~.\label{AB}
}
The values of $\langle A \rangle$ and $\langle B \rangle$ are
proportional to the average number of particles $\langle N
\rangle$ and, thus, to the average size  of the system. These
quantities are extensive.  The quantities $\overline{\alpha}$,
~$\overline{\beta}$ and $\overline{\alpha^2}$,
~$\overline{\beta^2}$, ~$\overline{\alpha\,\beta}$ are the first
and second moments of the single-particle distribution
$P(\alpha,\beta)$. Within the IPM they are independent of $\langle
N \rangle$ and play the role of intensive quantities.

Using Eq.~(\ref{A}) the scaled variance $\omega[A]$ which
describes the state-by-state fluctuations of $A$ can be expressed
as:
 \eq{\label{omega-A}
  \omega[A]~\equiv~ \frac{\langle A^2\rangle
~-~\langle A \rangle ^2}{\langle
A\rangle}~=~\frac{\overline{\alpha^2}~-~\overline{\alpha}^2}
{\overline{\alpha}}~+~\overline{\alpha}~\frac{\langle N^2\rangle
~-~\langle N\rangle^2}{\langle N\rangle}
~\equiv~\omega[\alpha]~+~\overline{\alpha}~ \omega[N]~,
}
where $\omega[\alpha]$ is the scaled variance of the
single-particle quantity $\alpha$, and $\omega[N]$ is the scaled
variance of $N$.
A similar expression follows from Eq.~(\ref{B}) for the scaled
variance $\omega[B]$. The scaled variances $\omega[A]$ and
$\omega[B]$ depend on the fluctuations of the particle number via
$\omega[N]$. Therefore, $\omega[A]$ and $\omega[B]$ are not
strongly intensive quantities.

From Eqs.~(\ref{A}-\ref{AB}) one obtains expressions for
$\Delta[A,B]$ and $\Sigma[A,B]$, namely:
\eq{
\Delta[A,B]~&=~ \frac{\langle N \rangle}{C_{\Delta}}~
\Big[~\overline{\beta}~ \omega[\alpha] ~-~\overline{\alpha}
~\omega[\beta]~\Big]~, \label{IPM-D}\\
\Sigma[A,B]~&=\frac{\langle N
\rangle}{C_{\Sigma}}~\Big[~\overline{\beta}~\omega[\alpha]~+~\overline{\alpha}~
\omega[\beta] ~-~2\left(~\overline{\alpha\, \beta}
-\overline{\alpha}\cdot \overline{\beta}~\right)~\Big]~.
\label{IPM-S}
}
Thus, the requirement that
\eq{\label{DS=1}
\Delta[A,B]~ = ~\Sigma[A,B]~ =~ 1~,
}
within the IPM leads to:
\eq{
C_{\Delta}~&=~\langle N \rangle~ \Big[~\overline{\beta}~
\omega[\alpha] ~-~\overline{\alpha}
~\omega[\beta]~\Big]~, \label{C-D}\\
C_{\Sigma}~&=~\langle N
\rangle~\Big[~\overline{\beta}~\omega[\alpha]~+~\overline{\alpha}~
\omega[\beta] ~-~2\left(~\overline{\alpha\, \beta}
-\overline{\alpha}\cdot \overline{\beta}~\right)~\Big]~.
\label{C-S}
}

In the IPM the  $A$ and $B$ quantities are expressed in terms of
sums of the single particle variables, $\alpha$ and $\beta$. Thus
in order to calculate the normalization $C_{\Delta}$ and
$C_{\Sigma}$ factors one has to measure the single particle
quantities $\alpha$ and $\beta$. However, this may not  always be
possible within a given experimental set-up. For example, $A$ and
$B$ may be energies of particles measured by two calorimeters.
Then one can study fluctuations in terms of $\Delta[A,B]$ and
$\Sigma[A,B]$ but can not calculate the normalization factors which
are proposed above.

For illustrative purposes we consider in more detail a
specific pair of extensive
variables: the transverse momentum $A=P_T=p_T^{(1)}+\dots
p_T^{(N)}$, where $p^{(i)}_T$ is the absolute value of the
$i^{{\rm th}}$ particle transverse momentum, and the number of
particles $B=N$.
%
%
To simplify notations we use $X=P_T$ and
$x_i=p_T^{(i)}$. Note that our consideration is also valid for
other motional variables $X$, e.g. the system energy
$X=E=\epsilon_1+\dots+\epsilon_N$.

First, we consider the IPM
which is used as a reference model to fix the normalization of the
strongly intensive measures $\Delta$ and $\Sigma$. The IPM assumes
that $A$, the motional extensive variable, can be written as a sum of single particle terms
\eq{\label{X}
X~=~x_1~+x_2~+\ldots~+ x_{N}~,
}
and $B=N$ is the number of particles. Inter-particle
correlations are absent in the IPM, i.e. the probability of any
multi-particle state is a product of probability distributions
$F(x_j)$ of single-particle variables $x_j$, and these probability
distributions are the same for all $j=1,\ldots, N$ and independent
of the number of particles $N$~:
\eq{ \label{dist-X}
F_N (x_1,x_2, \dots, x_N)~ = ~{\cal P}(N)\times
F(x_1)\,F(x_2)\times \cdots \times F(x_N)~,
}
where ${\cal P}(N)$ is an arbitrary multiplicity distribution of
particles. The functions entering Eq.~(\ref{dist-X}) satisfy the
normalization conditions:
\eq{\label{norm-P}
\sum_N {\cal P}(N)~=~1~,~~~~~\int dx~F(x)~=~1~.
}
The averaging procedure for $k^{{\rm th}}$ moments of any
multi-particle observable $A$ reads:
\eq{\label{av}
\langle A^k \rangle ~=~\sum_{N}{\cal P}(N) \int dx_1dx_2\ldots dx_N  F(x_1)\,F(x_2)\times
\cdots \times F(x_N)~\Big[A(x_1,x_2,\ldots,x_N)\Big]^k~.
}
For the first and second moments of $X$ and $N$ one obtains:
\eq{\label{IPM-X}
& \langle X\rangle =
\overline{x}\cdot \langle N\rangle~,~~~~
\langle X^2\rangle =\overline{x^2}\cdot\langle N\rangle
+\overline{x}^2\cdot\left[\langle N^2\rangle -\langle
N\rangle\right]~,~~~~
\langle XN\rangle =\overline{x}\cdot \langle N^2\rangle~,
 }
where
\eq{
\langle N^k\rangle ~=~\sum_N {\cal P}(N)\,
N^k~,~~~~~\overline{x^k}~=~\int dx\, F(x)\, x^k~.
}
Note that the over-line denotes averaging over a single particle
inclusive distribution, whereas $\langle \dots \rangle$ represents
event averaging over multi-particle states of the system,
e.g. e-by-e averaging over hadrons detected in A+A collisions.

Using Eq.~(\ref{IPM-X}) one finds
\eq{\label{omega-X-IPM}
& \omega[X]\equiv~\frac{\langle X^2\rangle -\langle
X^2\rangle}{\langle
X\rangle}~=~\frac{\overline{x^2}-\overline{x}^2}{\overline{x}}~+~
\overline{x}\cdot \frac{\langle N^2\rangle -\langle N\rangle^2}{\langle N\rangle}
~\equiv~\omega[x]~+~\overline{x}\cdot \omega[N]~,\\
&\langle XN\rangle ~-~\langle X\rangle \,\langle N\rangle
~=~\overline{x}\cdot \Big[\langle N^2\rangle ~-~\langle
N\rangle^2\Big]~\equiv~\overline{x}\cdot \langle N\rangle~ \omega[N]~,\label{XN-IPM}
}
 and finally,
 \eq{\label{Delta-IPM}
 &\Delta[X,N]
 ~=~ \frac{1}{C_{\Delta}}~ \Big[ \langle N\rangle\,
      \omega[X] ~-~\langle X\rangle\, \omega[N] \Big]~
      =~\frac{\omega[x]\cdot \langle N\rangle}{C_{\Delta}}~, \\
&\Sigma[X,N]
 ~=~ \frac{1}{C_{\Sigma}} ~ \Big[ \langle N\rangle\,
      \omega[X] ~+~\langle X\rangle\, \omega[N] ~-~2\Big(\langle X\,N\rangle
      ~-~\langle X\rangle \langle N\rangle\Big)\Big]~
      =~\frac{\omega[x]\cdot \langle N\rangle}
      {C_{\Sigma}}~ .\label{Sigma-IPM}
}
The requirement that
\eq{\Delta[X,N]~=~\Sigma[X,N]~=~1 \label{DelSig=1}
}
thus leads for the IPM to the normalization factors
\eq{\label{norm}
 C_{\Delta}~=~C_{\Sigma}=~
\omega[x]\cdot \langle N\rangle~,~~~~~
 \omega[x]~\equiv~\frac{\overline{x^2}~-~\overline{x}^2}
{\overline{x}}~.
}
In the IPM the event quantity $X$ can be expressed as a sum of
single-particle quantities $x_i$, see Eq.~(\ref{X}).
The single-particle quantities are needed to calculate
the normalization factors $C_{\Delta}$ and $C_{\Sigma}$.
Note that the event quantity $X$ can be measured directly
without measurements of the $x_i$ quantities.

As a next example let us consider multiplicity fluctuations. Here $A$ and $B$
will denote multiplicities of hadrons of types $A$ and $B$, respectively (e.g.
kaons and pions). Particle identities were introduced in Ref.~\cite{MG1999} to
study fluctuations between particle types, so-called ``chemical'' fluctuations.
One defines the identities $w_A^{j}$ and $w_B^{j}$
as $w_A^{j}=1$ and $w_B^{j}=0$ if the $j^{th}$ particle is
of $A$ type, and $w_A^{j}=0$ and $w_B^{j}=1$
if the $j^{th}$ particle is of $B$ type.
For single-particle quantities $\alpha\equiv w_A$ and
$\beta\equiv w_B$ one obtains:
\eq{\label{wAB}
\overline{\alpha}=\overline{\alpha^2}=\frac{\langle A\rangle}{\langle A+B\rangle}~,~~~~~
\overline{\beta}=\overline{\beta^2}=\frac{\langle B\rangle}{\langle A+B\rangle}~,~~~~~
\overline{\alpha\,\beta}=0~.
}
From Eq.~(\ref{wAB}) one then finds
\eq{\label{omegawAB}
\omega[\alpha]=\overline{\beta}~,~~~~~\omega[\beta]=\overline{\alpha}~,~~~~~
\overline{\alpha\,\beta}-\overline{\alpha}\cdot\overline{\beta}=
-\overline{\alpha}\cdot\overline{\beta}~.
}
Using Eqs.~(\ref{wAB}) and (\ref{omegawAB}) one obtains from Eqs.~(\ref{C-D}) and
(\ref{C-S}):
\eq{\label{CAB}
C_{\Delta}=\langle B\rangle -\langle A\rangle~,~~~~C_{\Sigma}=\langle A\rangle+\langle B\rangle~.
}
As seen from Eq.~(\ref{CAB}) the normalization factors depend only on the first moments
of extensive quantities $A$ and $B$. In general, more information is needed to calculate
$C_{\Delta}$ and $C_{\Sigma}$ when the multiplicities $A$ and $B$ correspond to partly overlapping
sets of particles, e.g. $A={\rm K}^+ + {\rm K}^-$ and $B={\rm H}^-$,
where ${\rm H}^-$ means the number of all negatively charged hadrons.

The normalization factors (\ref{norm}) and (\ref{CAB}) are suggested to be used for the calculation
for $\Delta$ and $\Sigma$ both in theoretical models and for the analysis of
experimental data (see Ref.~\cite{GGP:2013} for further details of the normalization
procedure).

As was noted in Ref.~\cite{GG:2011} the $\Phi$ measure, introduced some time ago~\cite{GM:1992},
belongs to the $\Sigma$ family within the current classification scheme. The fluctuation measure $\Phi$
was introduced for the study of transverse momentum fluctuations.
In the general case, when $A=X$  represents
any motional variable and $B=N$ is the particle multiplicity, one gets:
\eq{\label{Phi}
\Phi_X~=~\Big[\overline{x}\,\omega[x]\Big]^{1/2}\,\Big[\sqrt{\Sigma[X,N]}
~-~1\Big]~.
}
For the multiplicity fluctuations of hadrons belonging to non-overlapping types $A$ and $B$
the connection between the $\Phi[A,B]$ and $\Sigma[A,B]$ measures reads:
\eq{\label{phiAB}
\Phi[A,B]~=~\frac{\sqrt{\langle A\rangle\,\langle B\rangle}}{\langle A\rangle +\langle B\rangle}\,
\Big[\sqrt{\Sigma[A,B]}~-~1\,\Big]~.
}

The IPM plays an important role as the {\it reference
model}. The deviations of real data from the IPM
results Eq.~(\ref{DelSig=1}) can be used to learn about the physical
properties of the system. This resembles the situation in
studies of particle multiplicity fluctuations.
In this case, one uses the Poisson distribution
$P(N)=\exp\left(-\,\overline{N}\right)\,\overline{N}^N/{N!}$~
with $\omega[N]=1$ as the reference model. The other reference value
$\omega[N]=0$ corresponds to $N={\rm const}$, i.e. the absence
of $N$-fluctuations.
Values of $\omega[N]>1$ (or $\omega[N]\gg 1$) correspond to ``large''
(or ``very large'') fluctuations of $N$, and $\omega[N]<1$
(or $\omega[N]\ll 1$) to ``small''
(or ``very small'') fluctuations.

\subsubsection{$\Delta$ and $\Sigma$ in the Multiple Independent Source (MIS) model}\label{MIP}

Next, we study the Multiple Independent Source (MIS) model for multi-particle production.
In this model the number of
sources, $N_S$, changes from event to event. The sources are
statistically identical and independent of each other. A famous
example of  the MIS is the wounded nucleon model \cite{WNM} for
A+A collisions . Two fluctuating extensive quantities $X$ and $N$
can be expressed as
\eq{\label{ind-sour}
X~=~X_1~+X_2~+\ldots~+ X_{N_S}~,~~~~~N~=~n_1~+n_2~+\ldots~+
n_{N_S}~,
}
where   $n_j$ denotes the number of particles emitted from the
$j^{{\rm th}}$ source ($j=1,\ldots,N_S$), and
$X_j=x_1+\dots+x_{n_j}$ is the contribution from the $j^{{\rm
th}}$ source to the quantity $X$.

The over-line notations are used for the averages connected to a
single source. The single-source quantities are independent of
$N_S$ and have the properties of intensive quantities. The
single-source distribution $F_S(X_S,n)$ is assumed to be
statistically identical for all sources, thus, for all
$j=1,\dots,N_S$ it follows:
\eq{\label{id-source}
\overline{X_j^k}~\equiv ~\overline{X_S^k}~,~~~~
\overline{n_j^k}~\equiv ~\overline{n^k}~,~~~
\overline{X_jn_j}~\equiv ~\overline{X_Sn}~,
}
where  $\overline{X_S^k}$, $\overline{n^k}$, and
$\overline{X_S\,n}$  (for $k=1,2$) are the first and second
moments of the distribution $F_S(X_S,n)$ for a single source.
The sources are assumed to be independent. This gives at $i\neq j$:
\eq{\label{ind-source}
\overline{X_iX_j}~\equiv ~\overline{X_S}^2~,~~~~
\overline{n_in_j}~\equiv ~\overline{n}^2~,~~~
\overline{X_in_j}~\equiv ~\overline{X_S}~\overline{n}~.
}
Using Eqs.~(\ref{id-source}) and (\ref{ind-source})
one finds for the event averages:
 \eq{
 &\langle X \rangle~=~\overline{X_S}\cdot\langle N_S\rangle~,~~~~
\langle X^2 \rangle~=~\overline{X_S^2}\cdot \langle N_S \rangle ~+~
 \overline{X_S}^2
 ~\left[\langle N_S^2\rangle ~-~\langle N_S\rangle\right]~,\label{WNM-X2}\\
&\langle N \rangle~=~\overline{n}~\langle N_S\rangle~,~~~~
\langle N^2 \rangle~=~\overline{n^2}\cdot \langle N_S \rangle ~+~
 \overline{n}^2\cdot
\left[\langle N_S^2\rangle ~-~\langle N_S\rangle\right]~,\label{WNM-N2} \\
 &\langle X\,N \rangle~=~
\overline{X_S\,n} ~\langle N_S \rangle ~+~
 \overline{X_S}~\overline{n}\cdot
 \left[\langle N_S^2\rangle ~-~\langle
 N_S\rangle\right]~.\label{WNM-XN}
}
The probability distribution ${\cal P}_S(N_S)$ of the number of
sources is needed to calculate $\langle N_S\rangle $ and $\langle
N_S^2\rangle$ and, in general, it is unknown.

Using Eqs.~(\ref{WNM-X2}-\ref{WNM-XN}) one obtains:
 \eq{\label{WNM-omega-X}
  \omega[X]&\equiv ~\frac{\langle X^2\rangle
-\langle X \rangle ^2}{\langle X\rangle}=\frac{
\overline{X_S^2}-\overline{X_S}^2}{\overline{X_S}}+\overline{X_S}
\cdot \frac{\langle N_S^2\rangle -\langle
N_S\rangle^2}{\langle N_S\rangle} ~\equiv~\omega[X_S]~+~\overline{X_S}\cdot
\omega[N_S]~,\\
\label{WNM-omega-N}
  \omega[N]&\equiv~ \frac{\langle N^2\rangle
~-~\langle N \rangle ^2}{\langle N\rangle}~=~\frac{
\overline{n^2}~-~\overline{n}^2}{\overline{n}}~+~\overline{n}
\cdot \frac{\langle N_S^2\rangle ~-~\langle
N_S\rangle^2}{\langle N_S\rangle} ~\equiv~\omega[n]~+~\overline{n}\cdot
\omega[N_S]~,
}
where $\omega[X_S]$ and $\omega[n]$ are the scaled variances for
quantities $X_S$ and $n$ referring to a single source.  The scaled
variances $\omega[X]$ and $\omega[N]$ are independent of the
average number of sources $\langle N_S\rangle$. Thus, $\omega[X]$
and $\omega[N]$ are intensive quantities. However, they depend on
the fluctuations of the number of sources  via $\omega[N_S]$ and,
therefore, they are not strongly intensive quantities.

From Eqs.~(\ref{WNM-XN}-\ref{WNM-omega-N}) it follows:
 \eq{
\Delta[X,N]~&=~\frac{1}{\omega[x]}~
\Big[\, \omega[X_S] ~-
~
\overline{x}\cdot \omega[n] \,\Big]~. \label{D-WNM}\\
\Sigma[X,N]~&=~
\frac{1}{\omega[x]}~
\Big[\,
\omega[X_S] ~+~
\overline{x}\cdot \omega[n]~-~
2\,\frac{\overline{X_S\, n} -\overline{x}~\overline{n}^2}{\overline{n}}\,\Big]~,
\label{S-WNM}
}
where the relations
\eq{\label{XS}
 \overline{x}~=~ \frac{\overline{X_S}}{\overline{n}}~=~\frac{\langle X\rangle}
{\langle N\rangle}~,
}
and the normalization factors (\ref{norm}) were used.

Note that the terms with $\langle N_S^2\rangle$, which are present
in the expressions (\ref{WNM-X2}-\ref{WNM-XN}) for the second
moments of $X$ and $N$, are canceled out in the final expressions
(\ref{D-WNM},\ref{S-WNM}). From three second moments $\langle
X^2\rangle$, $\langle N^2\rangle$, and $\langle X\,N\rangle$ just
two linear combinations independent of $\langle N_S^2\rangle$ can
be constructed. These are the strongly intensive
quantities $\Delta$ and $\Sigma$.
%


The IPM and MIS  have a similar structure. The difference is that
the number of particles $N$ in the IPM is replaced by
the number of sources $N_S$ in the MIS. Each source can produce many
particles, and the number of these particles varies from source to
source and from event to event. Besides, the physical quantity
$X_S$ for particles emitted from the same source may include
inter-particle correlations. Therefore, in general, the MIS does
not satisfy the assumptions of the IPM. Nevertheless, a formal
similarity between the two models can be exploited and gives the
following rule of one to one correspondence: all results for  the
IPM can be found from the expressions obtained for the MIS,
assuming artificially that each source always produces exactly one
particle. In this case one finds
\eq{
\overline{n}=1~,~~~~\omega[n]=0~,~~~~\omega[X_S]=\omega[x]~,~~~~
\overline{X_S\,n}=\overline{x}~,
}
and Eqs.~(\ref{D-WNM}-\ref{S-WNM}) are transformed to Eq.~(\ref{DelSig=1}).

If particles are independently emitted from a single source, one
obtains
\eq{\label{PSx}
F_S(X_S,n)~=~{\cal P}_S(n)\times F_S(x_1)\times \dots \times F_S(x_n)~,
}
with the probability distributions  $F_S(x_i)$ which are the same
for all $i=1,\ldots, n$ and independent of the number of particles
$n$. Similar to Eqs.~(\ref{omega-X-IPM}) and (\ref{XN-IPM}) one
then finds:
\eq{
 \omega[X_S]~=~\omega[x]~+~\overline{x}\cdot \omega[n]~,~~~~
 \overline{X_S\,n}~-~\overline{X_S}\,\overline{n}~=~\overline{x}~\overline{n}\cdot \omega[n]~,
}
and Eqs.~(\ref{D-WNM}) and (\ref{S-WNM}) are again transformed to
Eq.~(\ref{DelSig=1}). Therefore, the MIS with independent particle
emission from each source is equivalent to the IPM.

\subsubsection{$\Delta$ and $\Sigma$ Evaluated in Specific Models}\label{SIQ_in_models}

The most popular model which satisfies the IPM assumptions  is the
ideal Boltzmann multi-component gas in the grand canonical
ensemble  formulation, which we refer to as the IB-GCE. In the
IB-GCE the probability of any microscopic state is equal to the
product of probabilities of single-particle states. These
probabilities are independent of particle multiplicity. Thus, the
IB-GCE satisfies the assumption (\ref{dist-alpha}) of  the IPM.

The IB-GCE predicts a specific form of the multiplicity
distribution ${\cal P}(N)$, namely the Poisson distribution, and
thus, $\omega[N]=1$. Moreover, it also predicts the specific form
of the single-particle probability in momentum space, namely the
Boltzmann distribution:
\eq{\label{Boltz}
f_B({\bf p})~=~ C\,\exp\left(-\,\frac{\sqrt{{\bf
p}^2+m^2}}{T}\right)~,
}
where ${\bf p}$ and $m$ are particle momentum and mass,
respectively, $T$ is the system temperature, and $C=\left[\int
d^3p \exp[-\,\sqrt{p^2+m^2}/T] \right]^{-1}$ is the normalization constant.

Note that by introducing quantum statistics one destroys the
correspondence between the GCE and the IPM. This is because of
(anti-)correlation between particles in the same quantum state for
the (Fermi) Bose ideal gas. Moreover, correlations between
particles are introduced if instead of resonances their decay
products are considered. Note that it is necessary to include
strong decays of resonances in order to compare the GCE
predictions to experimental results.
%

The correspondence between the IB-GCE and the IPM
remains valid even if the volume varies from micro-state to
micro-state\footnote{Statistical ensembles with volume
fluctuations were discussed in Ref.~\cite{volume,volume1}} but local
properties of the system, i.e., temperature and chemical
potentials are independent of the system volume. Let volume
fluctuations  be given by the probability density function
$F(V)$. The averaging over all micro-states includes the averaging
over the  micro-states with fixed volume and the averaging over
the volume fluctuations. The volume fluctuations broaden the
${\cal P}(N)$ distribution and increase its scaled variance:
\eq{\label{omega-j}
\omega[N] ~\equiv~\frac{\langle N^2\rangle~-~\langle
N\rangle^2}{\langle N\rangle } ~=~1~+~\frac{\langle
N\rangle}{\langle V\rangle}\cdot\frac{\overline{
V^2}~-~\overline{V}^2}{\overline{V}} ~,
}
where 
 $\overline{ V^k}~\equiv~\int dV~ F(V)~V^k$~ for $k=1,2$.
The first term on the right hand side  of Eq.~(\ref{omega-j})
corresponds to the particle number fluctuations in the IB-GCE at a
fixed volume $V$ (i.e., this is the scaled variance of the Poisson
distribution), and the second term is the contribution due to the
volume fluctuations.  Equation~(\ref{dist-alpha}) remains valid in
this example, therefore, the IB-GCE with arbitrary volume
fluctuations  satisfies the IPM assumptions.

Frequently used by experimentalists is the Mixed Event Model
employing a Monte Carlo procedure in order to create a sample of
artificial events in which correlations and fluctuations present
in the original ensemble of events are partly removed. The
original and mixed events are analyzed in the same way and the
corresponding results are compared in order to extract the
magnitude of a signal of interest, which by construction should be
present in the original events and absent in the mixed events. The
mixed event procedure is in particular popular in studies of
resonance production, particle correlations due to quantum
statistics and event-by-event fluctuations, see for examples
Ref.~\cite{mixed}.

There are many variations of the Mixed Event Model. Here we
describe the one which in the limit of infinite number of
original and mixed events gives results identical to the IPM.
The procedure to create a mixed event which corresponds to the
given ensemble of original events consists of two steps, namely:
\begin{enumerate}[(i)]
\setlength{\itemsep}{1pt} \item drawing a mixed event multiplicity, $N$,
from the set of multiplicities of all original events;
\item selecting randomly with replacement $N$ particles for the mixed event
from the set of all particles from all original events.
\end{enumerate}
Then the steps one and two are repeated to create the next mixed
event and the procedure is stopped when the desired number of
mixed events is reached.
In the limit of infinite number of original events, the
probability to have two particles from the same original event in
a single mixed event is zero and thus particles in the mixed
events are uncorrelated. Therefore, in this limit the mixed event
model satisfies the IPM assumptions. Note, for an infinite number
of mixed events, the first moments of all extensive quantities and
all single-particle distributions of the original and mixed events
are identical.

\begin{figure}[t]
\centering
\includegraphics[width=0.495\textwidth]{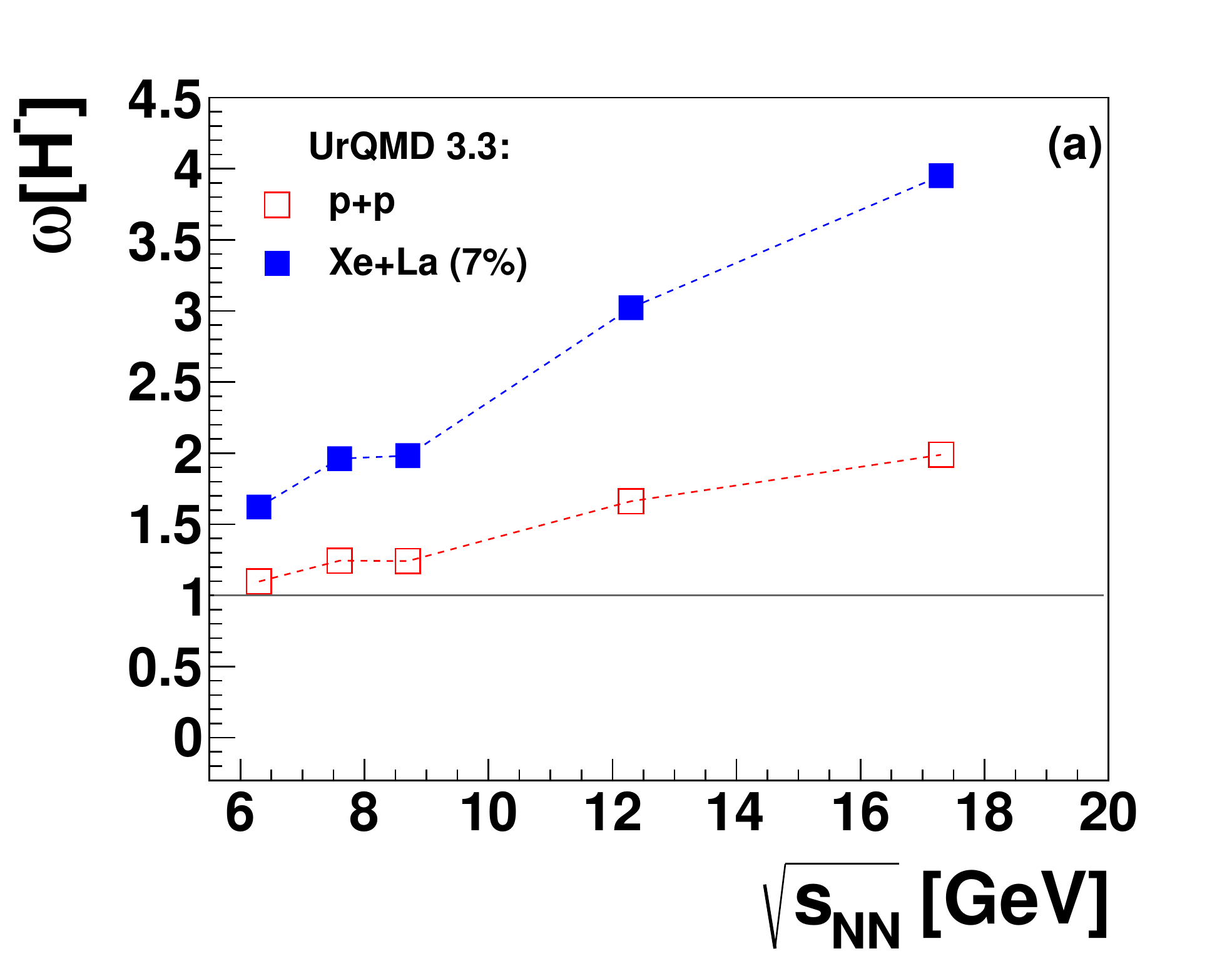}
\includegraphics[width=0.495\textwidth]{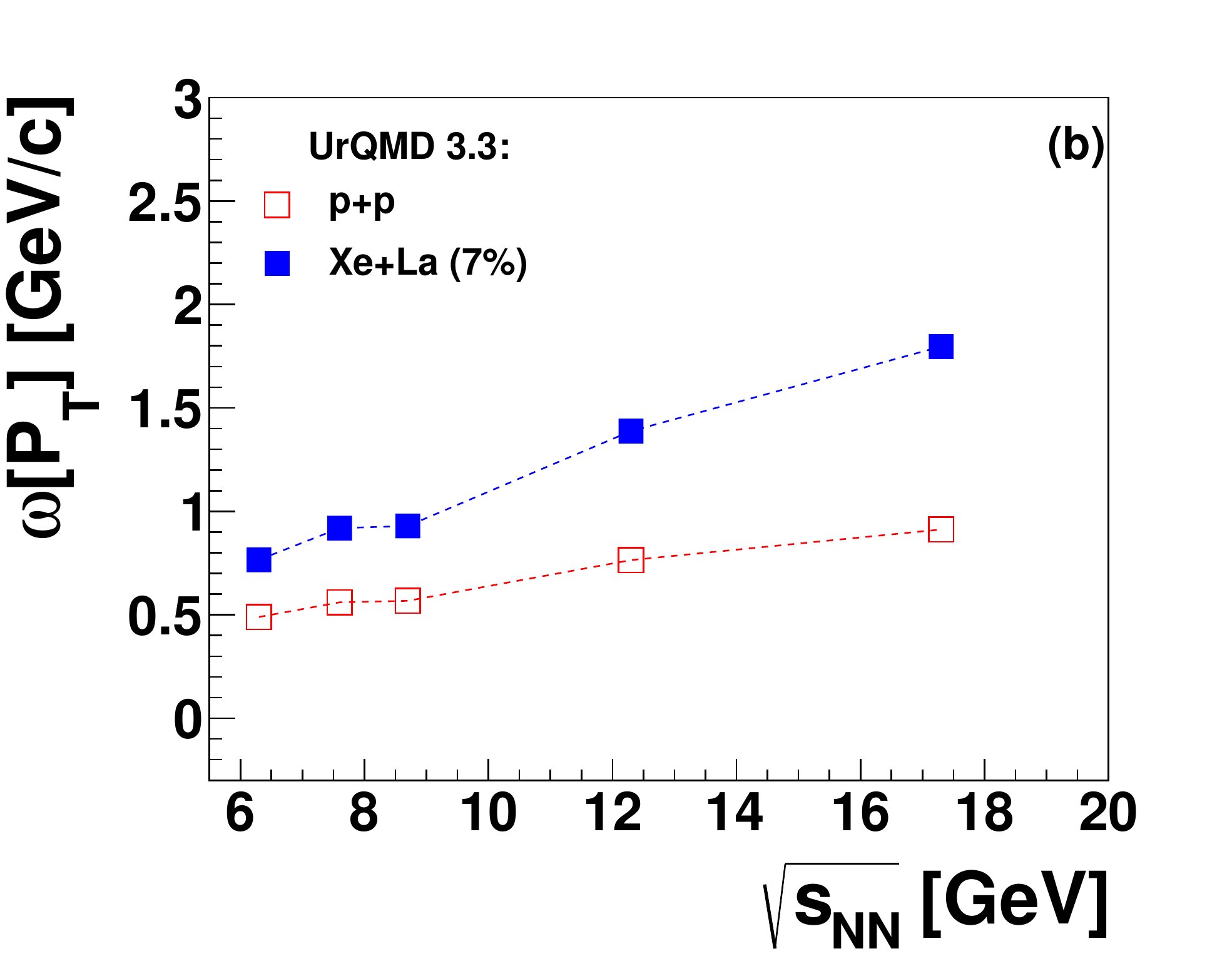}
\includegraphics[width=0.495\textwidth]{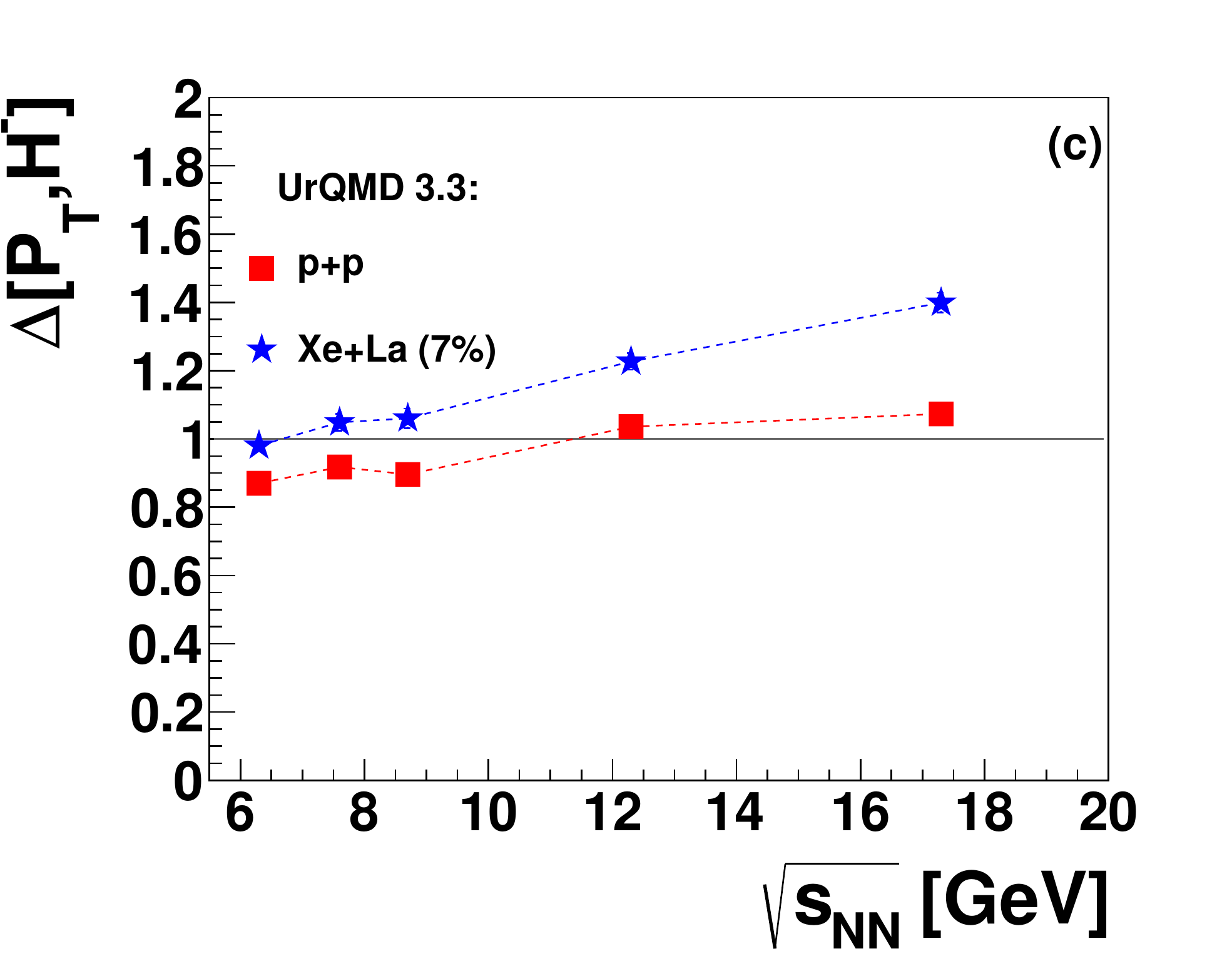}
\includegraphics[width=0.495\textwidth]{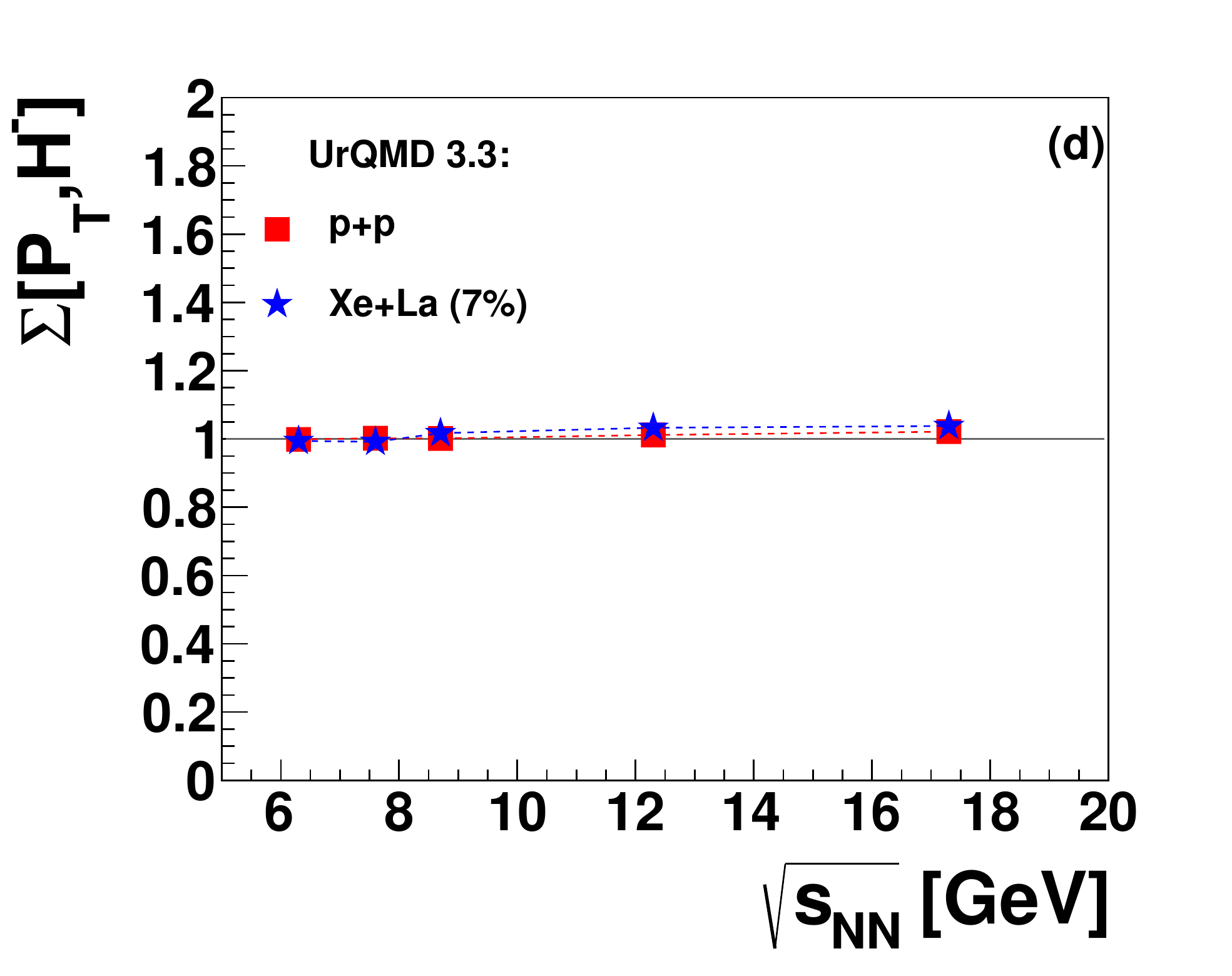}
\caption{ Fluctuation measures calculated within the UrQMD model
for negatively charged hadrons produced in inelastic p+p
interactions and the 7\% most central Xe+La collisions as
functions of collision energy in the CERN SPS
energy range~\cite{GGP:2013}. The
top plots show intensive measures of fluctuations, namely the
scaled variance of (a) the negatively charged hadron
multiplicity, $\omega[H^-]$, and (b) the sum of magnitudes of
their transverse momenta, $\omega[P_T]$. The bottom plots show the
corresponding strongly intensive measures (c) $\Delta[P_T,H^-]$
and (d) $\Sigma[P_T,H^-]$. Statistical uncertainties are smaller
than the symbol size and were calculated using the subsample
method. } \label{fig:pt}
\end{figure}

The UrQMD model~\cite{UrQMD} was used in Ref.~\cite{GGP:2013} to
illustrate the study of the $\Delta$ and $\Sigma$ measures by numerical results.
Figure~\ref{fig:pt} shows the collision energy dependence of these
measures in the CERN SPS energy range.
In this example, $A=P_T$ is the total transverse momentum of
negatively charged hadrons and $B=H^-$ is their multiplicity.
The UrQMD simulations were
performed for inelastic p+p interactions and for the 7\% most
central Xe+La collisions. This choice of reactions is motivated by
the experimental program of the NA61/SHINE
Collaboration~\cite{NA61} at the CERN SPS.
%
The top plots show intensive fluctuation measures, namely the
scaled variance of the negatively charged particle multiplicity
distribution, $\omega[H^-]$, and of the distribution of the sum of
the magnitudes of their transverse momenta, $\omega[P_T]$. The
bottom plots show the corresponding strongly intensive measures
$\Delta[P_T,H^-]$  and $\Sigma[P_T,H^-]$ normalized as proposed
in Eq.~(\ref{norm}).

The scaled variance of $H^-$ and $P_T$ is significantly larger in
central Xe+La collisions than in p+p interactions. To a large
extent this is due to fluctuations of the number of nucleons which
interacted (wounded nucleons), see Ref.~\cite{Be:2012} for a
detailed discussion of this issue. The advantages of the
$\Delta[P_T,H^-]$  and $\Sigma[P_T,H^-]$ quantities are obvious
from the results presented in the bottom plots. First, they are
not directly  sensitive to fluctuations of the collision geometry
(the number of wounded nucleons) in contrast to the scaled
variance. Second, they are dimensionless and expressed in units
common for all energies and reactions.
Due to the particular
normalization proposed in Ref.~\cite{GGP:2013}, they assume the value one
for the IPM and zero in the absence of event-by-event fluctuations.
The UrQMD results for $\Sigma[P_T,H^-]$ shown in Fig.~\ref{fig:pt}
are close to unity both for p+p interactions  and central Xe+La collisions.
This is not the case for the $\Delta[P_T,H^-]$ measure.
One observes both a deviation from the IPM prediction $\Delta[P_T,H^-]=1$
and a difference between results for p+p interactions and central Xe+La collisions.
Therefore, one concludes for the UrQMD model that the measure $\Delta[P_T,H^-]$
is more sensitive to inter-particle correlations than the measure
$\Sigma[P_T,H^-]$.

The  strongly intensive fluctuation measures  $\Delta[P_T,N]$
and $\Sigma[P_T,N]$ were recently studied in
Ref.~\cite{GR:2013} for the ideal Bose and Fermi gas within the
GCE. As already noted above, the GCE for the Boltzmann
approximation satisfies the conditions of the IPM, i.e.
Eq.~(\ref{DS=1}) is valid in the IB-GCE. The following general
relations were found \cite{GR:2013}:
\eq{\label{BBF}
& \Delta^{{\rm Bose}}[P_T,N]~<~\Delta^{{\rm Boltz}}=1~<~\Delta^{{\rm Fermi}}[P_T,N]~,\\
& \Sigma^{{\rm Fermi}}[P_T,N]~<~\Sigma^{{\rm
Boltz}}=1~<~\Sigma^{{\rm Bose}}[P_T,N]~,\label{FBB}
}
i.e.  the Bose statistics makes $\Delta[P_T,N]$ smaller and
$\Sigma[P_T,N]$ larger than unity, whereas the Fermi statistics
works in the opposite way.
The Bose statistics of pions appears to be the main source of
quantum statistics effects in a hadron gas with a temperature
typical for the hadron system created in A+A collisions. It gives
about 20\% decrease of $\Delta[P_T,N]$ and 10\% increase of
$\Sigma[P_T,N]$, at $T\cong 150$~MeV with respect to the IPM
results (\ref{DS=1}). The Fermi statistics of protons modifies
insignificantly $\Delta[P_T,N]$ and $\Sigma[P_T,N]$ for
typical values of $T$ and $\mu_B$. Note that UrQMD takes into account
several sources of fluctuations and correlations, e.g. the exact
conservation laws and resonance decays. On the other hand, it does
not include the effects of Bose and Fermi statistics.

Other examples of Monte Carlo simulations and analytical model
results for $\Delta[P_T,N]$ and $\Sigma[P_T,N]$ were presented in Refs.
\cite{KG:APP2012,GorGr:2013}.

\subsubsection{First Experimental Results on $\Delta[P_T,N]$ and $\Sigma[P_T,N]$
 in p+p and Pb+Pb Collisions}\label{SIQ_experiment}

\begin{figure}[!htb]
\centering
\includegraphics[width=0.80\columnwidth]{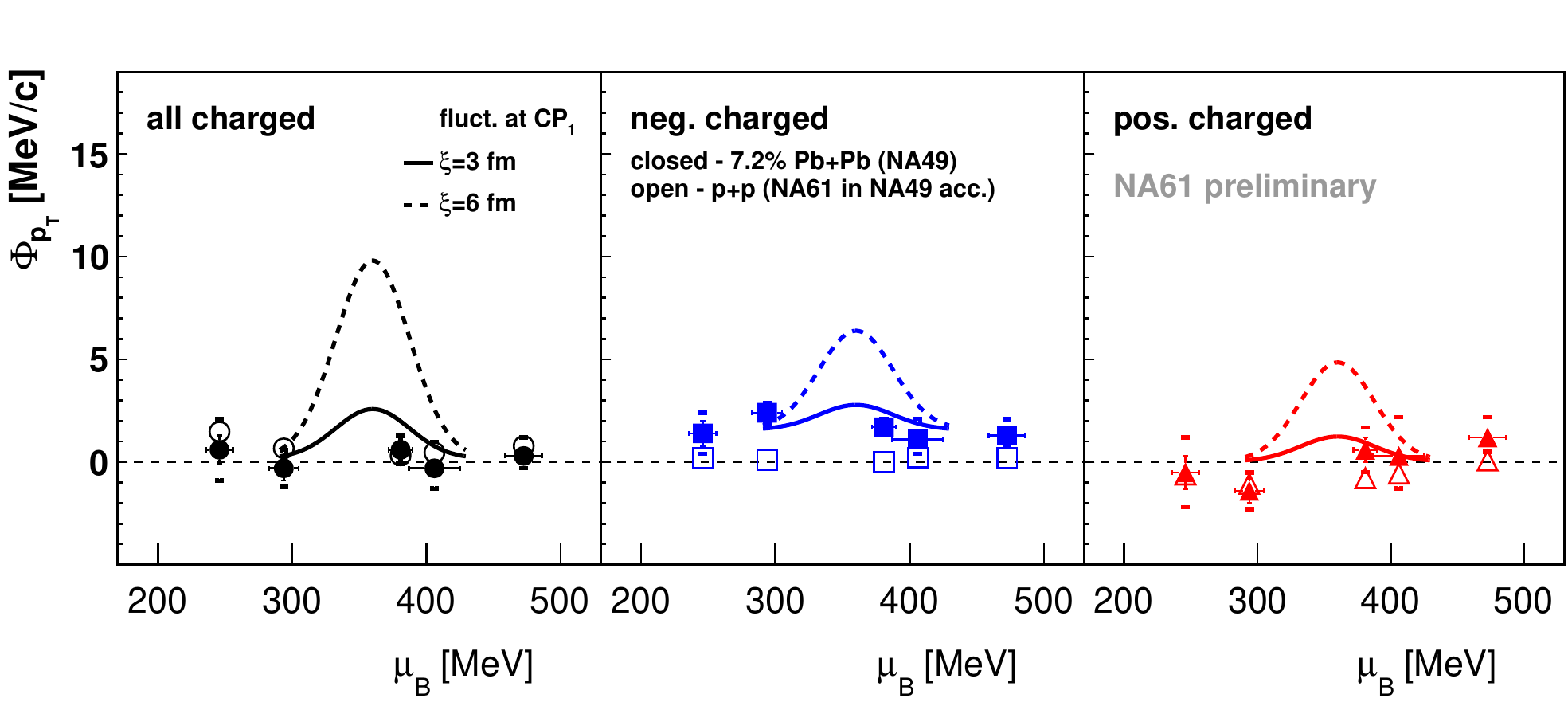}
\caption{Fluctuation measure $\Phi_{p_T}$ of the average transverse momentum of charged particles
versus baryochemical potential $\mu_B$ for the 7.2~\% most central Pb+Pb collisions
(full symbols, NA49~\cite{NA49_Phipt}) and inelastic p+p interactions
(open symbols, NA61/SHINE preliminary~\cite{ismd2013}).
Expectations for a critical point assuming correlation lengths of $\xi$ = 3 (6)~fm are shown
by full (dashed) curves. Results are for cms rapidity $1.1 < y < 2.6$
calculated assuming the pion mass.}
\label{Phipt}
\end{figure}

Fluctuations of the average transverse momentum of charged particles
have been investigated by the
NA49 experiment for some time~\cite{NA49_Phipt} employing the strongly intensive
measure $\Phi_{P_T}$ proposed in Ref.~\cite{GM:1992}.
Published results for the energy dependence of $\Phi_{P_T}$ in central Pb+Pb collisions
are shown in Fig.~\ref{Phipt} versus the baryochemical potential $\mu_B$. Also
plotted for comparison are preliminary results of NA61/SHINE from inelastic p+p interactions.
Curves show predictions~\cite{CP_Phipt,CP_Phipt1} for a critical point located
at $\mu_B$~=~360~MeV~\cite{CP_FK2004}. Clearly there is no evidence for the
expected maximum of fluctuations.

\begin{figure}[!htb]
\centering
\includegraphics[width=0.80\columnwidth]{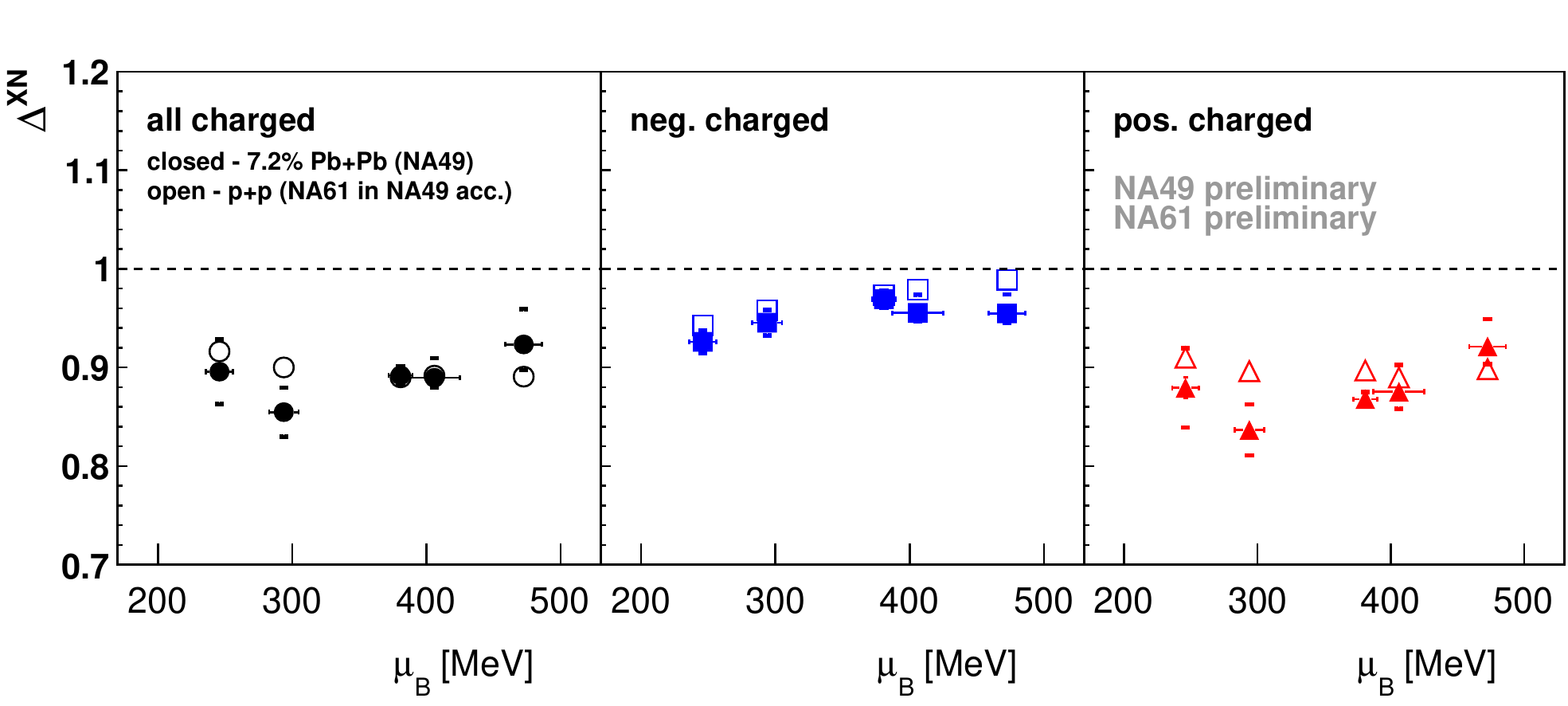}
\includegraphics[width=0.80\columnwidth]{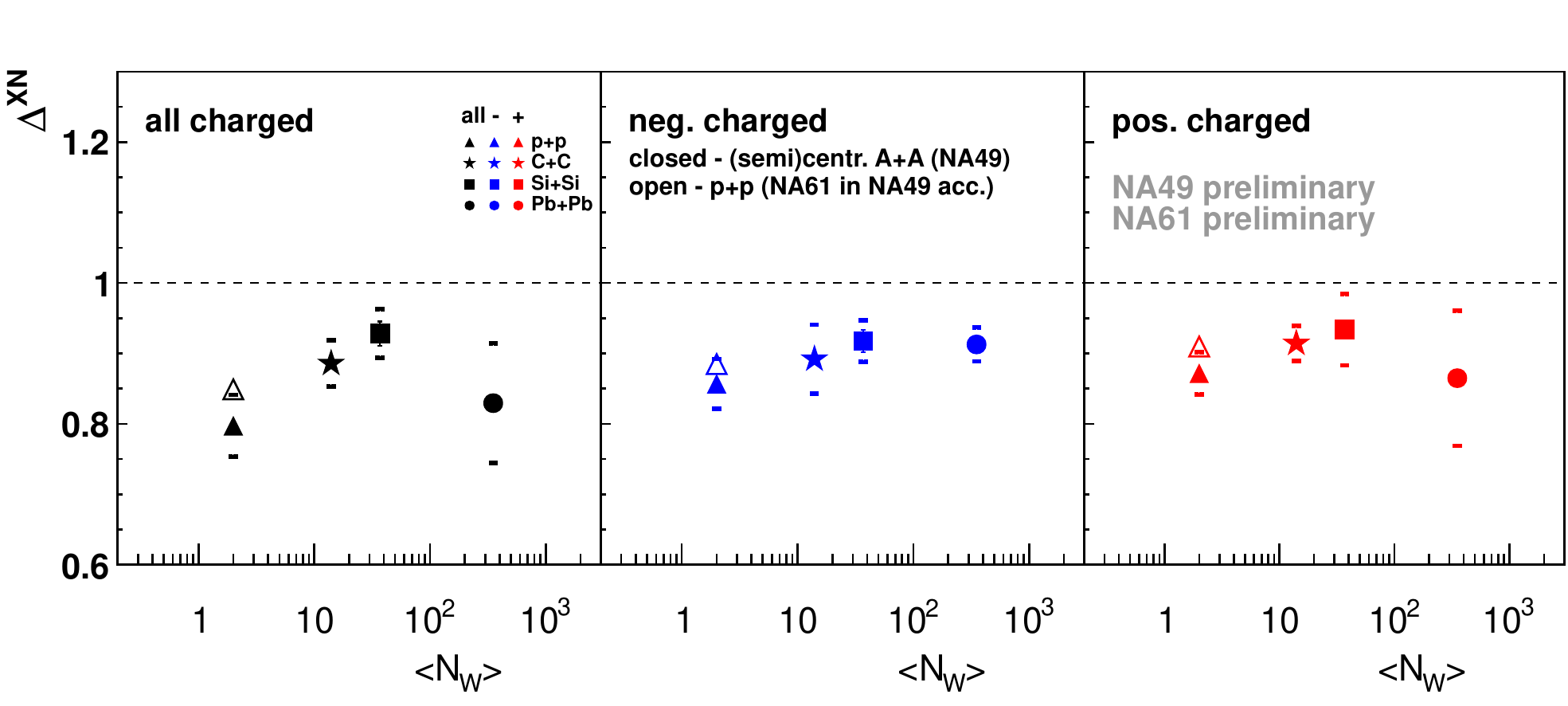}
\caption{Fluctuation measures $\Sigma[X,N]$~(top) and $\Delta[X,N]$~(bottom)
of the sum of transverse momenta $X$ of charged particles
versus $\mu_B$ for the 7.2~\% most central Pb+Pb collisions (full symbols)
and inelastic p+p interactions (open symbols).
Results are for cms rapidity $1.1 < y < 2.6$ assuming the pion mass
(NA49 and NA61/SHINE preliminary~\cite{ismd2013}).}
\label{Sigma_Delta_mub}
\end{figure}

For the same reactions preliminary results have also become available for the fluctuation quantities
$\Delta[X,N]$ and $\Sigma[X,N]$ introduced in Sec.~\ref{SIQ}, where $X$ is the sum of
absolute transverse momenta of the $N$ particles in the acceptance region of the analysis.
Figure~\ref{Sigma_Delta_mub} displays the results from Ref.~\cite{ismd2013}. Note that $\Phi_{p_T}$ is related to
$\Sigma[X,N]$ by Eq.~(\ref{Phi}) and for the IPM one expects for these quantities the values of 0 and 1 respectively.
The quantity $\Delta[X,N]$ is sensitive to fluctuations of $X$ and $N$ differently than $\Sigma[X,N]$.
Both measures show results which are close to the IPM expectations and display no distinct
features in their energy dependence.

\subsection{Fluctuation Studies with Incomplete Particle Identification}\label{identity}

Fluctuations of the chemical (particle-type) composition of
hadronic final states in A+A collisions are expected to be
sensitive to the phase transition between hadronic and partonic
matter. First experimental results on e-by-e chemical fluctuations
have already been published from the CERN SPS and BNL RHIC, and more
systematic measurements are in progress.
The e-by-e fluctuations of hadron multiplicities have been studied
theoretically in statistical models
(see, e.g. Ref.~\cite{CE,CE1,CE2,CE3})
and in dynamical transport models (see, e.g. review~\cite{HSD-rev}
and references therein).

Studies of chemical fluctuations in general require to determine the
number of particles of different hadron species (e.g. pions,
kaons, and protons) e-by-e.
%
A serious experimental problem in such measurements
is incomplete particle identification, i.e.
the impossibility to identify uniquely the type of each
detected particle.
The effect of particle mis-identification distorts the measured
fluctuation quantities. For this reason the analysis of chemical
fluctuations is usually performed in a small acceptance, where
particle identification is relatively reliable. However, an
important part of the information on e-by-e fluctuations in  full
phase space is then lost.

Although it is usually impossible to identify each detected
particle, one can nevertheless determine with high accuracy the
average multiplicities (averaged over many events) for different
hadron species.

\subsubsection{The Identity Method}\label{ident_method}
In Ref.~\cite{Ga:2011} a new experimental technique called the {\it
identity method} was proposed. It solved the misidentification
problem for one specific combination of the second moments in a
system of two hadron species (`kaons' and `pions').
In~Refs.~\cite{G:2011,RG:2012} this method was extended to show that
all the second moments as well as the higher moments of the joint
multiplicity distribution of particles of different types
can be uniquely reconstructed in spite of the effects of incomplete
identification. Notably, the identity method can be used for an
arbitrary number $k\geq 2$ of hadron species.
%
Following Ref.~\cite{Ga:2011} we assume that particle
identification is achieved by measuring the particle mass $m$.
Since any measurement is of finite resolution, we deal with
continuous distributions of observed masses  denoted as $\rho_j
(m)$ and normalized as ($j=1,\ldots,k\geq 2$)
\eq{ \label{norm-rho-i} \int dm \,\rho_j (m) = \langle N_j
\rangle~.
}
Note that for experimental data the functions $\rho_j(m)$ for particles
of type $j$ are obtained
from the inclusive distribution of the $m$-values for all particles from
all collision events. The identity variables $w_j(m)$ are
defined as
\eq{\label{wi}
w_j(m)~\equiv~\frac{\rho_j(m)}{\rho(m)}~,~~~~~ \rho (m) \equiv
\sum_{i=1}^k\rho_i (m) ~.
}
Complete identification (CI) of particles corresponds to
distributions $\rho_j (m)$ which do not overlap. In this case,
$w_j = 0$ for all particle species $i\neq j$ and $w_j = 1$ for the
$j$th species. When the distributions $\rho_j (m)$ overlap,
$w_j(m)$ can take the value of any real number from $[0,1]$.
We introduce the quantities
 \eq{\label{Wj2-def}
W_j~\equiv~ \sum_{i=1}^{N(n)}w_j(m_i)~,~~~~W_j^2~\equiv~\Big(\sum_{i=1}^{N(n)}w_j(m_i)\Big)^2~,~~~~W_pW_q~\equiv~
\Big(\sum_{i=1}^{N(n)}w_p(m_i)\Big)\times
\Big(\sum_{i=1}^{N(n)}w_q(m_i)\Big)~,
}
with $j=1,\cdots,k$ and $1\le p<q\le k$,
and define their event averages  as
\eq{
  \langle W_j^2 \rangle~ =~
\frac{1}{N_{\rm ev}} \sum_{n=1}^{N_{\rm ev}} W_j^2~,~~~~
\langle W_pW_q \rangle ~=~ \frac{1}{N_{\rm ev}} \sum_{n=1}^{N_{\rm
ev}} W_pW_q~,
\label{Wj2-av}
}
where $N_{\rm ev}$ is the number of events, and
$N(n)=N_1(n)+\cdots+N_k(n)$ is the total multiplicity in the $n$th
event. Each experimental event is characterized by a set of
particle masses $\{m_1,m_2,\ldots,m_N\}$, for which one can
calculate the full set of identity variables:
$\{w_j(m_1),w_j(m_2),\ldots,w_j(m_N)\}$, with $j=1,\ldots,k$.
Thus, the quantities $W_j$, $W_j^2$, and $W_{p}W_q$ are completely defined
for each event, and their average values (\ref{Wj2-av}) can be
found experimentally by straightforward e-by-e averaging.
The main idea is to find the relations between these $W$-quantities
and the unknown moments of the multiplicity distribution $\langle N_j^2\rangle$
and $\langle N_pN_q\rangle$.
In the
case of CI, one finds $W_j=N_j$,
thus,
Eq.~(\ref{Wj2-av}) yields
\eq{\label{WW}
\langle W_j^2\rangle~=~\langle N_j^2\rangle~,~~~~\langle
W_pW_q\rangle~=~\langle N_pN_q\rangle~.
}
The quantities $\langle W_j^2 \rangle$ and $\langle W_q
W_p\rangle$ can be calculated as follows
\eq{
&\langle W_j^2\rangle ~=~ \sum_{N_1=0}^\infty \sum_{N_2=0}^\infty
\ldots \sum_{N_k=0}^\infty {\cal P}( N_1,\ldots, N_k) \int dm_1^1
P_1 (m_1^1)\ldots  \int dm_{N_1}^1 P_1 (m_{N_1}^1)
 \nonumber\\
 &\times \int dm_1^2 P_2 (m_2^2)\ldots
 \int dm_{N_2}^k P_2(m_{N_2}^2)\times \ldots \times
\int dm_1^k P_k (m_1^k)\ldots \int dm_{N_k}^k P_k (m_{N_k}^k)
\nonumber \\
& \times \Big[w_j(m_1^1) +\cdots  w_j(m_{N_1}^1) + w_j(m_{1}^2) +
\cdots + w_j(m_{N_2}^2)+\ldots + w_j(m_1^k) + \cdots +
w_j(m_{N_k}^k) \Big]^2~ \nonumber \\
& = \sum_{i=1}^k\langle N_i \rangle \big[
u_{ji}^2~-~(u_{ji})^2\big]
 + \sum_{i=1}^k\langle N_i^2\rangle (u_{ji})^2
+ 2\sum_{1\leq i<l\leq k}\langle N_i N_l \rangle
u_{ji} u_{jl}~,\label{Wj2}
}
\eq{
&\langle W_p W_q\rangle ~=~ \sum_{N_1=0}^\infty
\sum_{N_2=0}^\infty \ldots \sum_{N_k=0}^\infty {\cal P}(N_1,\ldots
,N_k) \int dm_1^1 P_1 (m_1^1)\ldots  \int dm_{N_1}^1 P_1
(m_{N_1}^1)
 \nonumber\\
 &\times \int dm_1^2 P_2 (m_2^2)\ldots
 \int dm_{N_2}^k P_2(m_{N_2}^2)\times \ldots \times
\int dm_1^k P_k (m_1^k)\ldots \int dm_{N_k}^k P_k (m_{N_k}^k)
\nonumber \\
& \times \Big[w_p(m_1^1) +\cdots  w_p(m_{N_1}^1) + w_p(m_{1}^2) +
\cdots + w_p(m_{N_2}^2)+\ldots + w_p(m_1^k) + \cdots +
w_p(m_{N_k}^k) \Big]~ \nonumber \\
& \times \Big[w_q(m_1^1) +\cdots  w_q(m_{N_1}^1) + w_q(m_{1}^2) +
\cdots + w_q(m_{N_2}^2)+\ldots + w_q(m_1^k) + \cdots +
w_q(m_{N_k}^k) \Big]~ \nonumber \\
& = \sum_{i=1}^k\langle N_i \rangle \Big[u_{pqi}
~ -~u_{pi} u_{qi}\Big]
 ~+ ~\sum_{i=1}^k\langle N_i^2\rangle u_{pi}
 u_{ki}
 ~ + ~\sum_{1\leq i<l\leq k}\langle N_i N_l \rangle
\Big[u_{pi} u_{ql}~ +~ u_{pl} u_{qi}\Big]~.\label{WpWq}
}
In Eqs.~(\ref{Wj2}) and (\ref{WpWq}), ${\cal P}(N_1,\ldots ,N_k)$
is the multiplicity distribution, $P_i (m) \equiv \rho_i(m)/
\langle N_i \rangle $ are the mass probability distributions of
the $i$th species, and ($s=1,2$)
\eq{\label{j-av}
u_{ji}^s ~\equiv~ \frac{1}{\langle N_i \rangle } \int dm \,
w_j^s(m)~ \rho_i (m)  ~,~~~~ u_{pqi}~\equiv~\frac{1}{\langle
N_i\rangle} \int dm\, w_p(m)\, w_q(m)~\rho_i(m)~.
}

In the case of CI,  when the distributions $\rho_j (m)$ do not
overlap, one finds that
\eq{\label{CI-j}
u_{ji}^{s} ~ =~\delta_{ji}~,~~~~~ u_{pqi}~ =~ 0~,
}
and Eqs.~(\ref{Wj2}) and (\ref{WpWq}) reduce then to
Eq.~(\ref{WW}). The incomplete particle identification transforms
the second moments $\langle N_j^2\rangle$ and $\langle
N_pN_q\rangle$ to the quantities $\langle W_j^2\rangle$ and
$\langle W_p W_q\rangle$, respectively. Each of the later
quantities contains  linear combinations of all the first and
second moments, $\langle N_i\rangle$ and $\langle N_i^2\rangle$,
as well as all the correlation terms $\langle N_i N_l \rangle$.
Having introduced the notations
\eq{
\langle W_j^2 \rangle  - \sum_{i=1}^k\langle N_i \rangle \big[
u_{ji}^2~-~ (u_{ji})^2\big] ~\equiv~b_j~,
~~~~ \langle W_pW_q\rangle -\sum_{i=1}^k\langle N_i \rangle \big[
u_{pqi}~ -~ u_{pi} u_{qi} \big] ~\equiv~b_{pq}~, \label{bpq}
}
one can transform Eqs.~(\ref{Wj2}) and (\ref{WpWq}) to the
following form:
\eq{
&\sum_{i=1}^k\langle N_i^2\rangle~ u_{ji}^2
+ 2\sum_{1\leq i<l\leq k}\langle N_i N_l \rangle ~ u_{ji} u_{jl}
~=~b_j~,~~~~j=1,2,\ldots,k~,\label{bj1}\\
& \sum_{i=1}^k\langle N_i^2\rangle ~u_{pi} u_{qi}
+ \sum_{1\leq i<l\leq k}\langle N_i N_l \rangle
\Big(u_{pi}u_{ql}~+~ u_{pl} u_{qi}\Big)~=~b_{pq}~,~~~~ 1\leq
p<q\leq k~.\label{bpq1}
}
The right-hand side of Eqs.~(\ref{bj1}) and (\ref{bpq1}) defined
by Eq.~(\ref{bpq}) are experimentally measurable quantities. The
same is true for the coefficients  $u_{ji}^{s}$ (with $s=1$ and
$2$) entering the left-hand side of Eqs.~(\ref{bj1}) and
(\ref{bpq1}).
Therefore, Eqs.~(\ref{bj1}) and (\ref{bpq1}) represent a system of
$k+ k(k-1)/2$ linear equations for the $k$ second moments $\langle
N_j^2\rangle$ with $j=1,\ldots,k$ and $k(k-1)/2$ correlators
$\langle N_pN_q\rangle$ with $1\leq p< q\leq k$.
In order to solve Eqs.~(\ref{bj1}) and (\ref{bpq1}) we introduce
the $[k+k(k-1)/2]\times[k+k(k-1)/2]$ matrix $A$
\eq{\label{AA}
A~=~
\begin{pmatrix}
a_{1}^{1} & \ldots & a_{1}^{k} &|&
a_{1}^{12}
&\ldots  & a_{1}^{(k-1)k}\\
. & . & . &|&
 . &  .  & .\\
. & . & . &|&
 . &  .  & .\\
 a_{k}^{1} & \ldots & a_{k}^{k} & | &
 a_{k}^{12}
 &\ldots &  a_k^{(k-1)k}  \\
--- & --- &--- &|&---& ---&---
\\
a_{12}^1 &\ldots  & a_{12}^k  & | & a_{12}^{12}
& \ldots & a_{12}^{(k-1)k}\\
. & . & . &|&
. & . & .  \\
. & . & . &|&
 . & . & .  \\
a_{12}^{k}& \ldots & a_{(k-1)k}^k  & | &   a_{(k-1)k}^{12} &
\ldots & a_{(k-1)k}^{(k-1)k}
 \end{pmatrix}~,
}
where
\eq{
a^{i}_{j}~&\equiv ~ u_{ji}^2~,~~1\leq i,j\leq k~;~~~~~ a_{i}^{pq}
\equiv ~2 u_{ip}u_{iq}~,~~ 1\leq p< q\leq k~,~~
i=1,\ldots ,k ~~;\\
a_{pq}^{i}~&\equiv ~u_{pi}u_{qi}~,~~1\leq p< q\leq k~,~~i=1,\ldots
,k~;
\\
 a_{pq}^{lm}~&\equiv~u_{pl}u_{qm}+u_{ql}u_{pm}~,~~
1\leq p<q\leq k~,~~1\leq l<m\leq k~.
}
The solution of Eqs.~(\ref{bj1}) and (\ref{bpq1}) can be presented
by Cramer's formulas in terms of the determinants
\eq{\label{Nj2-sol}
\langle N_j^2\rangle ~=~ \frac{{\rm det}~A_j}{{\rm
det}~A}~,~~~~~\langle N_pN_q\rangle ~=~ \frac{{\rm
det}~A_{pq}}{{\rm det}~A}~,~
}
where the matrices $A_j$ and $A_{pq}$ are obtained by substituting
in the matrix $A$ the column $ a_{1}^{j},\ldots, a_{k}^{j},
a_{12}^j,\ldots, a_{(k-1)k}^j$ and the column $a_1^{pq},\ldots ,
a_k^{pq}, a^{pq}_{12},\ldots , a_{(k-1)k}^{pq}$, respectively, for
the column $b_1,\ldots, b_k,$ $b_{12}, \ldots, b_{(k-1)k}$.
Therefore, if ${\rm det}A\neq 0$, the system of linear equations
(\ref{bj1}) and (\ref{bpq1}) has a unique solution (\ref{Nj2-sol})
for all the second moments.
In the case of CI (\ref{CI-j}), one finds ${\rm det}A=1$, ${\rm
det}A_j=b_j$,  and ${\rm det}A_{pq}=b_{pq}$. The solution
(\ref{Nj2-sol}) reduces then to Eq.~(\ref{WW}).

Introducing the  $[k+k(k-1)/2]$-vectors
\eq{\label{vectors}
{\cal N}~\equiv~
\begin{pmatrix}
\langle N_1^2\rangle  \\
\ldots \\
\langle N_k^2\rangle \\
\langle N_1N_2\rangle \\
\ldots\\
\langle N_{k-1}N_k \rangle
\end{pmatrix}
~,~~~~~~
 {\cal B}~\equiv~
\begin{pmatrix}
b_1  \\
\ldots \\
b_k \\
b_{12} \\
\ldots\\
b_{(k-1)k}
\end{pmatrix}~,
}
one can write Eqs.~(\ref{bj1}) and (\ref{bpq1}) in the matrix form
$A{\cal N} ={\cal B}$. The solution (\ref{Nj2-sol}) can be then
rewritten as
\eq{\label{AN}
{\cal N}~=~A^{-1}~{\cal B}~,
}
where $A^{-1}$ is the inverse matrix of $A$.
For two particle species, $k=2$, this solution takes the form
\eq{\label{A2}
\begin{pmatrix}
\langle N_1^2 \rangle \\
\langle N_2^2 \rangle \\
\langle N_1 N_2 \rangle
\end{pmatrix}
~=~
\begin{pmatrix} u_{11}^2 ~~~~& u_{12}^2~~~~& 2 u_{11} u_{12} \\
 u_{21}^2~~~~& u_{22}^2~~~~
&2 u_{21} u_{22} \\
u_{11} u_{21} ~~~~~ & u_{12} u_{22}~~~~ & u_{11} u_{22} + u_{12}
u_{21}
\end{pmatrix} ^{-1}~
\begin{pmatrix}
b_1\\
b_2\\
b_{12}
\end{pmatrix}
~.
}
Then Eq.~(\ref{A2}) yields
\eq{
 \langle N_1^2\rangle
~&=~ \frac{b_1u_{22}^2~+~b_2u_{12}^2~-~2b_{12}
u_{12}u_{22}}{\big(u_{11}u_{22}~-~u_{12}u_{21}\big)^2}~,
\label{N1-sol}
\\
 \langle N_2^2\rangle ~& = ~
 \frac{b_2u_{11}^2~+~
 b_1u_{21}^2 ~-~ 2b_{12}u_{21}u_{11}}
 {\big( u_{11}u_{22}~-~u_{12}u_{21}\big)^2}~,\label{N2-sol}\\
\langle N_1N_2\rangle ~& = ~
\frac{b_{12}\big(u_{11}u_{22}+u_{12}u_{21}\big) -b_1u_{22}u_{21}-
b_2u_{11}u_{12}}{\big(u_{11}u_{22}~-~u_{12}u_{21}\big)^2}~.
\label{N12-sol}
}
The above procedure eliminates the effect of misidentification and
provides the values of all the second moments $\langle
N_j^2\rangle$ and $\langle N_pN_q\rangle$ in a model-independent
way, as they would be obtained in an experiment in which each
particle is uniquely identified.
Recently the identity method was generalized
to determine third and higher moments of the
multiplicity distributions in events consisting of an arbitrary
number of different particle species~\cite{RG:2012}. As emphasized
in Refs.~\cite{Step,Step1}, measurements of the third and higher moments of
event-by-event fluctuations are, in fact, expected to be more
sensitive for the search of the critical point in nucleus-nucleus collisions.

\subsubsection{First Results on Fluctuations Based on the Identity Method}\label{ident_experiment}

\begin{figure}[!htb]
\begin{center}
\includegraphics[width=0.6\textwidth]{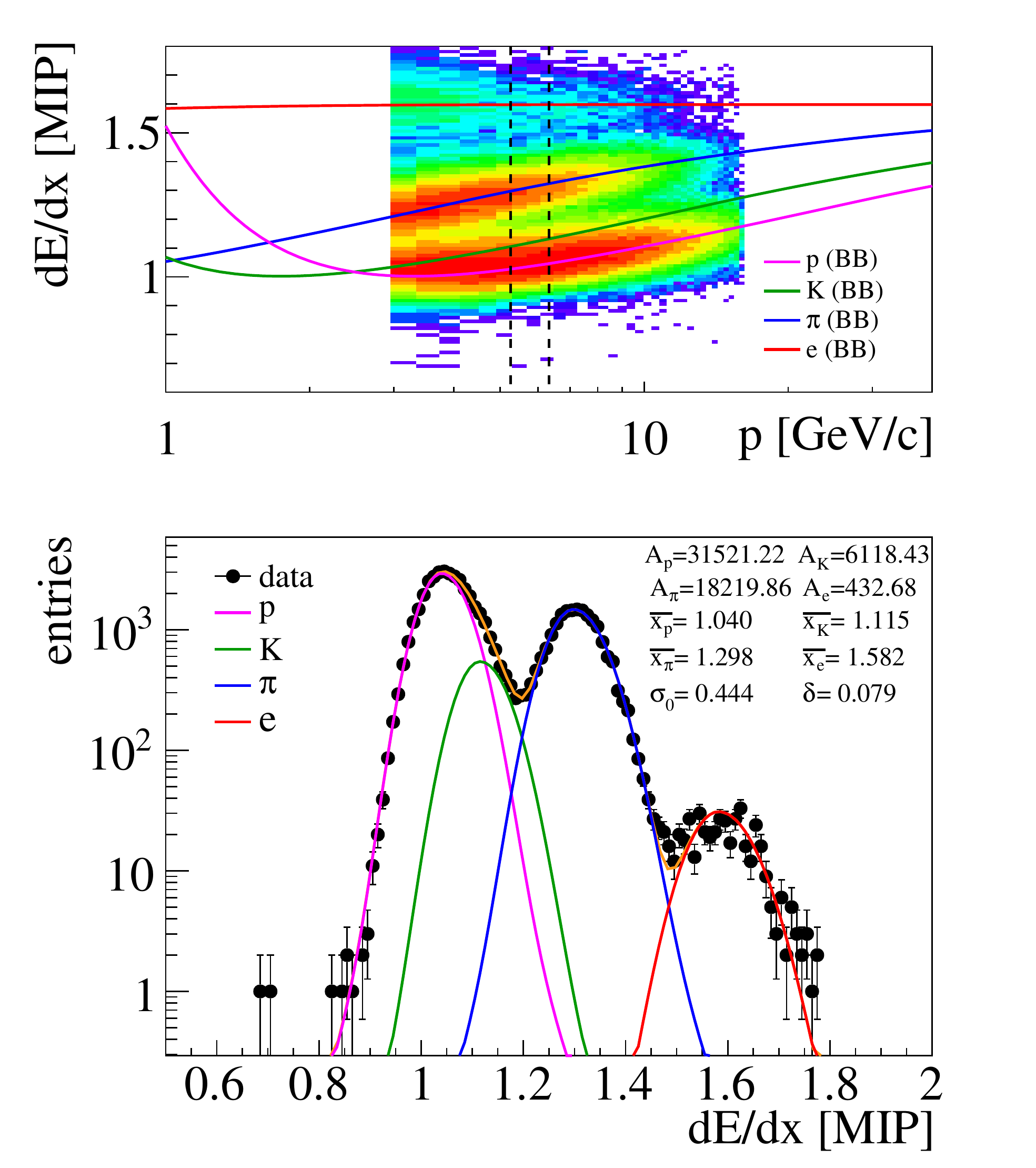}
\end{center}
\caption{dE/dx distributions for Pb+Pb collisions
at 20$A$~GeV from NA49~\cite{NA49_identity}.
Upper panel: Measured \textit{dE/dx} values as function of reconstructed momenta for the
phase space region 0.4 $< p_{\bot}$ [GeV/c] $<$ 0.6 and 135 $ < \phi$ [$^{\text{o}}$]  $<$ 180.
Lines correspond to calculations with the Bethe-Bloch (BB) formula for
different particle types.
Lower panel:  Projection of the upper plot onto the vertical axis in the momentum interval
 5.2 $ < p$ [GeV/c] $<$ 6.4 indicated by vertical dashed lines.
Colored lines represent the \textit{dE/dx} distribution functions of different particles.
}
\label{dEdx_20}
\end{figure}

Experiments NA49 and NA61/SHINE identify particles by measuring their
ionisantion energy loss dE/dx in the TPC tracking chambers. From
a typical inclusive distribution of the obtained dE/dx values,
as shown in Fig.~\ref{dEdx_20}, it is evident that the contributions
of protons, kaons, pions and electrons overlap. For measuring the
fluctuations of their numbers one can try to fit the dE/dx distributions
event-by-event as was done by NA49 previously~\cite{NA49_Rfluct,NA49_Rfluct1}.
However, this method has serious statistical limitations and can only
be applied to high-multiplicity events such as produced in Pb+Pb collisions.
To overcome this problem and be able to study multiplicity fluctuations
for the full range of nuclei from p to Pb the NA49 and NA61/SHINE collaborations
are now employing the identity method.

\begin{figure}[!htb]
\begin{center}
\includegraphics[width=0.7\textwidth]{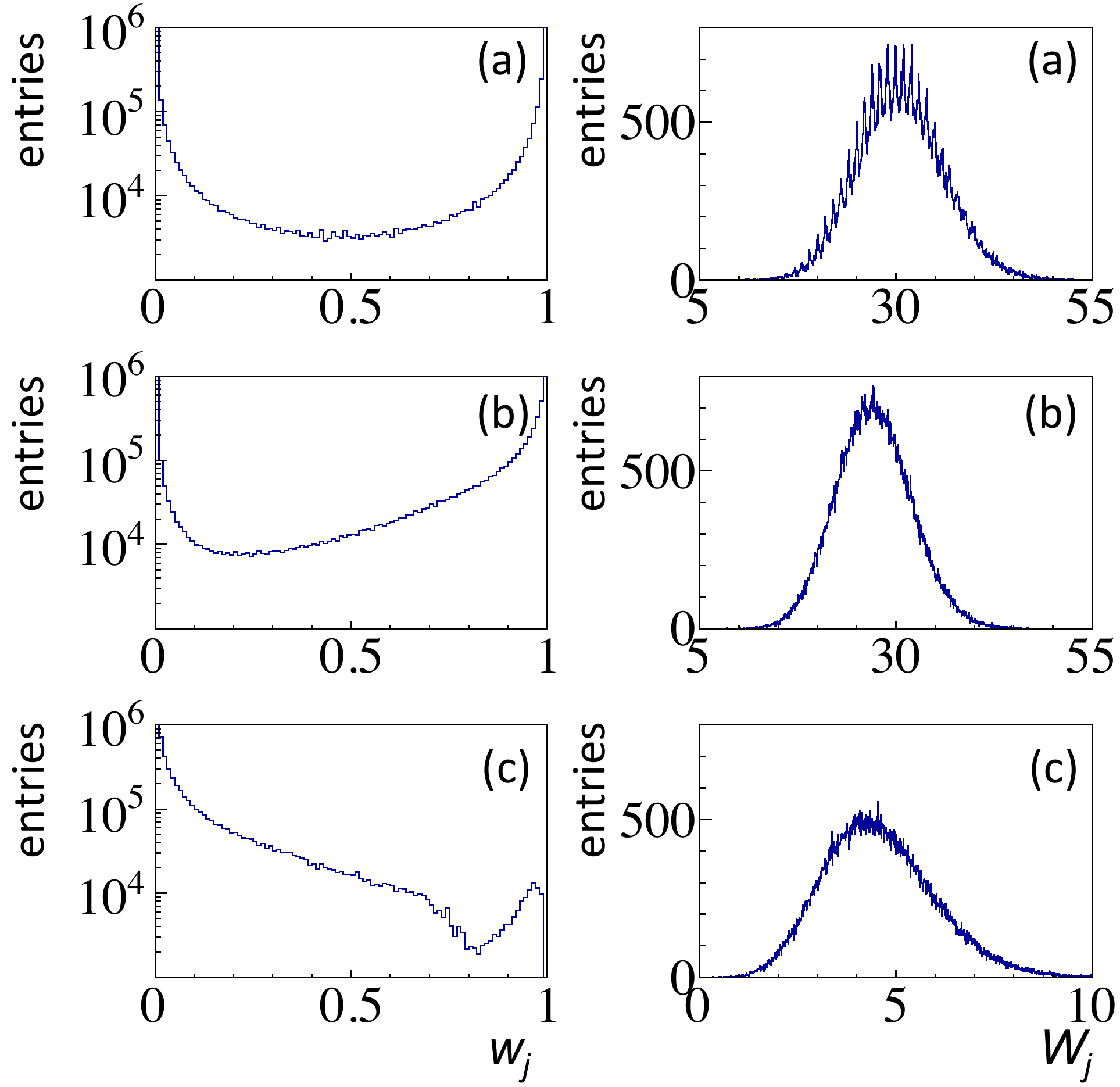}
\end{center}
\caption{Left panel: Distributions of $w_j$ of Eq.~(\ref{wi}) for (a) pions, (b) protons and (c) kaons for
central Pb+Pb collisions at 20$A$~GeV.
Right panel: Corresponding distributions of $W_j$
of Eq.~(\ref{Wj2-def})~\cite{NA49_identity}.
}
\label{w_dist}
\end{figure}

The distributions $\rho_j(m\equiv\text{dE/dx})$ of Eq.~(\ref{norm-rho-i}) are determined from
precise fits to the inclusive distributions of dE/dx (see curves in Fig.~\ref{dEdx_20} as an
example). Using the $\rho_j$ one then derives the distributions of of identities
$w_j(m\equiv\text{dE/dx})$ of Eq.~(\ref{wi}) (see Fig.~\ref{w_dist}, left)
which correspond to the probabilities for a particle with $m$ = dE/dx to be of type $j$.
The analogue $W_j$ of the multiplicity distribution of particles of type $j$,
calculated according to Eqs.~(\ref{Wj2-def},~\ref{Wj2-av})
(see Fig.~\ref{w_dist}, right),  is continuous for incomplete particle identification.
Moreover the multiplicities of particles of types $j$ and $k$ are in general correlated.
One obtains from from Eq.~(\ref{Wj2-av}) the pure and mixed second moments of the
smeared multiplicity distributions from which the moments of the true multiplicity distributions
are found by the matrix inversion procedure of the identity method~\cite{G:2011}.

\begin{figure}[!htb]
\begin{center}
\includegraphics[width=0.9\textwidth]{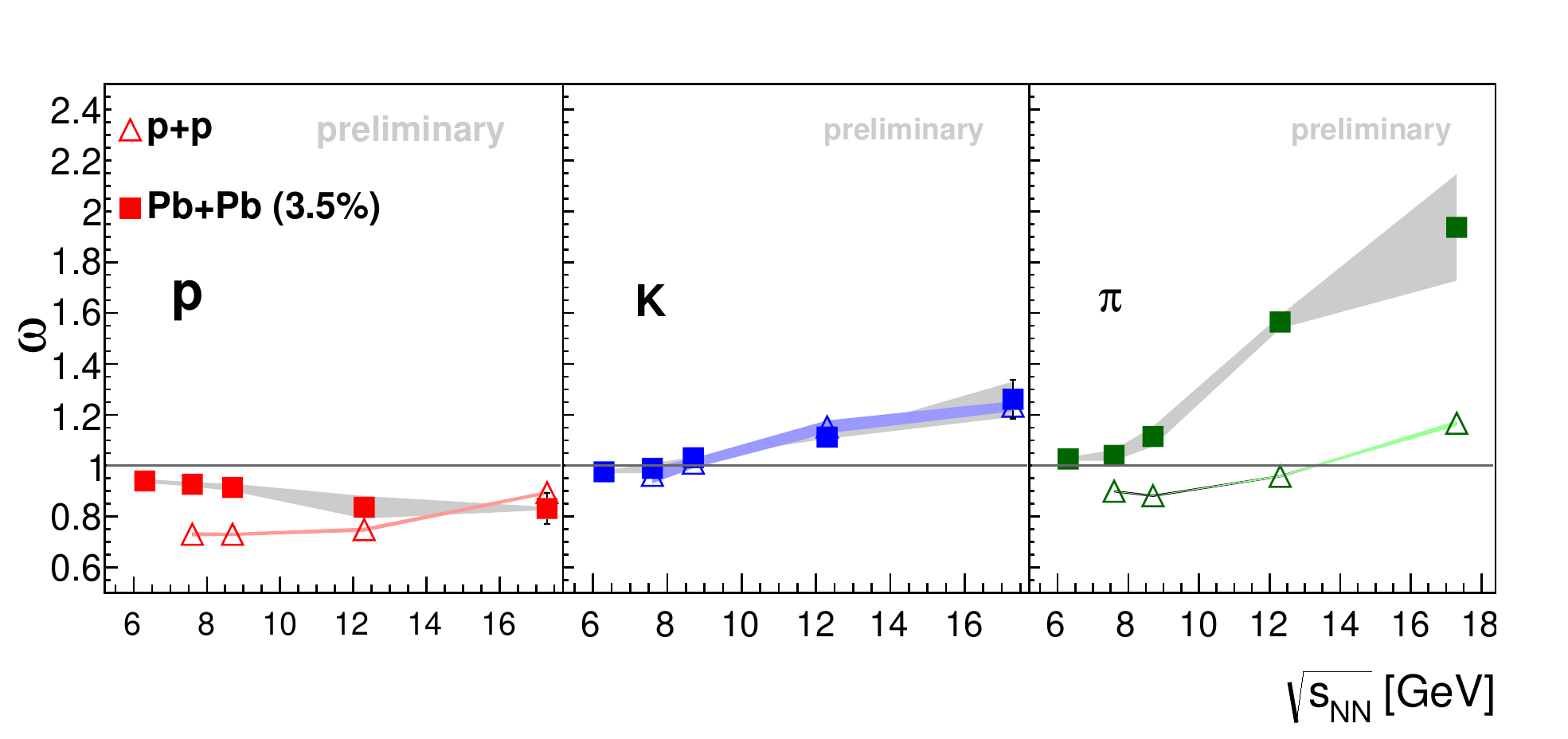}
\end{center}
\caption{Scaled variance $\omega$ of the multiplicity distributions of
p+$\bar{\text{p}}$, K$^+$+K$^-$
and $\pi^+$+$\pi^-$ in inelastic p+p and the 3.5~\% most central Pb+Pb collisions
versus the collision energy $\sqrt{s_{\text{NN}}}$~\cite{ismd2013}.
}
\label{omega_pKpi}
\end{figure}

The moments corrected for the effects of incomplete particle identification can then be  used to
compute the scaled variance $\omega$ of the multiplicity distributions of p+$\bar{\text{p}}$, K$^+$+K$^-$
and $\pi^+$+$\pi^-$. The results for inelastic p+p interactions
and the 3.5~\% most central Pb+Pb collisions plotted versus
the collision energy $\sqrt{s_{\text{NN}}}$ are shown in Fig.~\ref{omega_pKpi} (see Ref.~\cite{ismd2013}). One observes no
peak structure which might be attributed to the effects of a critical point. A similar conclusion was
reached for $\omega$ of unidentified charged particles in Ref.~\cite{NA49_omega}. The stronger
increase of $\omega$ for pions seen in Fig.~\ref{omega_pKpi} is mostly due to the more
relaxed centrality selection for Pb+Pb collisions and the fact that the scaled variance is only an
intensive and not a strongly intensive quantity.

\begin{figure}[!htb]
\begin{center}
\includegraphics[width=0.9\textwidth]{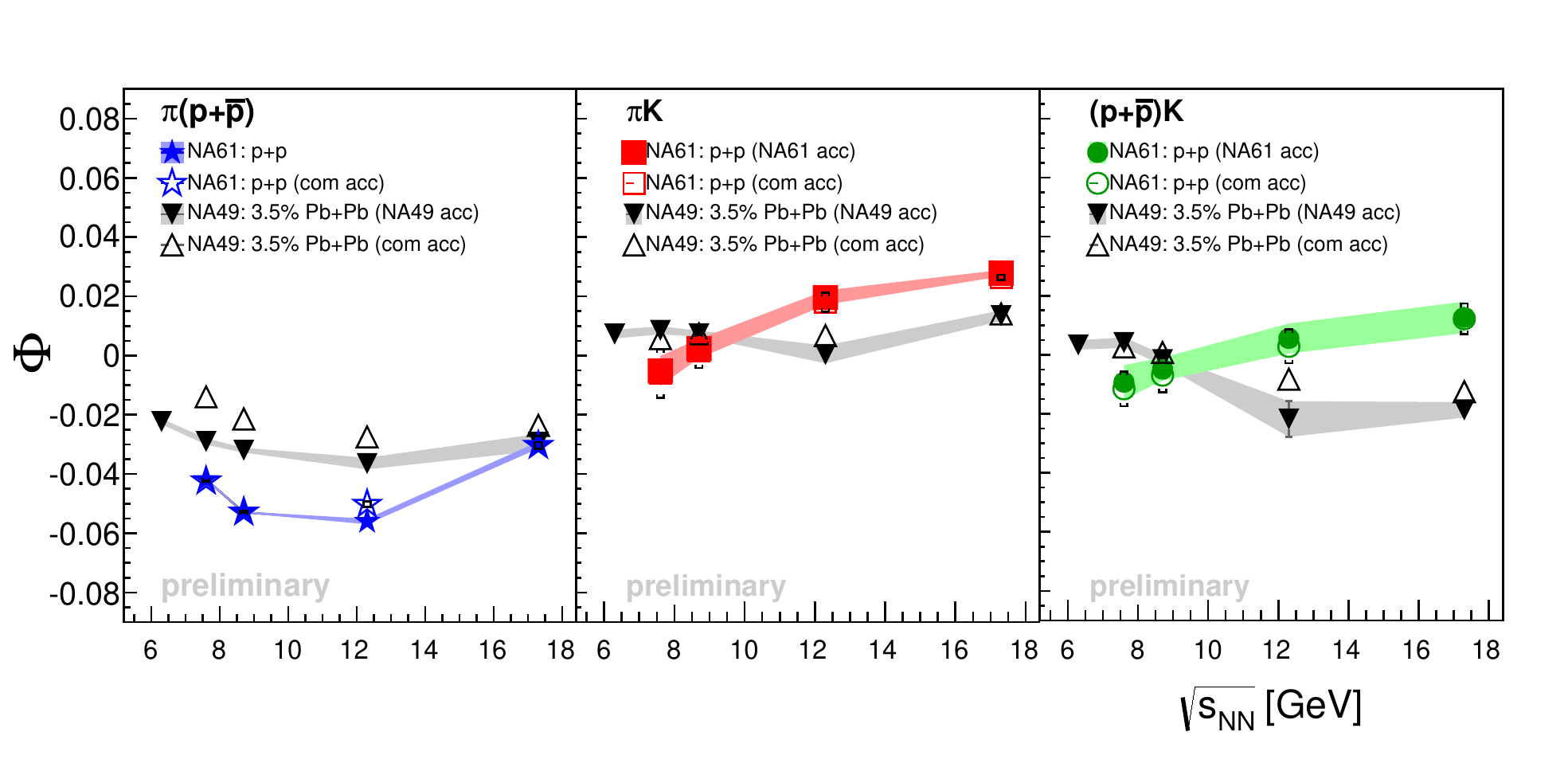}
\end{center}
\caption{Strongly intensive fluctuation quantity $\Phi$ measured for identified particle pairs
in inelastic p+p and the 3.5~\% most central Pb+Pb collisions versus
collision energy $\sqrt{s_{\text{NN}}}$~\cite{Mackowiak_CPOD2013}.
}
\label{Phi_pKpi}
\end{figure}

As the identity method allows to determine all pure and mixed first and second moments of the multiplicity
distributions of p+$\bar{\text{p}}$, K$^+$+K$^-$ and $\pi^+$+$\pi^-$, one is able to
derive the values of the strongly intensive quantity $\Phi[A,B]$ defined in Eq.~(\ref{phiAB}).
$\Phi[A,B]$ measures fluctuations between particle types, so-called chemical fluctuations.
The results for the three possible pair combinations are displayed in Fig.~\ref{Phi_pKpi} for
inelastic p+p interactions from NA61/SHINE and central Pb+Pb collisions
from NA49 (see Ref.~\cite{Mackowiak_CPOD2013}).
The values of $\Phi$ are close to that resulting from the IPM,
i.e. the the fluctuations are mostly statistical.
However, one observes some systematic energy dependence and changes of sign,
the origin of which need to be understood.

\clearpage

\section{Closing Remarks}

Experimental study of strongly interacting matter started
in the 1970s with experiments at JINR Dubna and LBL Berkeley
recording first collisions of relativistic heavy ions.
Since then rich experimental data on
Pb+Pb and Au+Au collisions were collected at energies
ranging from several GeV to several TeV.
Single particle spectra and mean multiplicities
were precisely measured in fixed target experiments,
whereas the most comprehensive results on correlations
in azimuthal angle of hadrons produced at mid-rapidity
were obtained in collider experiments.

These results
clearly indicate that a system of
strongly interacting particles created in
heavy ion collisions at high energies is close to, at least, local
equilibrium. At freeze-out the system occupies a volume which is
much larger than a volume of  individual hadron. The latter conclusion
is based on the failure of dynamical and the success of
statistical~\cite{Becattini:2009sc} and
hydrodynamical models~\cite{Florkowski_textbook}.
Thus, one concludes that strongly interacting matter
is created in
heavy ion collisions.

The phase transition of strongly interacting matter to the QGP -
the onset of deconfinement -
was discovered with the energy scan program of NA49 at
the CERN SPS~\cite{:2007fe,:2007fe1}.
The program was motivated by the predictions of the statistical model
of the early stage of collisions.
The discovery is based on the observation
that several basic hadron production properties measured
in heavy ion collisions rapidly change their dependence on
collision energy in a common energy
domain~\cite{review}.

The experimental discoveries of strongly interacting matter
and the onset of deconfinement ask for further studies,
in particular systematic results on event-by-event fluctuations
are still missing.
The fluctuations are difficult to measure as one cannot correct
for the limited acceptance of detectors using forward-backward
and rotational symmetries.  Here large acceptance detectors
are needed.  Moreover, precise control is required over fluctuations
of the number of nucleons which participate in the collisions.
The measurement of strongly intensive quantities will further
mitigate the effects of remaining participant number fluctuations.

These challenges are currently addressed by NA61/SHINE
at the CERN SPS and STAR BES II at the BNL RHIC with the help of
significant upgrades of the legacy detectors.
Furthermore, the methods to study event-by-event fluctuations
have recently advanced significantly.
All this should lead to new data on event-by-event fluctuations
which will soon shed more light on properties of the onset of deconfinement
and will have potential for the discovery of the critical point.

The third generation experiments, MPD at JINR NICA and
CBM at FAIR SIS, are originally planned to use multi-purpose
detectors. They are mostly optimized for high statistics
measurements of low cross-section processes.
Thus new future experiments dedicated
to measurements of event-by-event fluctuations are needed.
Their detectors should measure almost all charged hadrons
with a low background and allow for a precise determination of
the number of spectator nucleons ideally from both colliding
nuclei. The highest event rate is not the top priority in the study of
bulk properties of strongly interacting matter via analysis
of event-by-event fluctuations.

\vspace{1cm}
{\bf Acknowledgements}

\vspace{0.2cm}
This work was supported by
the National Science Centre of Poland (grant
UMO-2012/04/M/ST2/00816),
the German Research Foundation (grants GA 1480\slash 2-1, GA 1480\slash 2-2),
and the State Agency of Science, Innovations and
Informatization of Ukraine (contract F58/384-2013).


\begin{thebibliography}{100}

\bibitem{Stephanov_2006}
M.~Stephanov, PoS~LAT2006, {\bf 24} (2006).

\bibitem{Becattini_2004}
F.~Becattini {\it et al.}, Phys.~Rev. {\bf C69}, 024905 (2004); {\bf C73}, 044905 (2006).

\bibitem{Cleymans_2006}
J.~Cleymans {\it et al.}, Phys.~Rev. {\bf C73}, 034905 (2006).

\bibitem{qgp_sps}
  U.~W.~Heinz and M.~Jacob,
  arXiv:nucl-th/0002042.

\bibitem{qgp_signals}
  J.~Rafelski and B.~M\"uller,
  Phys.\ Rev.\ Lett.\  {\bf 48}, 1066 (1982)
  [Erratum-ibid.\  {\bf 56}, 2334 (1986)].

\bibitem{qgp_signals_1}
 T.~Matsui and H.~Satz,
  Phys.\ Lett.\ B {\bf 178}, 416 (1986).

\bibitem{qgp_signals_2}
  F.~Becattini, L.~Maiani, F.~Piccinini, A.~D.~Polosa and V.~Riquer,
  Phys.\ Lett.\ B {\bf 632}, 233 (2006).



\bibitem{GaGo}
M. Gazdzicki and M. I. Gorenstein,
Acta Phys. Polon. {\bf B30}, 2705 (1999).

\bibitem{:2007fe}
  C.~Alt {\it et al.}  [NA49 Collaboration],
  Phys.\ Rev.\  C {\bf 77}, 024903 (2008).

 \bibitem{:2007fe1}
 S.~V.~Afanasiev {\it et al.}  [NA49 Collaboration],
  Phys.\ Rev.\ C {\bf 66}, 054902 (2002).


\bibitem{review}
  M.~Gazdzicki, M.~Gorenstein, P.~Seyboth,
  Acta Phys.\ Polon.\  {\bf B42}, 307 (2011).


\bibitem{GaRo1}
M. Gazdzicki and D. R\"ohrich,
Z. Phys. {\bf C65}, 215 (1995).

\bibitem{GaRo2}
M. Gazdzicki and D. R\"ohrich,
Z. Phys. {\bf C71}, 55 (1996).

\bibitem{Fe}
E. Fermi, Prog. Theor. Phys. {\bf 5}, 570 (1950).

\bibitem{La}
L. D. Landau, Izv. Akad. Nauk SSSR Ser. Fiz. {\bf 17},
51 (1953).

\bibitem{Ga}
M. Gazdzicki, Proceedings of
NATO Advanced Research Workshop: "Hot
Hadronic Matter: Theory and Experiment", Divonne-les-Bains,
France, June 27 -- July 1 (1994),
edited by J. Letessier, H. H. Gutbrod and
J. Rafelski, NATO ASI Series B: Physics Vol. 346, Plenum Press (1995)
215.

\bibitem{Ga1}
M. Gazdzicki,
Z. Phys. {\bf C66}, 659 (1995).

\bibitem{Ga2}
M. Gazdzicki,
J. Phys. {G23}, 1881 (1997).

\bibitem{na49_add1}
J. B\"achler {\it et al.} [NA49 Collaboration],
{\it Searching for QCD Phase Transition},
Addendum--1 to Proposal SPSLC/P264, CERN/SPSC 97 (1997).




\bibitem{NA49_Rfluct1}
  T.~Anticic {\it et al.}  [NA49 Collaboration],
  Phys.\ Rev.\ C {\bf 83}, 061902 (2011).

\bibitem{NA49_Phipt}
  T.~Anticic {\it et al.}  [NA49 Collaboration],
  Phys.\ Rev.\ C {\bf 79}, 044904 (2009).

 \bibitem{NA49_Rfluct}
  C.~Alt {\it et al.}  [NA49 Collaboration],
  Phys.\ Rev.\ C {\bf 79}, 044910 (2009).

\bibitem{na49results3}
  C.~Alt {\it et al.}  [NA49 Collaboration],
  Phys.\ Rev.\ C {\bf 78}, 044907 (2008).

\bibitem{na49results4}
  C.~Alt {\it et al.}  [NA49 Collaboration],
  Phys.\ Rev.\ C {\bf 78}, 034918 (2008).

\bibitem{NA49_omega}
  C.~Alt {\it et al.}  [NA49 Collaboration],
  Phys.\ Rev.\ C {\bf 78}, 034914 (2008).

\bibitem{na49results6}
  C.~Alt {\it et al.}  [NA49 Collaboration],
  Phys.\ Rev.\ C {\bf 77}, 064908 (2008).

\bibitem{na49results7}
  C.~Alt {\it et al.}  [NA49 Collaboration],
  Phys.\ Rev.\ C {\bf 76}, 024914 (2007).

\bibitem{na49results8}
  C.~Alt {\it et al.}  [NA49 Collaboration],
  Phys.\ Rev.\ C {\bf 73}, 044910 (2006).

\bibitem{na49results9}
  T.~Anticic {\it et al.}  [NA49 Collaboration],
  Phys.\ Rev.\ C {\bf 69}, 024902 (2004).


\bibitem{vHove}
L.~Van~Hove, Phys.~Lett. {\bf B118}, 138 (1982).

\bibitem{Shuryak}
C.~Hung and E.~Shuryak, Phys.~Rev.~Lett. {\bf 75}, 4003 (1995).


\bibitem{HGM}
J.~Cleymans and K.~Redlich,
Phys.\ Rev.\ C {\bf 60}, 054908 (1999).

\bibitem{RQMD}
H.~Sorge, H.~St\"ocker and W.~Greiner,
Nucl.\ Phys.\ A {\bf 498}, (1989) 567C.

\bibitem{UrQMD}
S.~A.~Bass {\it et al.},
Prog.\ Part.\ Nucl.\ Phys.\  {\bf 41}, 225 (1998).

\bibitem{HSD}
W. Cassing and E.L. Bratkovskaya, Phys. Rep. {\bf 308}, 65 (1999).

\bibitem{HSD1}
W.~Cassing, E.~L.~Bratkovskaya, and S.~Juchem,
Nucl.\ Phys.\ A {\bf 674}, 249 (2000).

  \bibitem{Gorenstein:2003cu}
  M.~I.~Gorenstein, M.~Gazdzicki and K.~A.~Bugaev,
  Phys.\ Lett.\ B {\bf 567}, 175 (2003).

\bibitem{brasil}
M.~Gazdzicki, M.~I.~Gorenstein, F.~Grassi, Y.~Hama, T.~Kodama and O.~J.~Socolowski,
Braz.\ J.\ Phys.\  {\bf 34}, 322 (2004).

\bibitem{brasil1}
Y.~Hama, F.~Grassi, O.~Socolowski, T.~Kodama, M.~Gazdzicki and M.~Gorenstein,
  Acta Phys.\ Polon.\ B {\bf 35}, 179 (2004).


\bibitem{fischer}
H.~G.~Fischer, slides from the open 74th Meeting of the SPSC,
November 15, 2005,
http://indico.cern.ch/conferenceDisplay.py?confId=a057199.

\bibitem{cole}
  B.~A.~Cole {\it et al.},
  Phys.\ Lett.\ B {\bf 639}, 210 (2006).

\bibitem{bialas}
  A.~Bialas and W.~Czyz,
  Acta Phys.\ Polon.\ B {\bf 36}, 905 (2005).

\bibitem{mixing}
  M.~Gazdzicki and M.I.~Gorenstein,
  Phys.\ Lett.\ B {\bf 640}, 155 (2006).

\bibitem{bialkowska}
  H.~Bialkowska, M.~Gazdzicki, W.~Retyk and E.~Skrzypczak,
  Z.\ Phys.\ C {\bf 55}, 491 (1992).

\bibitem{EoI}
M.~Gazdzicki {\it et al.} [NA61/SHINE Proto-Collaboration],
\emph{A New Experimental Programme with Nuclei
and Proton Beams at the CERN SPS},
CERN-SPSC-2003-038; SPSC-EOI-001

\bibitem{mg_fluct}
  M.~Gazdzicki, M.~I.~Gorenstein and S.~Mrowczynski,
  Phys.\ Lett.\ B {\bf 585}, 115 (2004).

\bibitem{mg_fluct1}
  M.~I.~Gorenstein, M.~Gazdzicki and O.~S.~Zozulya,
  Phys.\ Lett.\ B {\bf 585}, 237 (2004).

\bibitem{Kolb:2000sd}
  J.-Y. Ollitrault, Phys. Rev. D {\bf 46}, 229 (1992).

\bibitem{Kolb:2000sd1}
  P.~F.~Kolb, J.~Sollfrank and U.~W.~Heinz,
  Phys.\ Rev.\ C {\bf 62}, 054909 (2000).

\bibitem{proposal}
  N.~Antoniou {\it et al.} [NA61/SHINE Collaboration],
  \emph{Study of the hadron production in hadron-nucleus
  and nucleus-nucleus collisions at CERN SPS,}
  CERN-SPSC-2006-034.

\bibitem{NA61facility}
  N.~Abgrall {\it et al.}  [NA61 Collaboration],
  arXiv:1401.4699 [physics.ins-det].

\bibitem{SR2013}
N.~Abgrall {\it et al.} [NA61 Collaboration],
CERN-SPSC-2013-028; SPSC-SR-124.

\bibitem{chic}
F.~Arleo {\it et al.} [CHIC Proto--Collaboration],
CERN-SPSC-2012-031; SPSC-EOI-008.



\bibitem{critRHIC}
  G.~S.~F.~Stephans,
  J.\ Phys.\ G {\bf 32}, S447 (2006).



\bibitem{cpodGO}
  G.~Odyniec,
  PoS CPOD {\bf 2013}, 043 (2013).

\bibitem{BES-II}
White paper [STAR Collaboration]: {\it Studying the Phase Diagram of QCD Matter at RHIC},
https://drupal.star.bnl.gov/STAR/starnotes/public/sn0598 (2014).


\bibitem{BES_v2}
L.~Adamczyk {\it et al.} [STAR Collaboration], Phys.~Rev.~C {\bf 88}, 014902 (2013).

\bibitem{BES_Rcp}
E.~Sangaline [STAR Collaboration], Nucl.~Phys.~A, {\bf 904-905}, 771c (2013).

\bibitem{Stephanov_netp}
Y.~Hatta and M.~Stephanov, Phys.~Rev.~Lett. {\bf 91}, 102003 (2003).

\bibitem{Step}
M.A. Stephanov, Phys. Rev. Lett. {\bf 107}, 052301 (2011).

\bibitem{Step1}
M.A. Stephanov, Phys.
Rev. Lett. {\bf 102}, 032301 (2009).

\bibitem{Step2}
C. Athanasiou, K. Rajagopal, and M. Stephanov, Phys. Rev. D {\bf
82}, 074008 (2010).

\bibitem{BES_netp}
L.~Adamczyk {\it et al.} [STAR Collaboration], Phys.~Rev.~Lett. {\bf 112}, 032302 (2014).




\bibitem{anar}
  A.~Rustamov, \url{https://indico.cern.ch/conferenceDisplay.py?confId=144745}

\bibitem{kumar}
  L.~Kumar [STAR Collaboration],
  J.\ Phys.\ G {\bf 38}, 124145 (2011).

\bibitem{kumar1}
  B.~Mohanty [STAR Collaboration],
  J.\ Phys.\ G {\bf 38}, 124023 (2011).

\bibitem{schukraft}
  J.~Schukraft {\it et al.} [ALICE Collaboration ],
  J.\ Phys.\ G {\bf 38}, 124003 (2011).

\bibitem{schukraft1}
  A.~Toia {\it et al.} [ALICE Collaboration ],
  J.\ Phys.\ G {\bf 38}, 124007 (2011).



\bibitem{AFTER}
  J.~P.~Lansberg, V.~Chambert, J.~P.~Didelez, B.~Genolini, C.~Hadjidakis,
  P.~Rosier, R.~Arnaldi and E.~Scomparin {\it et al.},
  PoS QNP {\bf 2012}, 049 (2012).


\bibitem{mpd}
  K.~.U.~Abraamyan, S.~V.~Afanasiev, V.~S.~Alfeev, N.~Anfimov,
  D.~Arkhipkin, P.~Z.~.Aslanyan, V.~A.~Babkin and M.~I.~Baznat {\it et al.},
  Nucl.\ Instrum.\ Meth.\ A {\bf 628}, 99 (2011).

\bibitem{cbm} P.~Senger [CBM Collaboration], Cent.~Eur.~J.~Phys., {\bf 10}, 1289 (2012);
see also http://www.fair-center.eu/for-users/experiments/cbm.html.

\bibitem{nica}
  A.~N.~Sissakian {\it et al.}  [NICA Collaboration],
  J.\ Phys.\ G {\bf 36}, 064069 (2009).

\bibitem{nica_wp}
NICA White Paper, January 24, 2014,
\url{http://theor.jinr.ru/twiki/pub/NICA/WebHome/WhitePaper_10.01.pdf}


\bibitem{fair} see http://www.fair-center.eu/public.html.

\bibitem{max_rhoB}
J.~Randrup and J.~Cleymans, Phys.~Rev.~C {\bf 74}, 047901 (2006).


\bibitem{Koch:2008ia}
V.~Koch,
in {\it Relativistic Heavy Ion Physics}, Landold-B\"ornstein
Volume I/23, edited by R. Stock (Springer, Berlin, 2010).


\bibitem{fluc2} I.~N.~Mishustin, Phys. Rev. Lett. {\bf 82}, 4779 (1999);
Nucl. Phys. A {\bf 681}, 56c (2001).

\bibitem{fluc2a}
H.~Heiselberg and A.~D.~Jackson, Phys. Rev. C {\bf 63}, 064904 (2001).

\bibitem{fluc3} M.~Stephanov, K.~Rajagopal, and E.~Shuryak,
Phys. Rev. Lett. {\bf 81}, 4816 (1998).

\bibitem{fluc3a}
 M.~Stephanov, K.~Rajagopal, and E.~Shuryak, Phys. Rev. D {\bf 60},
114028 (1999).

\bibitem{fluc3b} M.~Stephanov, Acta Phys. Polon. B {\bf 35}, 2939
(2004).

\bibitem{Koch:2005vg}
V.~Koch, A.~Majumder and J.~Randrup,
Phys.\ Rev.\ Lett.\  {\bf 95}, 182301 (2005).

\bibitem{Koch:2005pk}
V.~Koch, A.~Majumder and J.~Randrup,
Phys.\ Rev.\  C {\bf 72}, 064903 (2005).

\bibitem{BrFl}
W. Broniowski and W. Florkowski, Phys. Rev. C {\bf 65}, 024905 (2002).

\bibitem{hauer}
V.P. Konchakovski, M. Hauer, G. Torrieri, M.I. Gorenstein, and E.L. Bratovskaya,
Phys. Rev. C {\bf 79}, 034910 (2009).

\bibitem{KLGB}
V.~P.~Konchakovski, B.~Lungwitz, M.~I.~Gorenstein, and E.~L.~Bratkovskaya,
Phys. Rev. C {\bf 70}, 024906 (2008).

\bibitem{WNM}
A.~Bialas, M.~Bleszynski, and W.~Czyz,
Nucl. Phys. B {\bf 111}, 461 (1976).


\bibitem{Konch:2006}
 V.~P.~Konchakovski, S.~Haussler, M.~I.~Gorenstein,
        E.~L.~Bratkovskaya, M.~Bleicher, and H.~St\"ocker,
  Phys. Rev.  C  {\bf 73}, 034902 (2006).


\bibitem{size}
F. Becattini, J. Manninen, and M. Gazdzicki, Phys. Rev. C {\bf 73},
044905 (2006).

\bibitem{Ga:2009}
M.~Gazdzicki  [NA61/SHINE Collaboration],
J.\ Phys.\ G {\bf 36}, 064039 (2009).

\bibitem{RHIC-SCAN}
 G.~Odyniec  [STAR Collaboration],
J.\ Phys.\ G {\bf 35}, 104164 (2008);

\bibitem{GG:2011}
M.~I.~Gorenstein and M.~Gazdzicki, Phys. Rev. C {\bf 84}, 014904
(2011).


\bibitem{GGP:2013}
M.~Gazdzicki, M.~I.~Gorenstein, and M. Mackowiak-Pawlowska,
Phys. Rev. C {\bf 88}, 024907 (2013).

\bibitem{MG1999}
M. Gazdzicki, Eur. Phys. J. C {\bf 8}, 131 (1999).

\bibitem{GM:1992}
M.~Ga\'zdzicki and St.~Mrowczy\'nski,
Z.\ Phys.\  C {\bf 54}, 127 (1992).


\bibitem{volume}
 M.~I.~Gorenstein, J. Phys. G {\bf 35}, 125102
 (2008).

\bibitem{volume1}
  V.~V.~Begun, M.~Gazdzicki and M.~I.~Gorenstein,
  Phys.\ Rev.\ C {\bf 78}, 024904 (2008).





\bibitem{mixed}
S.~V.~Afanasev {\it et al.}  [NA49 Collaboration],
Phys.\ Rev.\ Lett.\  {\bf 86}, 1965 (2001).

\bibitem{NA61}
  N.~Antoniou {\it et al.}  [NA61/SHINE Collaboration],
  CERN-SPSC-2006-034.


\bibitem{Be:2012}
  V.~V.~Begun, V.~P.~Konchakovski, M.~I.~Gorenstein and E.~Bratkovskaya,
  J. Phys. G {\bf 40}, 045109 (2013).


\bibitem{GR:2013} M.~I.~Gorenstein and M.~Rybczynski,
Phys.~Lett.~B {\bf 730}, 70 (2014).



\bibitem{KG:APP2012}
K.~Grebieszkow,
Acta Phys.\ Polon.\ B {\bf 43}, 1333 (2012).


\bibitem{GorGr:2013}
M. I. Gorenstein and K.~Grebieszkow, Phys.~Rev.~C~{\bf 89}, 034903 (2014).





\bibitem{CP_Phipt}
M.~Stephanov, private communication.

\bibitem{CP_Phipt1}
K.~Grebieszkow [NA49 Collaboration], Nucl.~Phys.~A {\bf 830}, 547c (2009).

\bibitem{CP_FK2004}
Z.~Fodor and S.~Katz, JHEP {\bf 04}, 50 (2004).

\bibitem{ismd2013}
P.~Seyboth [NA49 and NA61/SHINE Collaborations], Proceedings Contribution arXiv:1402.4619;
see also slides at https://atlaswww.hep.anl.gov/ismd13/.


\bibitem{CE}
  V.~V.~Begun, M.~Gazdzicki, M.~I.~Gorenstein, and O.~S.~Zozulya,
  Phys.\ Rev.\  C {\bf 70},  034901 (2004).

\bibitem{CE1}
  V.V. Begun, M.I.~Gorenstein, M.~Hauer, V.P.~Konchakovski, and
  O.S.~Zozulya, Phys. Rev. C {\bf 74}, 044903 (2006).

\bibitem{CE2}
  V.V.~Begun, M.~Gazdzicki, M.I.~Gorenstein,
  M.~Hauer, V.P.~Konchakovski, and B.~Lungwitz,
  Phys. Rev.  C {\bf 76}, 024902 (2007).

\bibitem{CE3}
  M.~I.~Gorenstein and M.~Hauer,
  Phys. Rev. C {\bf 78}, 041902 (2008).

\bibitem{HSD-rev}
V.P.~Konchakovski, M.I.~Gorenstein, E.L.~Bratkovskaya, and
W.~Greiner,
J. Phys. G {\bf 37}, 073101 (2010).

\bibitem{Ga:2011}
M. Gazdzicki, K. Grebieszkow, M. Ma\' ckowiak, and S. Mr\'
owczy\' nski, Phys. Rev. C {\bf 83}, 054907 (2011).

\bibitem{G:2011}
  M.~I. Gorenstein,
 Phys. Rev. C {\bf 84}, 024902 (2011).

\bibitem{RG:2012}
A. Rustamov and M.I. Gorenstein, Phys. Rev. C {\bf 86}, 044906
(2012).





\bibitem{NA49_identity}
T.~Anticic {\it et al.} [NA49 Collaboration], arXiv:1310.3428 (2013).



\bibitem{Mackowiak_CPOD2013}
M.~Mackowiak-Pawlowska [NA49 and NA61/SHINE Collaborations], PoS (CPOD 2013), 048.


\bibitem{Becattini:2009sc}
  F.~Becattini,
  arXiv:0901.3643 [hep-ph].

\bibitem{Florkowski_textbook}
  W.~Florkowski,
  {\it Phenomenology of Ultra-Relativistic Heavy--Ion Collisions}
  World Scientific,
  ISBN: 9814280666,
  436 pages,
  2010.



\end{thebibliography}
\end{document}